\documentclass[12pt]{article}
\pdfoutput=1
\usepackage{jheppub}
\usepackage{epsfig}
\usepackage{amsmath}
\usepackage{amssymb}
\usepackage{amsfonts}
\usepackage{amsxtra}
\usepackage{amsthm}
\usepackage{mathrsfs}
\usepackage{makeidx}
\usepackage{graphics}
\usepackage{dsfont}
\usepackage{mathtools}
\usepackage{graphicx}
\usepackage{subcaption}
\usepackage{placeins}
\usepackage{bm}
\usepackage[capitalise]{cleveref}
\usepackage{empheq}
\usepackage{colortbl}
\usepackage{xcolor}
\usepackage{enumerate}
\usepackage{titlesec}
\usepackage{longtable}
\usepackage{float}
\usepackage{color}
\usepackage{tikz}
\usepackage{xfrac}
\usepackage{footnote}
\usepackage{rotating}
\usepackage{lscape}
\usepackage{makecell}
\usepackage{environ}
\usepackage{tabularx}
\usepackage{subfiles}
\usepackage[export]{adjustbox}
\usepackage{ytableau}
\usepackage{tikz-3dplot}
\usepackage{slashed}
\usepackage{pifont}
\usepackage{multirow}
\usepackage{mdframed}
\usepackage{bbm}
\usepackage{ulem}

\usetikzlibrary{positioning,trees,decorations.pathmorphing,decorations.markings,decorations.pathreplacing,calc,shapes,patterns,arrows,chains,arrows.meta,fit,fadings,decorations.markings,graphs,graphs.standard,quotes}

\titleformat{\paragraph}
{\normalfont\normalsize\bfseries}{\theparagraph}{1em}{}
\titlespacing*{\paragraph}{0pt}{3.25ex plus 1ex minus .2ex}{1.5ex plus .2ex}

\DeclareMathOperator{\U}{U}
\DeclareMathOperator{\SU}{SU}
\DeclareMathOperator{\SO}{SO}

\DeclareMathOperator{\USp}{USp}

\DeclareMathOperator{\rank}{rank}

\newcommand{\coma}{\, , \quad}
\newcommand{\fstop}{\, .}

\def\CC{{\mathbb{C}}}

\theoremstyle{definition}

\catcode`\@=11   
\newdimen\@rotdimen
\newbox\@rotbox  

\def\@vspec#1{\special{ps:#1}}%  passes #1 verbatim to the output
\def\@rotstart#1{\@vspec{gsave currentpoint currentpoint translate
   #1 neg exch neg exch translate}}% #1 can be any origin-fixing transformation
\def\@rotfinish{\@vspec{currentpoint grestore moveto}}% gets back in synch 
\def\@rotr#1{\@rotdimen=\ht#1\advance\@rotdimen by\dp#1%
   \hbox to\@rotdimen{\hskip\ht#1\vbox to\wd#1{\@rotstart{90 rotate}%
   \box#1\vss}\hss}\@rotfinish}
\def\@rotl#1{\@rotdimen=\ht#1\advance\@rotdimen by\dp#1%
   \hbox to\@rotdimen{\vbox to\wd#1{\vskip\wd#1\@rotstart{270 rotate}%
   \box#1\vss}\hss}\@rotfinish}%
\def\@rotu#1{\@rotdimen=\ht#1\advance\@rotdimen by\dp#1%
   \hbox to\wd#1{\hskip\wd#1\vbox to\@rotdimen{\vskip\@rotdimen
   \@rotstart{-1 dup scale}\box#1\vss}\hss}\@rotfinish}%
\def\@rotf#1{\hbox to\wd#1{\hskip\wd#1\@rotstart{-1 1 scale}%
   \box#1\hss}\@rotfinish}%
\def\rotate{\@ifnextchar[{\@rotate}{\@rotate[l\right]}}
\def\@rotate[#1]#2{\setbox\@rotbox=\hbox{#2}\@nameuse{@rot#1}\@rotbox}

\catcode`\@=12
\usetikzlibrary{positioning}
\usetikzlibrary{chains}
\usetikzlibrary{arrows, arrows.meta ,fit,decorations.pathreplacing}
\tikzstyle{every picture}+=[remember picture]
\tikzstyle{na} = [baseline]

\usetikzlibrary{arrows, decorations.markings, calc, fadings, decorations.pathreplacing, patterns, decorations.pathmorphing, positioning}
\tikzset{>={Latex[width=1.5mm,length=1.5mm]}}
\usetikzlibrary{graphs,graphs.standard,quotes}

\def\node#1#2{\overset{#1}{\underset{#2}{{\color{gray} \bullet}}}}

\def\node#1#2{\overset{#1}{\underset{#2}{\circ}}}

\tikzstyle{every picture}+=[remember picture]
\tikzstyle{na} = [baseline=-.5ex]

\newcommand{\eg}{e.g.}

\numberwithin{equation}{section}
\newcommand{\bes}[1]{\begin{equation} \begin{split} #1\end{split} \end{equation}}

\newcommand{\nn}{\nonumber}

\newcommand{\be}{\begin{equation}} \newcommand{\ee}{\end{equation}}
\newcommand{\bea}{\begin{equation} \begin{aligned}} \newcommand{\eea}{\end{aligned} \end{equation}}

\def\tilde{\widetilde}

\def\bar{\overline}

\def\rt2{\sqrt{2}}

\def\Tr{\mathop{\rm Tr}}

\def\CC{{\cal C}}

\def\CH{{\cal H}}

\def\CN{{\cal N}}

\def\CS{{\cal S}}
\def\CT{{\cal T}}

\def\1{{\ds 1}}

\newcommand{\fm}{\mathfrak{m}}
\newcommand{\fn}{\mathfrak{n}}

\def\SO{\mathrm{SO}}

\def\SU{\mathrm{SU}}

\def\fp{\mathfrak{p}}
\def\fr{\mathfrak{r}}

\def\fN{\mathfrak{N}}

\def\repa{\raise4pt\hbox{$\square$}\mkern-14mu\raise-4pt\hbox{$\square$}}
\def\repab{\overline{\raise4pt\hbox{$\square$}\mkern-14mu\raise-4pt\hbox{$\square$}\mkern-1mu}}

\def\smileface{\ensuremath{\hbox{\large$\bigcirc$}\mkern-15mu\raise-1pt\hbox{\scriptsize$\smallsmile$}%
\mkern-10mu\raise4pt\hbox{..}\mkern4mu}}
\def\frownface{\ensuremath{\hbox{\large$\bigcirc$}\mkern-15mu\raise-1pt\hbox{\scriptsize$\smallfrown$}%
\mkern-10mu\raise4pt\hbox{..}\mkern4mu}}

\newcommand{\ba}{\begin{array}}
\newcommand{\ea}{\end{array}}
\newcommand{\bi}{\begin{itemize}}
\newcommand{\ei}{\end{itemize}}
\def\vec#1{\bm{#1}}
\def\bea#1\eea{\allowdisplaybreaks \begin{align}#1\end{align}}
 \newcommand{\ben}{\begin{enumerate}}
\newcommand{\een}{\end{enumerate}}
\newcommand{\bean}{\begin{eqnarray*}}
\newcommand{\eean}{\end{eqnarray*}}
\newcommand{\eref}[1]{(\ref{#1})}

\newcommand{\PE}{\mathop{\rm PE}}

\newcommand{\BC}{\mathbb{C}}

\newcommand{\BZ}{\mathbb{Z}}

\newcommand{\diag}{\mathrm{diag}}

\definecolor{light-gray}{gray}{0.5}

\newcommand{\blue}{\color{blue}}

\newcommand{\red}{\color{red}}

\def\aup#1 {\overset{#1}{\uparrow} \, \overset{\tilde{#1}}{\downarrow}}

\tikzset{snake it/.style={decorate, decoration={snake, amplitude=.4mm, segment length=2mm,
                       post length=0mm,pre length=0mm}}}
                       
 \newcommand{\GCD}{\mathrm{GCD}}

\hypersetup{
	pdftitle={Conformal Manifolds and 3d Mirrors of Argyres-Douglas theories},    % title
	pdfauthor={\textcopyright\ Federico Carta, Simone Giacomelli, Noppadol Mekareeya, Alessandro Mininno},     % author
	pdfsubject={hep-th},   % subject of the document
	pdfcreator={pdfLaTex},   % creator of the document
	pdfproducer={LaTex}, % producer of the document
	pdfkeywords={},
	colorlinks=true,
}

\makeatletter
\newsavebox{\measure@tikzpicture}
\NewEnviron{scaletikzpicturetowidth}[1]{%
  \def\tikz@width{#1}%
  \begin{lrbox}{\measure@tikzpicture}%
  \BODY
  \end{lrbox}%
  \pgfmathparse{#1/\wd\measure@tikzpicture}%
  \BODY
}
\makeatother

\makeatletter
\def\squarecorner#1{
    \pgf@x=\the\wd\pgfnodeparttextbox%
    \pgfmathsetlength\pgf@xc{\pgfkeysvalueof{/pgf/inner xsep}}%
    \advance\pgf@x by 2\pgf@xc%
    \pgfmathsetlength\pgf@xb{\pgfkeysvalueof{/pgf/minimum width}}%
    \ifdim\pgf@x<\pgf@xb%
        \pgf@x=\pgf@xb%
    \fi%
    \pgf@y=\ht\pgfnodeparttextbox%
    \advance\pgf@y by\dp\pgfnodeparttextbox%
    \pgfmathsetlength\pgf@yc{\pgfkeysvalueof{/pgf/inner ysep}}%
    \advance\pgf@y by 2\pgf@yc%
    \pgfmathsetlength\pgf@yb{\pgfkeysvalueof{/pgf/minimum height}}%
    \ifdim\pgf@y<\pgf@yb%
        \pgf@y=\pgf@yb%
    \fi%
    \ifdim\pgf@x<\pgf@y%
        \pgf@x=\pgf@y%
    \else
        \pgf@y=\pgf@x%
    \fi
    \pgf@x=#1.5\pgf@x%
    \advance\pgf@x by.5\wd\pgfnodeparttextbox%
    \pgfmathsetlength\pgf@xa{\pgfkeysvalueof{/pgf/outer xsep}}%
    \advance\pgf@x by#1\pgf@xa%
    \pgf@y=#1.5\pgf@y%
    \advance\pgf@y by-.5\dp\pgfnodeparttextbox%
    \advance\pgf@y by.5\ht\pgfnodeparttextbox%
    \pgfmathsetlength\pgf@ya{\pgfkeysvalueof{/pgf/outer ysep}}%
    \advance\pgf@y by#1\pgf@ya%
}
\makeatother

\pgfdeclareshape{square}{
    \savedanchor\northeast{\squarecorner{}}
    \savedanchor\southwest{\squarecorner{-}}

    \foreach \x in {east,west} \foreach \y in {north,mid,base,south} {
        \inheritanchor[from=rectangle]{\y\space\x}
    }
    \foreach \x in {east,west,north,mid,base,south,center,text} {
        \inheritanchor[from=rectangle]{\x}
    }
    \inheritanchorborder[from=rectangle]
    \inheritbackgroundpath[from=rectangle]
}

\tikzset{stretch/.initial=1}
\newcommand\drawloop[4][]%
   {\draw[shorten <=0pt, shorten >=0pt,#1]
      ($(#2)!\pgfkeysvalueof{/tikz/stretch}!(#2.#3)$)
      let \p1=($(#2.center)!\pgfkeysvalueof{/tikz/stretch}!(#2.north)-(#2)$),
          \n1= {veclen(\x1,\y1)*sin(0.5*(#4-#3))/sin(0.5*(180-#4+#3))}
      in arc [start angle={#3-90}, end angle={#4+90}, radius=\n1]%
   }

\preprint{IFT-UAM/CSIC-21-55}
\title{Conformal Manifolds and 3d Mirrors of Argyres-Douglas theories}
\author[a]{Federico Carta,}
\author[b]{~Simone Giacomelli,}
\author[c,d,e]{~Noppadol Mekareeya,}
\author[f]{\\ and Alessandro Mininno}
\affiliation[a]{Department of Mathematical Sciences,
		Durham University, \\ Durham, DH1 3LE, United Kingdom}
\affiliation[b]{Mathematical Institute, University of Oxford, \\ Woodstock Road, Oxford, OX2 6GG, United Kingdom}
\affiliation[c]{INFN, sezione di Milano-Bicocca, \\ Piazza della Scienza 3, I-20126 Milano, Italy}
\affiliation[d]{Dipartimento di Fisica, Universit\`a di Milano-Bicocca, \\  Piazza della Scienza 3, I-20126 Milano, Italy}
\affiliation[e]{Department of Physics, Faculty of Science, Chulalongkorn University, \\ Phayathai Road,
Pathumwan, Bangkok 10330, Thailand}
\affiliation[f]{Instituto de F\'{\i}sica Te\'orica IFT-UAM/CSIC,\\
		C/ Nicol\'as Cabrera 13-15, 
		Campus de Cantoblanco, 28049 Madrid, Spain}
\emailAdd{federico.carta@durham.ac.uk}
\emailAdd{simone.giacomelli@maths.ox.ac.uk}
\emailAdd{n.mekareeya@gmail.com}
\emailAdd{alessandro.mininno@uam.es}
\abstract{Argyres-Douglas theories constitute an important class of superconformal field theories in $4$d. The main focus of this paper is on two infinite families of such theories, known as $D^b_p(\mathrm{SO}(2N))$ and $(A_m, D_n)$.  We analyze in depth their conformal manifolds. In doing so we encounter several theories of class $\mathcal{S}$ of twisted $A_{\text{odd}}$, twisted $A_{\text{even}}$ and twisted $D$ types associated with a sphere with one twisted irregular puncture and one twisted regular puncture. These models include $D_p(G)$ theories, with $G$ non-simply-laced algebras. A number of new properties of such theories are discussed in detail, along with new SCFTs that arise from partially closing the twisted regular puncture. Moreover, we systematically present the $3$d mirror theories, also known as the magnetic quivers, for the $D^b_p(\mathrm{SO}(2N))$ theories, with $p \geq b$, and the $(A_m, D_n)$ theories, with arbitrary $m$ and $n$. We also discuss the $3$d reduction and mirror theories of certain $D^b_p(\mathrm{SO}(2N))$ theories, with $p < b$, where the former arises from gauging topological symmetries of some $T^\sigma_\rho[\mathrm{SO}(2M)]$ theories that are not manifest in the Lagrangian description of the latter.}
\begin{document}
\maketitle

\newpage

\section{Introduction}

Superconformal field theories (SCFTs) in four spacetime dimensions and with eight supercharges attracted much interest over the past decades, as the conformal symmetry and the large amount of supersymmetry often make it possible to achieve exact results even in the strongly coupled regime.

One interesting class of such SCFTs are those of Argyres-Douglas (AD) type. The defining property of an AD SCFT is the presence of at least one Coulomb branch (CB) operator in the spectrum with fractional (non-integer) conformal dimension. The first examples of AD theories were found soon after the discovery of the Seiberg-Witten solutions~\cite{Seiberg:1994aj,Seiberg:1994rs}. It was then realized that at singular points in the Coulomb branch of gauge theories, mutually non-local BPS dyons become massless~\cite{Argyres:1995jj}. The low-energy dynamics of the system is thus captured by an intrinsically interacting non-Lagrangian theory. This initial set of AD theories was tremendously enlarged over the years (see, for example, \cite{Argyres:1995xn,Eguchi:1996vu,Eguchi:1996ds, Shapere:1999xr,Cecotti:2010fi,Bonelli:2011aa,Xie:2012hs,Wang:2015mra}). It was found that many AD theories can be realized inside the class $\mathcal{S}$ construction~\cite{Gaiotto:2009we,Gaiotto:2009hg}, and also many admit a geometric engineering description \cite{Shapere:1999xr,Cecotti:2010fi} as the IIB superstring compactified on a non-compact singular Calabi-Yau (CY) threefold. Some AD theories admit both descriptions.

Despite being interacting and non-Lagrangian, many AD theories are not isolated. They can admit exactly marginal operators in the spectrum, and therefore possess a conformal manifold. At weakly coupling cusps in the conformal manifold, the AD theory splits as a sector of vector multiplets gauging other matter sectors, which are themselves possibly non-Lagrangian SCFTs.

In this work, we consider AD SCFTs that can be realized either starting from the $6$d $\mathcal{N}=(2,0)$ theory of $D$-type compactified on a sphere with an irregular and a regular (possibly trivial) puncture or from the IIB geometrical engineering. Our focus is double: we systematically study the structure of the conformal manifold of this class of models, as well as derive their $3$d mirror theories \cite{Intriligator:1996ex} upon reduction on a circle. The latter class of theories turns out to be $3$d $\mathcal{N}=4$ gauge theories with the property that their Coulomb branch is identical to the Higgs branch (HB) of the original $4$d theory, and that their HB is identical to the CB of the $3$d reduction of the original $4$d theory on a circle.  Such mirror theories can be regarded as magnetic quivers, in the notation of \cite{Cremonesi:2015lsa, Ferlito:2017xdq, Hanany:2018uhm, Cabrera:2018jxt,Cabrera:2019izd}.

By using the geometric engineering picture, we uncover a complete and systematic pattern for the structure of the conformal manifold. We find that the $(A_n,D_m)$ and $D_p^b(\SO(2N))$ AD theories that admit marginal couplings can be described as an $\SO$ or $\USp$ gauging of matter sectors which can generically be realized with a twisted irregular (both $A$ and $D$ types can appear) plus a twisted regular puncture \cite{Wang:2018gvb}. In one class of cases, however, one of the matter sectors does not admit this type of class $\mathcal{S}$ description, and corresponds to theories of type VII or X in the notation of \cite{Xie:2015rpa}. We provide several checks of this proposal, by matching the IIB geometries, the CB spectrum, and the conformal central charges $a$ and $c$. To provide these checks, we first discuss what is the contribution to the CB spectrum of a partially closed twisted regular puncture in the presence of an irregular puncture. We also discuss how the contribution to the central charges of a regular (possibly twisted) puncture changes in the presence of an irregular puncture, compared to the contribution it would have had in the setup of regular punctures only.

The $3$d mirror theories that we find are quivers of orthosymplectic gauge nodes, as well as an overall $\mathbb{Z}_2$ quotient and a free sector. The presence of the free sector is explained by the fact that, at a generic point of the HB of the original $4$d theory, the low-energy effective theory involves a sector of strongly coupled AD theories each of which does not have a HB.\footnote{Note, however, that their CB is non-trivial.} In the same way as in \cite{Giacomelli:2020ryy}, we refer to such a sector as the non-Higgsable SCFTs. These theories dimensionally reduce to free twisted hypermultiplets \cite{Closset:2020afy,Nanopoulos:2010bv,Giacomelli:2020ryy}, which then correspond to the free hypermultiplets sector under the mirror map.  

Contrary to $D_p^b(\SU(N))$ theories, whose reduction to $3$d always leads to a Lagrangian theory \cite{Closset:2020afy,Giacomelli:2020ryy}, we do not have such a description for the class of theories studied in the present paper. As a result, the quiver description of each mirror theory in this paper is obtained by guesswork, based on information about the CB dimension, the value of $24(c-a)$ and the number of mass parameters of the original $4$d theory in question. Such conjectural quivers are then subject to several stringent tests, including matching of the HB and CB Hilbert series with those of the known theories in various special cases, as well as the Maruyoshi-Song (MS) flow \cite{Maruyoshi:2016tqk} of the $D^b_p(\SO(2N))$ theory. The latter deserves a more detailed discussion here, since it puts highly non-trivial constraints on the theory, as demonstrated in \cite{Giacomelli:2020ryy} for the $D^b_p(\SU(N))$ theory. Roughly speaking, this is a renormalization group (RG) flow from the $D_p(\SO(2N))$ theory to the $(A_{p-1},D_N)$ theory \cite{Giacomelli:2017ckh}, triggered by vacuum expectation values (VEVs) of certain flipping fields. This flow can be realized at the level of the $3$d mirror theory, in particular, it is possible to determine how the flipping fields turn into each component of the mirror theory. In this way, we can establish the constraints on the latter.  To achieve this, we need to extend the Flip-Flip duality of \cite{Aprile:2018oau} to the $T[\SO(2N)]$ theory, introduced in \cite{Gaiotto:2008ak}. Hence, we propose the Flip-Flip duality for $T[\SO(2N)]$ as a by-product of this study. We also comment on the possibility that the sector of free hypermultiplets could admit a discrete gauging if the defect group of the AD theory is nontrivial.

We find that the $3$d mirrors of $D_p^b(\SO(2N))$ theories organize themselves in five different qualitatively distinct classes, depending on whether $p$ is larger or smaller than $b$, depending on if $b=2N-2$ or $b=N$, and finally depending on whether $\GCD(b,p)$ is even or odd.\footnote{For $b=2N-2$ with $p\geq b$ and $\GCD(p,b)$ even and for $b=N$ with $p \geq b$, the mirror theories also depend on the parity of $\frac{b}{\GCD(b,p)}$.} We systematically discuss each of these classes in a different section, leaving the more involved case of $b=N$, $p<b$ for further study.

As a consequence of the analysis of the $3$d mirrors, we find that the dimensional reduction of $E_6$ Minahan-Nemeschansky (MN) theory~\cite{Minahan:1996cj} admits a UV completion as $\USp(4)$ with $5$ flavors plus a gauging of the $\U(1)$ topological symmetry. We argue that such topological symmetry is present from the existence of monopole operators of conformal dimension $1$. This provides an ADHM-like description of the moduli space of one $E_6$ instanton on $\BC^2$.

The paper is organized as follows. In Section \ref{sec:sec2} we review the properties of class $\mathcal{S}$ theories on the sphere with one irregular and one regular puncture. In Section \ref{sec:confManDtype} we discuss the structure of the conformal manifold of $D_p^b(\SO(2N))$ and $(A_n,D_n)$ theories. In Section \ref{sec:MS} we collect various facts that we use in the following sections in order to derive the $3$d mirrors. We discuss the Flip-Flip duality for $T[\SO(2N)]$, we review the supersymmetry enhancing MS flows for $D_p^b(\SO(2N))$, and we discuss the equivalence among some $3$d $\mathcal{N}=4$ Abelian quiver theories, where in one side hypermultiplets have charge one and on the other have charge two. In~\cref{sec:D2N-2SO2NGCDodd,sec:generalGCDeven,sec:DNpSO2NpgrN,sec:D2N-2SO2Npls2N-2} we discuss explicit examples of the $3$d mirror theories, as well as non-Higgsable SCFTs. In Appendix \ref{sec:HilbSer} we provide details of the HS computation supporting the claims in the latter four sections. In Appendix \ref{app:Featureshypesing} we discuss the number of marginal operators of the theories of Xie-Yau type VII. In Appendix \ref{app:nonHiggsableSCFTs} we provide tables listing the rank and $24(c-a)$ for all the non-higgsable $(G,G')$ theories, up to $n=100$ when either $G$ or $G'$ is $A_n, D_n$, computed using the program given in~\cite{Carta:2020plx}. We further conjecture that such tables can be read as a prediction for the non-vanishing genus-zero Gopakumar-Vafa invariants~\cite{Gopakumar:1998ii,Gopakumar:1998jq,Gopakumar:1998ki} of numerous non-toric Calabi-Yau manifolds.

For the sake of the readers, we summarize the $4$d theories studied in this paper together with their definitions in Table~\ref{tab:studiedtheories}.
\begin{table}[!htp]
	\centering
\renewcommand{\arraystretch}{1.25}
\begin{tabular}{c|c}
Theory & Definition \\
\hline
$D^b_p(\SO(2N))$ & \eref{defeq} and \eref{defeq2} \\
$(A_{p-1}, D_N)$  and $\SO(2N)^N[p]$ & \eref{defeq3} and \eref{defeq4}\\
$D^b_p(\USp(2N))$ & \eref{defeq5} and \eref{defeq6}\\
$D_p(\SO(2N+1))$ & \eref{defeq7} \\
$D_p\left(\USp'(2N)\right)$ & \eref{defeq8} \\
Hypersurface singularities of types VII and X & \eref{closeDpUSp2Nm2} and \eref{closeDpUSp2Nm2-2}\\
\end{tabular}
\caption{Summary of the $4$d theories studied in this paper with the reference to their definitions.}
\label{tab:studiedtheories}
\end{table}

\section{Class $\mathcal{S}$ theories with irregular punctures} 
\label{sec:sec2}

In this section, we will review the properties of four dimensional superconformal theories obtained by compactifying $6$d $\mathcal{N}=(2,0)$ theories on a sphere with one irregular and one regular puncture, which are the main focus of the present work. The case of untwisted punctures of type A has already been discussed in detail in \cite{Giacomelli:2020ryy} and here we will mainly consider the D case. As we will see, in the description of the conformal manifold of these models, the superconformal theories engineered by twisted irregular punctures (both of type A and D) introduced in \cite{Wang:2018gvb} play a key r\^ole, and we will therefore discuss these as well in detail.

\subsection{Review of $D_p^b(\SO(2N))$ theories}

As was discussed in \cite{Wang:2015mra}, there are two families of irregular punctures for $\mathcal{N}=(2,0)$ theories of type $D_N$. When combined with a full regular puncture, one family leads to the $D_p(\SO(2N))$ theories introduced in \cite{Cecotti:2013lda} and the other leads to the $D_p^{N}(\SO(2N))$ models discussed in detail in \cite{Giacomelli:2017ckh}. We will often refer to both families as $D_p^b(\SO(2N))$, with the parameter $b$ equal to $2N-2$ or $N$ respectively (and $p \geq 1$ in both cases). For our analysis it will be convenient to use also an alternative geometric realization of these theories as compactifications of Type IIB string theory on local Calabi-Yau threefolds. These are described by hypersurface singularities in $\mathbb{C}^3\times \mathbb{C}^*$ of the form 
\be\label{defeq}F(u,x,y,z)= u^2+x^{N-1}+xy^2+z^p=0\,;\quad \Omega=\frac{dudxdydz}{zdF}\coma\ee 
for $b=2N-2$ and 
\be\label{defeq2}F(u,x,y,z)= u^2+x^{N-1}+xy^2+yz^p=0\,;\quad \Omega=\frac{dudxdydz}{zdF}\fstop\ee 
for $b=N$. The $\mathbb{C}^*$ variable is $z$ and $\Omega$ denotes the holomorphic three-form. 

Both families enjoy a $\SO(2N)$ global symmetry (manifest in the class $\mathcal{S}$ realization of the theories since it is carried by the full puncture) which is sometimes enhanced for certain values of $p$ and $b$. Determining the rank of the global symmetry can be done by counting the number of deformation parameters of unit dimension. The result of this counting (see \cite{Giacomelli:2017ckh}) is %$N$ plus
\be\label{masscount1}
\begin{dcases}
  N+\GCD(N,p) &\text{for $\frac{N}{\GCD(N,p)}$ odd}\\
  N & \text{otherwise}
\end{dcases}\ee
when $b=N$, whereas for $b=2N-2$ the result is %$N$ plus 
\be\label{masscount2} \begin{dcases}
  N+\frac{\GCD(2N-2,p)}{2}+1 &\text{for $\frac{2N-2}{\GCD(2N-2,p)}$ odd}\\
  N+1 &\text{when both $p$ and $\frac{2N-2}{\GCD(2N-2,p)}$ are even}\\
  N &\text{when $p$ is odd}
\end{dcases}\ee
The rank of the theory (dimension of the Coulomb branch) can be obtained using the formula 
\be 2r+\text{rk}(G_F)=\frac{p\left(2N^2-2N\right)}{b}\,,\ee 
where $r$ indeed denotes the rank of the theory and $\text{rk}(G_F)$ is the rank of the global symmetry, which we already know how to determine. The VEVs of Coulomb branch operators correspond to deformation parameters for the hypersurface singularities of dimension larger than one. 

Finally, we know how to compute the 't Hooft anomalies of these models. The flavor central charge of the $\SO(2N)$ global symmetry is \be\label{Fcc}k_{\SO(2N)}=4N-4-\frac{2b}{p}\,,\ee 
and the a and c central charges can be determined as follows. The combination $2a-c$ is given by the Shapere-Tachikawa relation \cite{Shapere:2008zf}: 
\be\label{STformula} 2a-c=\frac{1}{4}\sum_i\left(2\Delta_i-1\right)\,,\ee 
where the sum runs over Coulomb branch operators and $\Delta_i$ denotes their scaling dimension. All the models discussed in this paper satisfy this relation.\footnote{This relation does not apply to theories with a discrete gauging, whenever the group being gauged acts nontrivially on Coulomb branch operators. See \cite{Aharony:2016kai}.} The central charge c is given instead by the formula 
\be\label{cDp}c=\frac{2N^2-N}{12b}(2pN-2p-b)-\frac{\text{rk}(G_F)-N}{12}\fstop\ee

For $p>b$ we can completely remove the $\SO(2N)$ global symmetry by turning on a principal nilpotent VEV for the associated moment map,\footnote{For $p\leq b$ the chiral ring relations of the theory prevents us from turning on a principal nilpotent VEV.} an operation which is usually referred to as closure of the puncture in the class $\mathcal{S}$ literature and results in removing the regular puncture completely. Upon closure of the puncture we find another class of theories, associated to a sphere with an irregular puncture only, which we denote as $\SO(2N)^b[p-b]$. When $b=2N-2$ we will always refer to these models as $(A_{p-b-1},D_N)$ theories. In addition, these models admit a description in Type IIB string theory in terms of threefold singularities very similar to \eqref{defeq} and \eqref{defeq2}. The main difference is that the hypersurface singularity is defined in $\mathbb{C}^4$ rather than $\mathbb{C}^3\times\mathbb{C}^*$ and the holomorphic three-form is different. More precisely, for $(A_{p-1}, D_N)$ theories 
\be\label{defeq3}F(u,x,y,z)= u^2+x^{N-1}+xy^2+z^p=0\,;\quad \Omega=\frac{dudxdydz}{dF}\,,\ee 
and for $\SO(2N)^N[p]$ models 
\be\label{defeq4}F(u,x,y,z)= u^2+x^{N-1}+xy^2+yz^p=0\,;\quad \Omega=\frac{dudxdydz}{dF}\,.\ee

\subsection{AD theories from twisted punctures}

In the study of the conformal manifold of $\SO(2N)^b[p]$ theories, we will come across AD models defined by a sphere with twisted punctures, one irregular and one regular. As was discussed in \cite{Wang:2018gvb} for all $6$d theories of type $A$ or $D$ there are two families of irregular $\mathbb{Z}_2$ twisted punctures, leading to two infinite sets of superconformal theories once the $6$d parent theory is given. In this section we will discuss their properties, focusing on the families of models relevant for the analysis of the conformal manifolds of $\SO(2N)^b[p]$ theories: both types of irregular punctures for twisted $D_N$ and only one type for twisted $A_{\text{odd}}$ and twisted $A_{\text{even}}$. We will first discuss the properties of theories with a full regular puncture following \cite{Wang:2018gvb} and then move to the analysis of theories with a generic regular puncture in Section \ref{partclosed}. This latter part is new.

\subsubsection{Theories with a full regular puncture}

We now review the properties of AD theories defined by a sphere with one irregular and one full regular punctures with a $\mathbb{Z}_2$ twist, focusing on the families relevant for the analysis of the conformal manifold of $\SO(2N)^b[p]$ theories. These models can still be described by hypersurface singularities in Type IIB as in~\cref{defeq,defeq2} with the defining equation describing a ADE singularity fibered over the $z$ plane. As it is well known, the versal deformations of the ADE singularity are in one-to-one correspondence with the Casimir invariants of the underlying Lie algebra. 

In the case of theories with twisted punctures, the ADE singularity is that of the parent $6$d theory and the deformations associated with Casimir invariants which transform non-trivially under the action of the $\mathbb{Z}_2$ outer automorphism are all proportional to half-integer powers of $z$. This accounts for the monodromy associated with the twist line. 

\subsubsection*{Twisted $D_{N+1}$ theories}
Let us start with the twisted $D_{N+1}$ theories. As we have mentioned, there are two families of irregular punctures. One family is engineered in Type IIB by the hypersurface singularity \be\label{defeq5} F(u,x,y,z)=u^2+x^N+xy^2+z^p=0\,;\quad \Omega=\frac{dudxdydz}{zdF}\,.\ee We will call the resulting theories $D^{2N}_p(\USp(2N))$,\footnote{Regarding the nomenclature for the $D^b_p(\USp(2N))$ theory, we choose $b$ in such a way that the scaling dimension of $z$ is $\Delta[z]=b/p$.} or simply $D_p(\USp(2N))$. The CB spectrum can be described as follows \cite[3rd row, Table 14]{Wang:2018gvb}:\footnote{According to \cite[Table 14]{Wang:2018gvb}, the values of $\Delta[z]$ are needed to compute the CB spectrum. In order to reproduce correctly the results in this article, $k'$ or $2k'$ in the expressions for $\Delta[z]$ in \cite[Table 1]{Wang:2018gvb} must be replaced by $p-b_t$, with $b_t$ given by \cite[Table 4]{Wang:2018gvb}.}
\begin{equation}
    \begin{dcases}
     N+1-\frac{2k+1}{2}\frac{2N}{p} & k\geq 0\\
     2j-k\frac{2N}{p} & j=1,\ldots,N\coma k\geq 1\coma
    \end{dcases}
\end{equation}
where $k$ is restricted to integer values such that the resulting scaling dimensions are strictly larger than one. The parameters of dimension $1$ correspond to mass parameters, and we denote their number as $f$.
The $\USp(2N)$ flavor symmetry\footnote{This $\USp(2N)$ symmetry is denoted by $C^{\text{\sout{$anom$}}}_N$ in \cite{Wang:2018gvb} due to the absence of the Witten anomaly associated with such a $\USp(2N)$ symmetry.} of the full puncture has central charge\footnote{We remark that the normalization of the flavor central charges adopted in this paper is {\it twice} of that adopted in \cite{Wang:2018gvb}.} \cite[(4.2)]{Wang:2018gvb}
\be k_{\USp}=2N\left(1-\frac{1}{p}\right)+2\,.\ee 
Regarding the a and c central charges, the combination $2a-c$ is still captured by the Shapere-Tachikawa formula \eqref{STformula} and the c central charge is given by the relation \cite[(4.4)]{Wang:2018gvb}
\be c=\frac{k_{\USp}(2N^2+N)}{12(2N+2-k_{\USp})}-\frac{f}{12}\,.\ee 

The second family of irregular punctures leads to a class of SCFTs which we call $D^{N+1}_p(\USp(2N))$ and are described in Type IIB by the hypersurface singularity 
\be\label{defeq6} F(u,x,y,z)=u^2+x^N+xy^2+yz^p=0\,;\quad \Omega=\frac{dudxdydz}{zdF}\,,\ee 
where $p$ is a half-integer. The CB spectrum is given by the formula \cite[3rd row, Table 14]{Wang:2018gvb}
\begin{equation}
    \begin{dcases}
      N+1-\frac{2k+1}{2}\frac{N+1}{p} & k\geq 0\\
     2j-k\frac{N+1}{p} & j=1,\ldots,N\coma k\geq 1\coma
    \end{dcases}
\end{equation}
with $k$ integer. Again, the range of $k$ is restricted by the constraint that the scaling dimensions are strictly larger than one. We denote again the number of mass parameters of dimension $1$ as $f$. The $\USp(2N)$ flavor central charge is (see \cite[(4.2)]{Wang:2018gvb})
\be \tilde{k}_{\USp}=2N+2-\frac{N+1}{p}\ee 
and the $c$ central charge is given by the expression \cite[(4.4)]{Wang:2018gvb}
\be c=\frac{\tilde{k}_{\USp}(2N^2+N)}{12(2N+2-\tilde{k}_{\USp})}-\frac{f}{12}\,.\ee 
The combination $2a-c$ is again given by the Shapere-Tachikawa relation.

\subsubsection*{Twisted $A_{\text{odd}}$ theories}
In the case of twisted $A_{\text{odd}}$ theories we consider only one class of irregular punctures, whose geometric realization in Type IIB is given by the hypersurface singularity 
\be\label{defeq7} F(u,v,x,z)=uv+x^{2N}+z^p=0\,;\quad \Omega=\frac{dudvdxdz}{zdF}\,.\ee 
In the following we call them $D_p(\SO(2N+1))$ where $p$ is a positive integer. The CB spectrum is given by the formula \cite[2rd row, Table 14]{Wang:2018gvb}:\footnote{Minors typos in \cite{Wang:2018gvb} have been corrected here.}
\begin{equation}
    \begin{dcases}
      2j+1-\frac{2k+1}{2}\frac{2N}{p} & j=1,\ldots,N-1\coma k\geq 0\\
     2j-k\frac{2N}{p} & j=1,\ldots,N\coma k\geq 1\coma
    \end{dcases}
\end{equation}
where again, the range of $k$ is constrained by the requirement that the CB operators have dimensions larger than one. The flavor central charge of the $\SO(2N+1)$ symmetry is (see \cite[(4.2)]{Wang:2018gvb})
\be k_{\SO}=2N\left(1-\frac{1}{p}\right)+2N-2\,.\ee 
The central charges a and c are given by the Shapere-Tachikawa relation and by the formula  \cite[(4.4)]{Wang:2018gvb}
\be c=\frac{k_{\SO}(2N^2+N)}{12(4N-2-k_{\SO})}-\frac{f}{12}\,,\ee 
where $f$ is again the number of mass parameters of dimension $1$.

\subsubsection*{Twisted $A_{\text{even}}$ theories}
Finally, for twisted $A_{\text{even}}$ trinions the geometric realization is given (for the class of irregular punctures we are interested in) by the hypersurface singularity
\be\label{defeq8} F(u,v,x,z)=uv+x^{2N+1}+z^p=0\,;\quad \Omega=\frac{dudvdxdz}{zdF}\ee 
and we call the resulting theories $D_p\left(\USp'(2N)\right)$. The value of $p$ in this case is half-integer. The CB spectrum is given by \cite[1st row, Table 14]{Wang:2018gvb}
\begin{equation}
    \begin{dcases}
      2j+1-\frac{2k+1}{2}\frac{2N+1}{p} & j=1,\ldots,N\coma k\geq 0\\
     2j-k\frac{2N+1}{p} & j=1,\ldots,N\coma k\geq 1\fstop
    \end{dcases}
\end{equation}
Again $k$ is bounded above for consistency with the unitarity bound. The $\USp'(2N)$ flavor\footnote{This $\USp'(2N)$ symmetry is denoted by $C^{\text{anom}}_N$ in \cite{Wang:2018gvb} due to the presence of the Witten anomaly \cite{Witten:1982fp}, as pointed out by \cite{Tachikawa:2018rgw}.} central charge is (see \cite[(4.2)]{Wang:2018gvb})
\be k_{\USp'}=(2N+1)\left(1-\frac{1}{2p}\right)+1\,.\ee 
The formula for the c central charge is, similarly to the previous cases \cite[(4.4)]{Wang:2018gvb}, 
\be c=\frac{k_{\USp'}(2N^2+N)}{12(2N+2-k_{\USp'})}-\frac{f}{12}\,.\ee 
As usual $f$ denotes the number of mass parameters.

\subsubsection{Closing the twisted regular puncture}\label{partclosed} 

In the rest of the paper, we will also need to consider the case of punctures labeled by partitions of the form $\left[a,1^b\right]$, with $b$ odd for twisted $A_{\text{odd}}$ and $b$ even in the other cases. Since for this class of punctures the global symmetry has always embedding index one in the global symmetry of the full puncture, we can compute the flavor central charge by starting from the central charge of the theory with a full puncture, which we know, and subtracting the contribution of Goldstone multiplets, whose fermions always have charge $-1$ under $\U(1)_r$. These are always organized into $a-1$ fundamentals of the global symmetry, which is $\SO(b)$ in the twisted $A_{\text{odd}}$ case and $\USp(b)$ for twisted $A_{\text{even}}$ and twisted $D_N$. The contribution of Goldstone multiplets to the flavor central charge is $2a-2$ when the symmetry is $\SO(b)$ and $a-1$ when the symmetry is $\USp(b)$.

We would now like to understand what sort of theories we get upon the complete closure of the regular puncture, starting from the models described above. 

\subsubsection*{Twisted $A_{2N-1}$ theories}
Let us start by discussing the twisted $A_{2N-1}$ theories. The regular puncture is labeled by a B-partition of $2N+1$ and the fully closed (i.e. minimal) puncture corresponds to the partition $[2N+1]$. We need to look at the contribution to the graded Coulomb branch of the regular puncture. In the case of the full puncture $\left[1^{2N+1}\right]$ we have (see \cite{Chacaltana:2012ch})
\be n_k^{\text{full}}=\frac{3k}{2}-\left\lfloor\frac{k}{2}\right\rfloor-1\text{ with } k=2,\dots, 2N\,,\ee 
where $\left\lfloor\frac{k}{2}\right\rfloor$ denotes the greatest integer less than or equal to $\frac{k}{2}$. For the minimal puncture $[2N+1]$ we have instead 
\be n_k^{\text{min}}=\frac{k}{2}-\left\lfloor\frac{k-1}{N}\right\rfloor\text{ with } k=2,\dots, 2N\,.\ee 
The full puncture gives us the $D_p(\SO(2N+1))$ theory, whose CB spectrum was described before. To find the spectrum of the theory with the minimal puncture, we have to remove the operators of the highest dimension for each value of $k$, to account for the difference $n_k^{\text{full}}-n_k^{\text{min}}$. Using this prescription we conclude that upon closure we land on the theory we get by removing the regular puncture from the theory $D_{p}^{p-N}(\SO(2p-2N))$, which is described in Type IIB by the hypersurface 
\bes{\label{IIBSO2pm2N}
F(u,x,y,z)=u^2+x^{p-N-1}+xy^2+yz^{N}=0\,;\quad \Omega=\frac{dudxdydz}{dF}\,.}
This relation is of course valid for $p\geq N+2$. We recognize here the defining equation \eqref{defeq4} of the $\SO(2p-2N)^{p-N}[N]$ theory.

Let us give one example. We consider the case $N=3$ and $p=8$ for definiteness. For $N=3$ the contribution to the CB from full and minimal twisted punctures is 
\begin{equation}
    \begin{dcases}
      \left(n_k^{\text{full}}\right)=\left(1,\frac{5}{2},3,\frac{9}{2},5\right)\\
      \left(n_k^{\text{min}}\right)=\left(1,\frac{3}{2},1,\frac{3}{2},2\right)
    \end{dcases} \Longrightarrow \left(n_k^{\text{full}}-n_k^{\text{min}}\right)=(0,1,2,3,3)\fstop
\end{equation}
We can easily describe the CB spectrum of the $D_8(\SO(7))$ theory using the results of the previous subsection: 
\be\renewcommand{\arraystretch}{1.25}\begin{array}{>{\centering\arraybackslash$} p{0.5cm} <{$}|l} 
\multicolumn{1}{c|}{k} & \multicolumn{1}{c}{\text{CB operators}}\\
\hline
2 & \frac{5}{4}\\
\hline 
3 & {\color{red}\frac{21}{8}},\frac{15}{8},\frac{9}{8}\\ 
\hline
4 & {\color{red}\frac{13}{4},\frac{10}{4}},\frac{7}{4}\\ 
\hline
5 & {\color{red}\frac{37}{8},\frac{31}{8},\frac{25}{8}},\frac{19}{8},\frac{13}{8}\\
\hline
6 & {\color{red}\frac{21}{4},\frac{18}{4},\frac{15}{4}},\frac{12}{4},\frac{9}{4},\frac{6}{4}\end{array}\ee 
We marked in red the operators which should be removed upon closing the regular puncture. It is easy to check that the resulting spectrum is the same as we get from the singularity 
$$F(u,x,y,z)=u^2+x^{4}+xy^2+yz^{3}=0\,;\quad \Omega=\frac{dudxdydz}{dF}\,.$$ 

\subsubsection*{Twisted $D_N$ theories}

Let us now consider the twisted $D_N$ case, namely the $D_p(\USp(2N-2))$ and $D^{N}_p(\USp(2N-2))$ theories. The contribution of the full $\left[1^{2N-2}\right]$ and minimal $[2N-2]$ punctures to the graded Coulomb branch is (see \cite{Chacaltana:2013oka})
\be \left(n_k^{\text{full}}\right)=\left(1,3,\dots,2N-3;\frac{2N-1}{2}\right) \text{ and }  \left(n_k^{\text{min}}\right)=\left(1,1,\dots,1;\frac{1}{2}\right)\,,\ee 
and therefore we find $$\left(n_k^{\text{full}}-n_k^{\text{min}}\right)=(0,2,4,\dots,2N-4;N-1)\,.$$ 
From the spectrum of the $D_p(\USp(2N-2))$ theory described before, we therefore conclude that upon closing the full puncture the CB spectrum becomes identical to that of the theory engineered by the hypersurface singularity 
\be \label{closeDpUSp2Nm2}
F(u,x,y,z)=u^2+x^{p+3-2N}+xy^{N-1}+yz^{2}=0\,;\quad \Omega=\frac{dudxdydz}{dF}\coma\ee  
which corresponds to the hypersurface singularity of type VII in~\cite[Table 2]{Xie:2015rpa}.  We will discuss some properties of these hypersurface singularities in Appendix~\ref{app:Featureshypesing}. 
Using the same method, we can also analyze the other class of twisted $D_N$ theories. Upon closure of the twisted $D_N$ puncture we get a $4$d SCFT described by the hypersurface singularity 
 \be F(u,x,y,z)=u^2+xy^2+yz^{p+\frac{1}{2}-N}+zx^{N-1}=0\,;\quad \Omega=\frac{dudxdydz}{dF}\coma\label{closeDpUSp2Nm2-2}\ee 
 that, instead is the hypersurface singularity of type X in~\cite[Table 2]{Xie:2015rpa}.

\subsubsection*{Twisted $A_{\text{even}}$ theories}

Finally, let us consider twisted $A_{\text{even}}$ theories (or $D_p\left(\USp'(2N)\right)$ models). We will now see that upon higgsing $D_p\left(\USp'(2N)\right)$ flows in the IR to the $\left(A_{2N},D_{\frac{2p+1-2N}{2}}\right)$ Argyres-Douglas theories. In this case the contribution from the regular punctures to the graded Coulomb branch has not been worked out yet and therefore we cannot just quote a result from the existing literature. However, we can bypass this difficulty with a trick. The contribution to the Coulomb branch is a local property of the puncture and therefore we can determine it by considering trinions with regular punctures only. Luckily, some $A_{\text{even}}$ trinions have already been studied in \cite{Beem:2020pry, Beratto:2020wmn} and we can exploit these results. We choose to compare the two trinions 
\be\label{trinions} \left[1^{2N+1}\right]\,, \left[2N\right]_t \,, \left[2N\right]_t \text{ and } \left[1^{2N+1}\right]\,, \left[1^{2N}\right]_t \,, \left[2N\right]_t \,, \ee 
where the subscript $t$ denotes the twisted puncture.  Note that the difference is one twisted puncture: minimal in the first case and full in the second. The result we are looking for simply follows by comparing the CB spectra of the two theories. The first is known to describe two copies of $D_2(\SU(2N+1))$ and therefore the CB spectrum is  $\frac{3}{2},\frac{3}{2},\frac{5}{2},\frac{5}{2},\dots,\frac{2N+1}{2},\frac{2N+1}{2}$. The spectrum of the second trinion has not been studied, but it is known that by replacing the full untwisted puncture $\left[1^{2N+1}\right]$ with the puncture $[N+1,N]$ we find one copy of $D_2(\SU(2N+1))$. Using now the known contribution to the graded CB from untwisted punctures, we find that the spectrum of the second trinion in \eqref{trinions} contains, besides the CB operators of $D_2(\SU(2N+1))$, operators of dimension $$3,4,5,5,6,6,7,7,7,\dots,\underbrace{2N+1,\ldots,2N+1}_{N}\,.$$ 
Overall, we conclude that for twisted $A_{even}$ punctures 
\be\label{closure} n_k^{\text{full}}-n_k^{\text{min}}=\left(0,0,1,1,\dots,\left\lfloor\frac{k}{2}\right\rfloor-1,\left\lfloor\frac{k}{2}\right\rfloor-1,\ldots,N-1,N-1\right)\coma\ee 
for $k=2,\ldots,2N+1$.
Furthermore, the presence of CB operators of fractional dimension in the first trinion of \eqref{trinions} suggests the presence of an a-type constraint for each odd value of $k$.\footnote{An a-type constraint means that one of the coefficients appearing in the deformation of the Type IIB singularity, or equivalently one of the coefficients which corresponds to a leading pole for the Hitchin field at a puncture in the class $\mathcal{S}$ setup, is the square of a more elementary gauge invariant parameter and we should regard the latter as a generator of the Coulomb branch. See \cite{Chacaltana:2012zy} for a general discussion about such constraints.} In practice this means that for $k$ odd, besides removing $\frac{k-3}{2}$ operators, we should also trade a dimension $k$ operator for an operator of dimension $\frac{k}{2}$. 

Let us consider one example in detail. We consider the case $N=2$ and $p=\frac{11}{2}$ (remember that $p$ is half-integer). According to \eqref{closure} we should remove one operator of dimension 4 and one of dimension 5. Furthermore, for $k=3,5$ we should also implement the a-constraint and therefore divide by two the scaling dimension of the parameter corresponding to the leading pole. The CB spectrum of the theory with a full twisted regular puncture is 
 \be\renewcommand{\arraystretch}{1.25}\begin{array}{>{\centering\arraybackslash$} p{0.5cm} <{$}|l} 
\multicolumn{1}{c|}{k} & \multicolumn{1}{c}{\text{CB operators}}\\
\hline
2 & \frac{12}{11}\\
\hline 
3 & {\color{blue}\frac{28}{11}},\frac{18}{11}\\ 
\hline
4 & {\color{red}\frac{34}{11}},\frac{24}{11},\frac{14}{11}\\ 
\hline
5 & {\color{red}\frac{50}{11}},{\color{blue}\frac{40}{11}},\frac{30}{11},\frac{20}{11}\end{array}\ee 
where we indicate in red the operators which should be dropped and in blue those whose dimension needs to be halved. The resulting spectrum is 
 \be\renewcommand{\arraystretch}{1.25}\begin{array}{>{\centering\arraybackslash$} p{0.5cm} <{$}|l} 
\multicolumn{1}{c|}{k} & \multicolumn{1}{c}{\text{CB operators}}\\
\hline
2 & \frac{12}{11}\\
\hline 
3 & \frac{18}{11}, {\color{blue}\frac{14}{11}}\\ 
\hline
4 & \frac{24}{11},\frac{14}{11}\\ 
\hline
5 & \frac{30}{11},\frac{20}{11}, {\color{blue}\frac{20}{11}}\end{array}\ee 
which reproduces the spectrum of $(A_4,D_4)$. Repeating this analysis in general, we find a match with $\left(A_{2N},D_{\frac{2p+1-2N}{2}}\right)$ theories, which are described by the hypersurface singularities 
\bes{ \label{IIBAD}
F(u,x,y,z)=u^2+x^{\frac{2p-2N-1}{2}}+xy^{2}+z^{2N+1}=0\,;\quad \Omega=\frac{dudxdydz}{dF}\,.
}

\subsection{Closure of the twisted regular puncture to a generic puncture}
Let us discuss how to generalize the procedure of closing the twisted regular puncture to another puncture, but this time not the minimal one. The technique might become involved when many constraints are present.  Indeed, working directly with the graded CB dimensions may fail when there are a-constraints~\cite{Chacaltana:2012ch} among the leading coefficients $c_{p_k}^{(k)}$ of the $k$-differentials near $z=0$, i.e.
\begin{equation}
    \phi_k(z)=\frac{c_{p_k}^{(k)}}{z^{p_k}}+\ldots
\end{equation}
Constraints usually arise whenever it is possible to express a leading coefficient using more basic gauge-invariant coefficients. In these cases, the pole structure changes. There are two kinds of constraints, following~\cite{Chacaltana:2012ch}: c-constraints, for which the local contribution to $n_k$ from the pole order $p_k$ is reduced by one; and a-constraints, for which the contribution to $n_{2k}$ is reduced by one, while the contribution to $n_k$ is raised by one. 

The presence of a-constraints modifies the graded CB dimensions by increasing the local contributions by one unit, and the previous procedure that worked directly at the level of the graded CB dimensions does not work anymore.

The correct procedure, instead, is to obtain the contribution to the graded CB dimensions of the punctures from the pole structure. Using Eq. (2.13) of~\cite{Chacaltana:2012ch}, it is possible to compute the contribution $n_k$ for a twisted puncture. That equation already keeps into account the constraints, so it is possible to obtain the pole structure using Section 2.4 of~\cite{Chacaltana:2012ch}. We call $p_k^{\text{full}}=n_{k}$ the pole structure from the full twisted puncture,\footnote{Notice that it does not change, since there are no constraints for the full puncture.} and the arriving puncture will have a pole structure $p_{k}^{\text{fin.}}$. In general $p_k^{\text{fin.}}$ is different from $n_{k}^{\text{fin.}}$ due to the constraints. We now subtract the two pole structures, i.e. $\left(p_k^{\text{full}}-p_{k}^{\text{fin.}}\right)$ and these are the number of CB operators that must be subtracted from the original CB spectrum. Only now we can apply the constraints, reducing or increasing the spectrum according to the rules in~\cite{Chacaltana:2012ch}. 

To clarify this point, we are going to make an explicit example. Let us consider $D_5(\SO(9))$ and we want to close the maximal puncture $\left[1^9\right]$ to $\left[3,1^6\right]$. If we compute the contribution to the graded CB for the two punctures, we get
\begin{equation}
    n_k^{\text{full}}=\left(1,\frac{5}{2},3,\frac{9}{2},5,\frac{13}{2},7\right) \coma n_k^{\text{fin.}}=\left(1,\frac{5}{2},4,\frac{9}{2},5,\frac{11}{2},6\right)\fstop
\end{equation}
The puncture $\left[3,1^6\right]$ has an a-constraint for $k=8$. If we subtract directly the two contributions, we get
\begin{equation}
    \left(n_k^{\text{full}}-n_k^{\text{fin.}}\right)=(0,0,-1,0,0,1,1)\coma
\end{equation}
that means that we should add a CB operator for $k=4$, and the resulting CB spectrum is
\begin{equation}
\renewcommand{\arraystretch}{1.25}\begin{array}{>{\centering\arraybackslash$} p{0.5cm} <{$}|l} 
\multicolumn{1}{c|}{k} & \multicolumn{1}{c}{\text{CB operators}}\\
\hline
2 & \\
\hline 
3 & \frac{11}{5}\\ 
\hline
4 & {\color{blue}{\frac{12}{5}}},\frac{12}{5}\\ 
\hline
5 & \frac{21}{5},\frac{13}{5}\\
\hline 
6 & \frac{22}{5},\frac{14}{5},\frac{6}{5}\\
\hline 
7 & {\color{red}{\frac{31}{5}}},\frac{23}{5},3,\frac{7}{5}\\
\hline 
8 & {\color{red}{\frac{32}{5}}},\frac{24}{5},\frac{16}{5},\frac{8}{5}
\end{array}
\end{equation}
where, in blue we show the operator that has been added at $k=4$.

However, the correct procedure is to work directly with the pole structure. Indeed, we have
\begin{equation}
    p_k^{\text{full}}=\left(1,\frac{5}{2},3,\frac{9}{2},5,\frac{13}{2},7\right)  \coma p_k^{\text{fin.}}=\left(1,\frac{5}{2},3,\frac{9}{2},5,\frac{11}{2},7\right)\fstop
\end{equation}
If we subtract the two contributions, we get
\begin{equation}
    \left(p_k^{\text{full}}-p_k^{\text{fin.}}\right)=(0,0,0,0,0,1,0)\fstop
\end{equation}
We can now subtract from the original spectrum, the operator with dimension $\frac{31}{5}$, but applying the a-constraint for $k=8$ we need to cancel the operator $\frac{32}{5}$ and add an operator with dimension $\frac{16}{5}$ at $k=4$. The final spectrum is 
\begin{equation}
\renewcommand{\arraystretch}{1.25}\begin{array}{>{\centering\arraybackslash$} p{0.5cm} <{$}|l} 
\multicolumn{1}{c|}{k} & \multicolumn{1}{c}{\text{CB operators}}\\
\hline
2 & \\
\hline 
3 & \frac{11}{5}\\ 
\hline
4 & {\color{blue}{\frac{16}{5}}},\frac{12}{5}\\ 
\hline
5 & \frac{21}{5},\frac{13}{5}\\
\hline 
6 & \frac{22}{5},\frac{14}{5},\frac{6}{5}\\
\hline 
7 & {\color{red}{\frac{31}{5}}},\frac{23}{5},3,\frac{7}{5}\\
\hline 
8 & {\color{red}{\frac{32}{5}}},\frac{24}{5},\frac{16}{5},\frac{8}{5}
\end{array}
\end{equation}
The two procedures indeed lead to different results, and it is important to work with the pole structure when we perform the closing of the puncture to avoid ambiguities, and later apply the constraints to obtain the graded CB.

\section{The conformal manifold of $D$ type Argyres-Douglas theories}
\label{sec:confManDtype}

\subsection{The conformal manifold of $D_p(\SO(2N))$} \label{sec:confmfoldDpSO2N}

Let us now analyze the conformal manifold of $D_p(\SO(2N))$ theories. We focus on the case $N>2$ since for $N=2$ the theory is equivalent to two copies of $D_p(\SU(2))=(A_1,D_p)$.
Starting from \eqref{defeq}, we can easily see that the allowed deformations are 
\be\label{deform1} u^2+x^{N-1}+xy^2+z^p+\sum_{i,j}u_{ij}x^iz^j+yP(z)\,,\ee 
where $P(z)$ is a polynomial in $z$. If we set \be\label{gcdparam}m=\GCD(N-1,p)\,;\quad N-1=mn\,;\quad p=mq\,,\ee 
we easily see that all terms of the form $x^{N-1-nk}z^{qk}$ for $1\leq k\leq m-1$ describe marginal deformations. We therefore conclude that the conformal manifold is at least $m-1$ dimensional. There is potentially one extra marginal deformation of the form $yz^k$. Marginality implies that $k=\frac{pN}{2N-2}$, which is possible only if $pN$ is a multiple of $2N-2$ and this in turn requires either $p$ to be a multiple of $2N-2$ or $N$ to be even and $p$ a multiple of $N-1$.
In conclusion, we find that the dimension of the $D_p(\SO(2N))$ conformal manifold is
\bes{ \label{dimcmDpSO2N}
\begin{dcases} 
m &\text{if $Np$ is multiple of $2N-2$}\\ 
m-1 &\text{otherwise.}
\end{dcases}
}
Suppose we turn on a marginal deformation of the form $x^{N'}z^{2p'}$ with $p'$ integer. If we redefine $y\rightarrow z^{p'}y$ and $u\rightarrow z^{p'}u$ and then divide everything by $z^{2p'}$. The resulting geometry is 
$$ F(u,x,y,z)=\frac{x^{N-1}}{z^{2p'}}+\dots+u^2+x^{N'}+xy^2+z^{p-2p'}=0\,;\quad \Omega=\frac{dudxdydz}{zdF}\,,$$ 
which clearly describes, as explained in \cite{Tachikawa:2011yr}, a $\SO(2N'+2)$ gauge theory. We can also deduce that the $\SO(2N'+2)$ vector multiplet is coupled to a $D_{p-2p'}(\SO(2N'+2))$ theory at $z=\infty$ and another superconformal sector at $z=0$. The CB operators of the sector at infinity are obtained by collecting $u_{ij}$ parameters with $j>2p'$ and the terms in $P(z)$ with degree larger than $p'$. If instead $p'$ is half-integer, via the same steps we conclude that the gauge group is $\USp(2N')$ and the sector at $z=\infty$ is a $D_{p-2p'}(\USp(2N'))$, which is engineered by twisted $D_{N'+1}$ punctures. The Coulomb branch sector indeed confirms this interpretation. 

Notice that whenever the conformal manifold has dimension $m$, one of the allowed marginal deformations is $xz^{p-q}$. Furthermore, $p-q$ is necessarily even which means that the sector at infinity is $D_q(\SO(4))$ and the corresponding vectormultiplet is $\SO(4)$, which provides two marginal couplings instead of one. Conversely, whenever we have a $\SO(4)$ gauging the conformal manifold has dimension $m$. We therefore understand the r\^ole of the extra marginal deformation: it provides the second marginal coupling associated with an $\SO(4)$ gauging.

\subsection{The conformal manifold of $(A,D)$ Argyres-Douglas theories} \label{sec:confmfoldAD}

As we have already discussed, the geometric engineering of $(A_{p-1},D_N)$ is given by \eqref{defeq3}: 
\be\label{defeq2A} 
F(u,x,y,z)= u^2+x^{N-1}+xy^2+z^p=0\,;\quad \Omega=\frac{dudxdydz}{dF}\,.
\ee 
The allowed deformations are as in \eqref{deform1}, apart from the fact that the terms proportional to $z^{p-1}$ can be dropped. From this, we conclude that the conformal manifold of $(A_{p-1},D_N)$ has the same dimension as that of $D_p(\SO(2N))$. There are two exceptions to this rule: When $p$ is a divisor of $N-1$ (i.e. $q=1$ in \eqref{gcdparam}) the dimension of the conformal manifold is reduced by $1$ with respect to the  $D_p(\SO(2N))$ theory. We should also subtract $1$ when $p=\frac{2N-2}{N-2}$, since in this case the marginal deformation $yz^k$ of the $D_p(\SO(2N))$ theory is missing (because $k=p-1$). 

Our goal now is to exhibit a weakly coupled cusp for each exactly marginal deformation $x^{N-1-nk}z^{qk}$ with $1\leq k\leq m-1$.\footnote{The number of weakly coupled cusps is, in general, greater than or equal to the number of exactly marginal deformations.  The former is equal to the number of duality frames that the theory in question admits. In this article, we explore only some cusps of the conformal manifold but do not attempt to classify all of them.}

\subsubsection*{The case of $qk\geq 2N-2-2nk$}
Let us start from the case $qk\geq 2N-2-2nk$. The term $x^{N-1-nk}z^{qk}$ then describes an exactly marginal deformation associated with (in a suitable duality frame) a $\SO(2N-2nk)$ or $\USp(2N-2nk-2)$ gauge group depending on whether $qk$ is even or odd respectively. To see this, let us consider the change of variables $x'=xz^2$; $y'z=y$, which brings the defining equation of the hypersurface \eqref{defeq3} to the form 
\be \frac{x'^{N-1}}{z^{2N-2}}+u^2+x'^{N-1-nk}z^{qk+2nk+2-2N}+x'y'^2+z^p=0\,;\quad \Omega=\frac{dudx'dy'dz}{zdF}\,.\ee 
Using now the same change of variables as in the previous section: $y'\rightarrow z^{a+nk+1-N}y'$ and $u\rightarrow z^{a+nk+1-N}u$ (where $a$ is the integer part of $qk/2$) we find that whenever $qk$ is even the geometry becomes 
\be \frac{x'^{N-1}}{z^{qk+2nk}}+u^2+x'^{N-nk-1}+x'y'^2+z^{p+2N-2-2nk-qk}=0\,;\quad \Omega=\frac{dudx'dy'dz}{zdF}\,,\ee 
which manifestly describes a $\SO(2N-2nk)$ gauging with {\it matter sector at $z=\infty$} given by $D_{p+2N-2-2nk-qk}(\SO(2N-2nk))$. 
When instead $qk$ is odd we have a $\USp(2N-2nk-2)$ gauging of $D_{p+2N-2-2nk-qk}(\USp(2N-2nk-2))$. 

Let us now discuss the other sector {\it near $z=0$}. 
For $qk$ even it can be obtained starting from $D_{nk+qk/2}(\SO(qk+1))$ by partially closing the regular puncture. Specifically, the full puncture $\left[1^{qk+1}\right]$ is replaced by $\left[qk+2nk+1-2N,1^{2N-2nk}\right]$. We can notice that the Coulomb branch is consistent with this claim and the beta function of the gauged $\SO(2N-2nk)$ vanishes automatically with this matter content. The above discussion applies to the case $qk=2N-2-2nk$ as well, although with one caveat: in this case the gauge group is $\SO(2N-2nk-1)$.
For $qk$ odd instead we conjecture that the sector at $z=0$ is a higgsing of $D_{nk+qk/2}\left(\USp'(qk-1)\right)$ with regular puncture $\left[qk+2nk+1-2N,1^{2N-2nk-2}\right]$.

\subsubsection*{The case of $qk\leq 2N-2-2nk$}
For $qk\leq 2N-2nk-2$ the sectors at $z=0$ and $z=\infty$ are interchanged with respect to the previous case, in the sense that the sector at $z=\infty$ has a partially closed puncture whereas the matter sector at $z=0$ has a full puncture. Specifically,
for $qk$ even the sector at infinity is a descendant of $D_{p+2N-2nk-qk-2}(\SO(2N-2nk))$, with regular puncture labeled by the partition $\left[2N-2nk-qk-1,1^{qk+1}\right]$. The sector at $z=0$ instead is $D_{nk+qk/2}(\SO(qk+1))$ and the gauge group is $\SO(qk+1)$. Again, the matter content ensures the vanishing of the beta function. 
For $qk$ odd instead we have at $z=0$ a $D_{nk+qk/2}\left(\USp'(qk-1)\right)$ theory. The gauge group is $\USp(qk-1)$ and the SCFT at $z=\infty$ is a higgsing of $D_{p-qk+2N-2nk-2}(\USp(2N-2nk-2))$. The regular puncture is labeled by the partition $\left[2N-2nk-qk-1,1^{qk-1}\right]$. 

\subsubsection*{Summary}
\noindent Let us summarize the results we have found.  The theories we studied above can be described schematically as follows:
\bes{ \label{gluespheres}
\scalebox{1}{
\begin{tikzpicture}[baseline=0, font=\footnotesize]
\tikzstyle{every node}=[minimum size=2.5cm]
\node[draw, circle] (c1) at (-2,0) {};
\node[draw=none] (starleft) at (-3,0) {{\blue \Large $*$}};
\node[draw=none] (starleft) at (-1.2,0) {$P$};
\node[draw, circle] (c2) at (2,0) {};
\node[draw=none] (starleft) at (3,0) {{\red \Large $*$}};
\node[draw=none] (starleft) at (1.2,0) {$F_G$};
\node[draw=none] at (0,0) {$\longleftarrow \,\, G \,\, \longrightarrow$};
\node[draw=none] at (-2,-1.6) {$D_p(G')$};
\node[draw=none] at (2,-1.6) {$D_p(G)$};
\end{tikzpicture}}
}
where each sphere denotes a class $\CS$ theory whose name is indicated below; ${\blue *}$ and ${\red *}$ denote corresponding irregular punctures; $F_G$ denotes the full puncture whose symmetry is $G$; $P$ denotes a regular puncture resulting from partially closing of the full puncture of $G'$, where $P$ is labelled by a partition of the form $[a, 1^b]$ and the associated symmetry is $G$; and $\leftarrow G \rightarrow$ denotes gauging of symmetry $G$.  The details are as follows:
\begin{itemize} 
\item For $qk> 2N-2-2nk$ and $qk$ even 
$$\begin{array}{l} D_{nk+qk/2}(\SO(qk+1))\leftarrow \SO(2N-2nk)\rightarrow D_{p+2N-2-2nk-qk}(\SO(2N-2nk))\\
\downarrow \\
\text{Puncture labeled by $P=\left[qk+2nk+1-2N,1^{2N-2nk}\right]$}\end{array}$$
\item For $qk> 2N-2-2nk$ and $qk$ odd 
$$\begin{array}{l} D_{nk+qk/2}\left(\USp'(qk-1)\right)\leftarrow \USp(2N-2nk-2)\rightarrow D_{p+2N-2-2nk-qk}(\USp(2N-2nk-2))\\
\downarrow \\
\text{Puncture labeled by $P=\left[qk+2nk+1-2N,1^{2N-2nk-2}\right]$}\end{array}$$
\item For $qk\leq 2N-2-2nk$ and $qk$ even 
$$\begin{array}{r} D_{nk+qk/2}(\SO(qk+1))\leftarrow \SO(qk+1)\rightarrow D_{p+2N-2-2nk-qk}(\SO(2N-2nk))\\
\downarrow \\
\text{Puncture labeled by $P=\left[2N-2nk-qk-1,1^{qk+1}\right]$}\end{array}$$
\item For $qk< 2N-2-2nk$ and $qk$ odd 
$$\begin{array}{r} D_{nk+qk/2}\left(\USp'(qk-1)\right)\leftarrow \USp(qk-1)\rightarrow D_{p+2N-2-2nk-qk}(\USp(2N-2nk-2))\\
\downarrow \\
\text{Puncture labeled by $P=\left[2N-2nk-qk-1,1^{qk-1}\right]$}\end{array}$$
\end{itemize}

\subsubsection*{Comments about the Witten $\mathbb{Z}_2$ anomaly} 
As it was pointed out in \cite{Tachikawa:2018rgw}, the full twisted $A_{\text{even}}$ puncture carries a global $\mathbb{Z}_2$ anomaly analogous to the one discussed in \cite{Witten:1982fp} which prevents us from gauging the corresponding $\USp$ symmetry, unless we also couple to the vectormultiplet another matter sector with the same property. In this way, the anomaly cancels and the theory is consistent. This fact potentially implies that the gaugings we have discussed in this section are not consistent whenever a $D_p\left(\USp'(2n)\right)$ matter sector (or a descendant thereof) is involved. This happens in the second and fourth cases we have just discussed, namely, for $qk$ odd. 

We will now argue that the anomaly always cancels and the gauging is consistent for all models at hand. To do this, we exploit the fact that all theories with a $\USp$ global symmetry we have discussed are labeled by a special family of regular punctures labeled by C-partitions of the form $\left[a,1^b\right]$, with both $a$ and $b$ {\it even} and we are gauging the $\USp(b)$ global symmetry. The main point is that, due to the structure of these punctures, we can also construct these theories by starting from the matter sector with a full puncture $\left[1^{a+b}\right]$, gauge a $\USp(b)$ subgroup of the flavor symmetry and then activate the nilpotent VEV for the moment map which partially closes the puncture. In a sense, for this class of punctures the operations of gauging and higgsing do not interfere. 

Exploiting this fact, we start from the matter sector with a full puncture and gauge the $\USp(b)$ global symmetry. If the matter sector involved is of $D_p\left(\USp'(a+b)\right)$ type, in order to get a consistent (not anomalous) theory we should add an odd number of half hypermultiplets in the fundamental of $\USp(b)$. If instead the matter sector is of $D_p^b\left(\USp(a+b)\right)$ type (and therefore belongs to the twisted D family), we can only add an even number of half hypermultiplets. If, once we have gauged, we turn on the nilpotent VEV for the moment map, in the IR we find, besides the matter sector labeled by the puncture $\left[a,1^b\right]$, $a-1$ (which is {\it odd}) Goldstone multiplets in the fundamental of $\USp(b)$ (see Section \ref{partclosed}). Their effect clearly changes the parity of $\USp(b)$ fundamentals for any $a>0$. 

Since the consistency of the theory is not affected by the higgsing, we come to the following conclusion: $D_p\left(\USp'(a+b)\right)$ is affected by the $\mathbb{Z}_2$ global anomaly whereas $D_p^b\left(\USp(a+b)\right)$ is not. If instead we consider the descendants labeled by the puncture $\left[a,1^b\right]$ with $a>0$ the opposite is true, namely we have no anomaly for twisted $A_{\text{even}}$ and we do have it for twisted $D$ theories. Overall, we find that for $qk>2N-2-2nk$ both matter sectors are not anomalous and therefore the gauging is consistent whereas for $qk<2N-2-2nk$ both matter sectors are affected by the $\mathbb{Z}_2$ anomaly and the gauging is still consistent. This confirms that all of our models are anomaly free.

\subsection{Theories with $b=N$}

\subsubsection{The conformal manifold of $D_p^N(\SO(2N))$ theories} 

Let us now discuss marginal deformations of $D_p^N(\SO(2N))$ theories. 
We focus again on the case $N>2$ and the allowed deformations are 
\be\label{deff} u^2+x^{N-1}+xy^2+yz^p+\sum_{i,j}u_{ij}x^iz^j+yP(z)\,,\ee 
where $P(z)$ is a polynomial in $z$. Since the defining equation \eqref{defeq2} imposes the ring relation $xy=z^p$, we can trade all terms proportional to $xy$ for monomials of the form $x^mz^n$, possibly with $n>p$. This is convenient for identifying marginal deformations since the polynomial $P(z)$ cannot provide any. We can therefore focus on the parameters $u_{ij}$ in \eqref{deff}. If we set \be\label{gcdparam2}m=\GCD(N,p)\,;\quad N=mn\,;\quad p=mq\,,\ee 
we can easily see that a parameter $u_{ij}$ describes a marginal deformation (and has therefore dimension 0) only if the following equation holds: 
\be\label{margcond}2N-2-2i=j\frac{n}{q}\,.\ee 
We should therefore find all the solutions of \eqref{margcond} with $i$ and $j$ both integer. We can immediately notice that, since $n$ and $q$ are coprime, $j$ has to be a multiple of $q$. Once this constraint is satisfied, if $n$ is even there is always exactly one value of $i$ which solves \eqref{margcond}, at least as long as $j<2mq$. We therefore find $2m-1$ marginal deformations. If instead $n$ is odd, we need to impose the constraint that $j$ is an even multiple of $q$ smaller than $2mq$, leading to $m-1$ solutions. We therefore come to the conclusion that the dimension of the $D_p^N(\SO(2N))$ conformal manifold is 
\bes{ \label{dimcmDNpSO2N}
\begin{dcases} 
2m-1 &\text{if $n$ is even}\\
m-1 &\text{otherwise.}
\end{dcases}
}
The same counting also applies to the $\SO(2N)^N[p-N]$ theory, since in this case closing the regular puncture does not affect at all, the dimension of the conformal manifold.

Let us now briefly discuss the weakly coupled cusps. If the marginal deformation is parametrized by $u_{ij}$, we have as in the $b=2N-2$ case a $\SO(2i+2)$ or $\USp(2i)$ gauge group depending on whether $j$ is even or odd respectively. For $j$ even the matter sector at $z=\infty$ is $D_{p-j/2}^{i+1}(\SO(2i+2))$. This can be seen by performing the redefinition $u\rightarrow uz^{j/2}$ and $y\rightarrow yz^{j/2}$ in \eqref{deff} and then dividing the defining equation by $z^j$. It is easy to see that if we keep only deformations proportional to positive powers of $z$ in the resulting expression, we recover the spectrum of the $D_{p-j/2}^{i+1}(\SO(2i+2))$ theory. 

For $j$ odd, we find instead a new class of twisted AD theories. The matter sector at $z=\infty$ in this case involves a twisted $D_n$ model with $b_t=2n$ in the notation of \cite{Wang:2018gvb}, which is specified in Type IIB by the hypersurface singularity \eqref{defeq6} \be\label{Dtwist2} u^2+x^{n-1}+xy^2+yz^p=0\,,\ee with $p$ half integer. These are indeed identified with the $D_p^n(\USp(2n-2))$ theory. Specifically, for $j$ odd the matter sector at infinity is $D_{p-j/2}^{i+1}(\USp(2i))$. Again, this can be derived by considering the change of variables $u\rightarrow uz^{j/2}$ and $y\rightarrow yz^{j/2}$ in \eqref{deff} and dividing the defining equation by $z^j$.

\subsubsection{The conformal manifold of $\SO(2N)^N[p]$ theories} \label{sec:conformalSO2NNp}

Let us analyze weakly coupled cusps in the conformal manifold of $\SO(2N)^N[p]$ theories. Consider again the marginal deformation $u_{ij}x^iz^j$. We have to consider the two cases $j\leq 2i$ and $j> 2i$. 

\subsubsection*{The case of $j\leq 2i$}
For $j\leq 2i$ the matter sector at $z=\infty$ can be described 
for $j$ even as a higgsing of $D_{p+i+1-j/2}^{i+1}(\SO(2i+2))$. More precisely, the higgsing is implemented by partially closing the regular puncture from $\left[1^{2i+2}\right]$ to $\left[2i+1-j,1^{j+1}\right]$. The matter sector at $z=0$ is instead $D_{N+j/2-i-1}(\SO(j+1))$. The two sectors are coupled via an $\SO(j+1)$ gauging. 
For $j$ odd the matter sector at $z=\infty$ is a higgsing of $D_{p+i+1-j/2}^{i+1}(\USp(2i))$, with regular puncture labeled by the partition $\left[2i-j+1,1^{j-1}\right]$. The sector at $z=0$ is instead $D_{N+j/2-i-1}\left(\USp'(j-1)\right)$ and the gauging in this case is $\USp(j-1)$. It can be easily checked both for $j$ even and odd that with this matter content the beta function vanishes. To see this, one should use the relation 
\be\label{margcond2} 2p=\frac{Nj}{N-i-1}\,,\ee 
which is implied by the marginality of the term $u_{ij}x^iz^j$.

\subsubsection*{The case of $j> 2i$}
For $j> 2i$ the r\^oles of the two matter sectors are interchanged as in the $b=2N-2$ case. The matter sector at $z=\infty$ can be described 
for $j$ even as the $D_{p+i+1-j/2}^{i+1}(\SO(2i+2))$ theory and the matter sector at $z=0$ is instead a higgsing of $D_{N-i-1+j/2}(\SO(j+1))$. The relevant twisted $A_{j-1}$ puncture is labeled by the partition $\left[j-2i-1,1^{2i+2}\right]$. The two sectors are coupled via a $\SO(2i+2)$ gauging. 
For $j$ odd instead the matter sector at $z=\infty$ is the $D_{p+i+1-j/2}^{i+1}(\USp(2i))$ theory and the sector at $z=0$ is a higgsing of $D_{N+j/2-i-1}\left(\USp'(j-1)\right)$. The corresponding twisted $A_{j-1}$ puncture is labeled by the partition $\left[j-2i-1,1^{2i}\right]$. The gauging in this case is $\USp(2i)$. Again, it can be easily checked using \eqref{margcond} that, both for $j$ even and odd, with this matter content the beta function vanishes. 

\subsubsection*{Summary}
The results can be represented schematically by \eref{gluespheres}, where the details are summarized as follows:
\bi
\item For $j\leq 2i$ and $j$ even, 
\bes{
& D_{p+i+1-j/2}^{i+1}(\SO(2i+2)) \,\, \leftarrow \SO(j+1) \rightarrow \,\, D_{N+j/2-i-1}(\SO(j+1)) \\
& \downarrow \\
&\text{Puncture labeled by $P=\left[2i+1-j,1^{j+1}\right]$}
}
\item For $j\leq 2i$ and $j$ odd,
\bes{
& D_{p+i+1-j/2}^{i+1}(\USp(2i)) \,\, \leftarrow \USp(j-1) \rightarrow \,\, D_{N+j/2-i-1}\left(\USp'(j-1)\right) \\
& \downarrow \\
&\text{Puncture labeled by $P=\left[2i-j+1,1^{j-1}\right]$}
}
\item For $j> 2i$ and $j$ even, 
\bes{
D_{p+i+1-j/2}^{i+1}(\SO(2i+2)) \,\, \leftarrow \SO(2i+2) \rightarrow \,\, & D_{N-i-1+j/2}(\SO(j+1)) \\
& \downarrow \\
&\text{Puncture labeled} \\
&\text{by $P=\left[j-2i-1,1^{2i+2}\right]$}
}
\item For $j> 2i$ and $j$ odd, 
\bes{
D_{p+i+1-j/2}^{i+1}(\USp(2i)) \,\, \leftarrow \USp(2i) \rightarrow \,\, &D_{N+j/2-i-1}\left(\USp'(j-1)\right) \\
& \downarrow \\
&\text{Puncture labeled} \\
& \text{by $P=\left[j-2i-1,1^{2i}\right]$}
}
\ei 
The cancellation of the global $\mathbb{Z}_2$ anomaly of \cite{Tachikawa:2018rgw} works as in Section \ref{sec:confmfoldAD} and therefore we do not repeat the argument.

\section{RG flows with supersymmetry enhancement, $3$d mirrors and Abelian quivers in $3$d}
\label{sec:MS}

In this section, we collect several results and observations which are instrumental for constructing the $3$d mirrors of $D_p^b(\SO(2N))$ theories. We will first discuss a class of RG flows exhibiting supersymmetry enhancement and how they affect the $3$d mirror theory. In passing, we will propose a new duality for $T[\SO(2N)]$. We will then discuss our conventions for the $3$d quivers and discuss several nontrivial equivalences among Abelian theories in $3$d which we need to compare our findings with the results of \cite{Giacomelli:2020ryy}.

\subsection{The Maruyoshi-Song flow for $D_p^b(\SO(2N))$ theories}  

As we have explained, $D_p(\SO(2N))$ theory with $p>2N-2$ flows to $(A_{p-2N+1},D_N)$ upon closure of the regular puncture, or equivalently by giving to the $\SO(2N)$ moment map a principal nilpotent VEV which completely breaks the global symmetry. As was pointed out in \cite{Giacomelli:2017ckh}, we can also flow to $(A_{p-1},D_N)$ theory with a different procedure, introduced in \cite{Agarwal:2014rua, Maruyoshi:2016tqk, Maruyoshi:2016aim, Agarwal:2016pjo, Agarwal:2017roi}, which we refer to as MS flow: We couple to the $\SO(2N)$ moment map a chiral multiplet (flipping field) in the adjoint of the global symmetry and we turn on a principal nilpotent VEV for this field. As we flow to the IR, $N$ Coulomb branch operators of $D_p(\SO(2N))$ hit the unitarity bound and decouple from the theory. 

We can analyze this RG flow at the level of the $3$d mirror theory, as it was done for $D_p(\SU(N))$ theories in \cite{Giacomelli:2020ryy} (see also \cite{Benvenuti:2017lle, Benvenuti:2017kud, Benvenuti:2017bpg}). In the unitary case, it was argued that the introduction of the flipping field can be implemented by flipping the HB moment map of the $T[\SU(N)]$ tail and the nilpotent VEV is introduced simply by removing the tail. Furthermore, the decoupling of the $N$ CB operators hitting the unitarity bound can be implemented by removing from the theory the Cartan components of the flipping field. Here we will follow the same procedure in the $\SO$ case. As we will see, this will provide a powerful constraint on the structure of the $3$d mirror theory. 

\subsubsection*{Flip-Flip duality for $T[\SO(2N)]$} 

Let us now discuss a $3$d duality for $T[\SO(2N)]$ which will play a key r\^ole in understanding the effect of the MS flow on the $3$d mirror of $D_p^b(\SO(2N))$. This duality was introduced for $T[\SU(N)]$ in \cite{Aprile:2018oau} (see also \cite{Hwang:2020wpd}) and states that the theory is infrared equivalent to a variant of the model, in which both the HB and CB moment maps are flipped. We therefore introduce by hand two chiral multiplets in the adjoint of $\SU(N)$ $M_C$ and $M_H$ and couple them via superpotential terms to the CB and HB moment maps $\mu_{C,H}$ respectively. The duality states that the moment maps of $T[\SU(N)]$ are mapped to the corresponding flipping fields in the dual theory: 
\be\label{flip}\begin{array}{rll}
T[\SU(N)] && \text{Flip-Flip $T[\SU(N)]$}\\ 
\mathcal{W}_{\mathcal{N}=4} && \mathcal{W}=\mathcal{W}_{\mathcal{N}=4}+\Tr(M_C\mu_C)+\Tr(M_H\mu_H)\\
\mu_C & \leftrightarrow& M_C\\
\mu_H & \leftrightarrow& M_H\\
\end{array}\ee 
We claim that this duality also applies to $T[\SO(2N)]$ and we would now like to provide a stringy argument based on the Hanany-Witten brane realization of the theory \cite{Cremonesi:2014uva}. It would be interesting to also find a field theoretic derivation of this statement. We can engineer $T[\SO(2N)]$ in Type IIB on the worldvolume of $N$ D3 branes parallel to an O$3^-$ plane and suspended between half D5 branes on one side and half NS5 on the other. There is exactly one D3 brane ending on each 5-brane. The branes are oriented as follows: 
\be\label{branes}
\renewcommand{\arraystretch}{1.25}
\begin{array}{c|cccccccccc}
\text{Branes} & 0&1&2&3&4&5&6&7&8&9\\
\hline 
\text{NS5}& \times&\times&\times&\times&\times&\times&&&&\\
\text{D5}& \times&\times&\times&&&&&\times&\times&\times \\ 
\text{D3}& \times&\times&\times&&&&\times&&& \\
\text{O3}^-& \times&\times&\times&&&&\times&&& 
\end{array}\ee
The theory is known to be self-mirror and this is reflected in the brane system being invariant under S-duality. 

We now exploit an observation made in \cite{Giacomelli:2017vgk} (see also \cite{Xie:2013gma}) that rotating the D5 branes into $\text{D5}'$ extending along directions 012457 has the effect of flipping the HB moment map. If we now perform S-duality we send $\text{D5}'$ branes to $\text{NS5}'$, reaching a configuration in which, starting from the $T[\SO(2N)]$ brane system, we rotate NS5 branes until they extend along directions 012389. On the other hand, we also know that S-duality implements mirror symmetry and since the brane system with rotated $\text{D5}'$ describes a flipping of the HB moment map, we conclude that rotating the NS5 branes has the effect of flipping the CB moment map. 

Now, starting from the brane system in \eqref{branes}, let us rotate both D5 and NS5 branes. According to what we have said so far, this is expected to describe the Flip-Flip $T[\SO(2N)]$ theory. However, it is easy to see from \eqref{branes} that this operation simply amounts to a rotation of the brane system (we are just interchanging the planes $4-5$ and $8-9$) and therefore we recover the brane system describing $T[\SO(2N)]$. We therefore conclude that $T[\SO(2N)]$ is equivalent to its flipped-flipped version. 

The reason this is relevant for describing the effect of the MS flow at the level of the $3$d mirror is the same as in the unitary case discussed in \cite{Giacomelli:2020ryy}. The first step is to flip the $\SO(2N)$ moment map of the $4$d theory, which maps (at least for $p>b$) to the Coulomb branch moment map of the $T[\SO(2N)]$ tail in the $3$d mirror theory. Since a flipping of the CB moment map is hard to describe, it is convenient to use the Flip-Flip duality to map this operation to a flipping of the HB moment map of the $T[\SO(2N)]$ tail, which is described by an ordinary superpotential interaction (we will denote the flipping field of the HB moment map in the $3$d mirror by $M$). This is therefore the duality frame we will focus on. The last step is to notice that the flipping field in $4$d maps, after the Flip-Flip duality, to the CB moment map in the $3$d mirror and therefore the effect of the nilpotent VEV is simply to remove the $T[\SO(2N)]$ tail from the quiver. 

Finally, we have to take into account the fact that $N$ CB operators in $4$d hit the unitarity bound and decouple. To implement this decoupling in $3$d, we should flip the corresponding operators in $3$d \cite{Benvenuti:2017lle}. These are HB generators in the $3$d mirror. In \cite{Giacomelli:2020ryy} it was argued that, after the Flip-Flip duality, this operation amounts to flipping all Cartan components of $M$, therefore removing them from the spectrum. Here we will assume the same is true in the $\SO$ case we are interested in.

\subsubsection*{MS flow and $3$d mirrors}

Let us start by describing the effect of the flipping operation on the $3$d mirror theory. For $p>2N-2$ the $3$d mirror always involves a $T[\SO(2N)]$ tail coupled to a collection of $\SO(2)$ nodes. Let us label the Abelian nodes as $\SO(2)_i$ and denote the multiplicity of the $\USp(2N-2)\times \SO(2)_i$ bifundamental as $n_i$. We also allow the presence of $\USp(2N-2)$ fundamental hypermultiplets and denote their number as $F$. Indeed, we have the constraint $$\sum_in_i+F=N$$ and the vector representation of $\SO(2N)$ decomposes as $${\bf 2N}\rightarrow \sum_i n_i{\bf 2}_i+{\bf 2F}\,.$$ 
We can similarly work out the decomposition of the adjoint of $\SO(2N)$, which is the representation in which the flipping field transforms. This will tell us how the MS flow affects the quiver. We will now describe the decomposition in detail. 
\begin{itemize}
\item We find an adjoint of $\SO(2F)$ and, since this symmetry is ungauged in the $3$d mirror, the corresponding components of the flipping field will become $F(F-1)$ free hypermultiplets when we remove the tail. This counting takes into account the fact that the Cartan components of the flipping field have been removed from the spectrum. This is done to account for the CB operators in $4$d which hit the unitarity bound and decouple.
\item We also get bifundamentals of $\SO(2n_i)\times \SO(2F)$, which become $n_iF$ hypermultiplets which transform as doublets of $\SO(2)_i$ upon removal of the $T[\SO(2N)]$ tail.  We shall refer to these as $2n_i F$ hypermultiplets carrying charge $1$ under $\U(1)_i \cong \SO(2)_i$.
\item The components in the bifundamental of $\SO(2n_i)\times \SO(2n_j)$ with $i\neq j$ become $n_in_j$ half hypermultiplets in the $({\bf 2}_i,{\bf 2}_j)$ of $\SO(2)_i\times \SO(2)_j$. 
\item Finally, we have the adjoint of $\SO(2n_i)$ for every $i$. Once the $T[\SO(2N)]$ tail is removed, this provides $n_i(n_i-1)$ hypermultiplets (always taking into account the fact that the Cartan components have been removed from the spectrum). It turns out that half of them become free hypermultiplets, uncharged under the gauge group of the theory, whereas the other half provide $\frac{n_i(n_i-1)}{2}$ flavors of $\SO(2)_i$.
\end{itemize} 
The last statement about the $\SO(2n_i)$ adjoint requires some further explanations. Out of the $\SO(2n_i)$ global symmetry, we are gauging a $\SO(2)$ subgroup which is generated by the diagonal combination of the Cartan generators in $\SO(2n_i)$. Said differently, the $\SO(2)$ action is described by a $2n_i\times 2n_i$ matrix of the form $R_{\theta}\otimes I_{n_i}$, where $I_{n_i}$ is the $n_i\times n_i$ identity matrix and 
\be R_{\theta}=\left(\begin{array}{cc}\cos\theta & -\sin\theta \\ \sin\theta & \cos\theta\end{array}\right)\,.\ee  
The components of the flipping field transforming in the adjoint of $\SO(2n_i)$ can now be conveniently organized into $\frac{n_i(n_i-1)}{2}$ $2\times 2$ matrices $M_{i}^{J}$ (with $J=1,\dots, \frac{n_i(n_i-1)}{2}$) and the gauged $\SO(2)$ acts on each one of these matrices as 
\be\label{trans2x2}M_i^{J}\longrightarrow R_{\theta}M_i^J R_{-\theta}\,.\ee 
It is now convenient to rewrite each $M_i^J$ as a linear combination of the Pauli matrices $\sigma_{1,2,3}$ and the $2\times 2$ identity matrix. From \eqref{trans2x2} it is easy to see that the components proportional to the identity matrix and to $\sigma_2$ (the antisymmetric Pauli matrix) are invariant under $\SO(2)$ and together they provide a free hypermultiplet upon implementing the MS flow. The other two components, namely, those that are proportional to $\sigma_1$ and $\sigma_3$,  become instead a flavor of $\SO(2)_i$ as we have claimed before. Since 
\bes{
R_\theta (\sigma_1 \pm i \sigma_3) R_{-\theta} = e^{\pm 2 i \theta} (\sigma_1 \pm i \sigma_3)~,
}
we shall refer to the above linear combinations of the latter two components as the hypermultiplets carrying charge $2$ under $\U(1)_i \cong \SO(2)_i$.

Overall, we find the following result: Starting from $D_p(\SO(2N))$, upon closure of the regular $\SO(2N)$ puncture, we get an Abelian quiver with bifundamentals and flavors plus a collection of free hypermultiplets. If instead we consider the MS flow, which amounts to increasing $p$ by $2N-2$, i.e.
\bes{ \label{deltap}
\delta p = 2N-2\,,
}
we find a similar quiver with the same number of Abelian nodes, where the number of free hypermultiplets is increased by 
\be\label{deltaHfree}
\delta H_{\text{free}}=\sum_i\frac{n_i(n_i-1)}{2}+F(F-1)\,.\ee 
The number of bifundamental half-hypermultiplets increases by $n_in_j$ and the number of flavors at the $i$-th Abelian node increases as follows: we get $2n_iF$ hypermultiplets with charge 1 and $\frac{n_i(n_i-1)}{2}$ hypermultiplets with charge 2. This procedure can be easily generalized to the $D^N_p(\SO(2N))$ theory simply by taking 
\bes{
\delta p = N~.
}
We use these facts to constrain the mirror theories discussed in the next sections.

\subsection{Notations} \label{sec:notation}
To describe the mirror theory for $D_p(\SO(2N))$ in the subsequent sections, we adopt the following notations for the quiver diagrams.  

The $R$ copies of half-hypermultiplets in the representation $[\mathbf{2N-2}; \mathbf{2}]$ of the gauge group $\USp(2N-2)\times SO(2)$ are denoted by
\bes{ \label{rededge}
\USp(2N-2) \begin{tikzpicture}[baseline] \draw[draw,solid,red,thick] (0,0.1)--(1,0.1) node[midway, above] {\red \scriptsize $R$}; \end{tikzpicture} \SO(2)\fstop
}
It gives rise to an $\SU(R)$ flavor symmetry.  To make the Cartan elements of $\SU(R)$ manifest, we should interpret \eref{rededge} as denoting the half-hypermultiplets in the following representation of $\USp(2N-2) \times \U(1) \times \SU(R)$, where $\U(1)\cong \SO(2)$,
\bes{\label{reprededge}
[\mathbf{2N-2}; +1; \mathbf{\bar{R}}] \oplus [\mathbf{2N-2}; -1; \mathbf{R}]~. 
}

The $F$ flavors of hypermultiplets carrying charge $2$ under $\U(1) \cong \SO(2)$ are denoted by
\bes{ \label{wiggleline}
\SO(2)  \begin{tikzpicture}[baseline] \draw[draw,solid,black,snake it] (0,0.1)--(1,0.1) node[midway, above] {}; \end{tikzpicture} [F]_2\coma
}
where the wiggle line and subscript 2 emphasize the charge $2$ under the $\U(1)$ gauge group.  This gives rise to an $\SU(F)$ flavor symmetry.  In other words, \eref{wiggleline} denotes the chiral multiplets in the following representation of $\U(1) \times \SU(F)$:
\bes{
[+2; \mathbf{\bar{F}}] \oplus [-2; \mathbf{F}]~.
}

An edge connecting two $\SO(2)$ gauge nodes with multiplicity $M$ is denoted by  
\bes{ \label{blueedge}
\SO(2) \begin{tikzpicture}[baseline] \draw[draw,solid,blue,thick] (0,0.1)--(1,0.1) node[midway, above] {\blue \scriptsize $M$}; \end{tikzpicture} \SO(2)\fstop
}
This represents $M$ copies of half-hypermultiplets in the representation $[\mathbf{2}; \mathbf{2}]$ of the gauge group $\SO(2) \times \SO(2)$. It gives rise to a $\U(M)^2/\U(1)$ flavor symmetry, whose algebra is isomorphic to $\SU(M) \times \SU(M) \times \U(1)$. To make the Cartan elements of the latter manifest, we should interpret \eref{blueedge} as denoting the half-hypermultiplets in the following representation of $\left[\U(1) \times \U(1)\right] \times \SU(M) \times \SU(M) \times \U(1)$, where each of the first two $\U(1)$ factor is isomorphic to each $\SO(2)$ gauge group:
\bes{ \label{repblueedge}
&[+1; +1; \bar{\mathbf{M}}; \mathbf{1}; -1] \oplus [-1; -1; \mathbf{M}; \mathbf{1}; +1 ] \\
&\oplus [+1; -1;  \mathbf{1}; \mathbf{M}; +1]\oplus [-1; +1;  \mathbf{1}; \bar{\mathbf{M}}; -1]~.  
}

To save space, we sometimes use the following abbreviations in the quiver diagrams: $\SO(2N)=D_N$, $\USp(2N)=C_N$ and $\SO(2N+1)=B_N$.  We also denote by $/\BZ_2$ the diagonal $\BZ_2$ quotient of the gauge symmetry. We shall emphasize the latter again in the context.  

\subsection{Correspondences between certain Abelian gauge theories}

In this section, we discuss some correspondences between certain Abelian gauge theories.  On one side, we consider theories with $\SO(2)$ gauge groups with hypermultiplets in vector representations and possibly with those carrying charge two under $\U(1) \cong \SO(2)$. On the other side of the correspondence, the involved theory has $\U(1)$ gauge groups with hypermultiplets carrying charge one.

The first example we would like to present is $\fm$ copies of half-hypermultiplets in the representation $[\mathbf{2}; \mathbf{2}]$ of the gauge group $(\SO(2) \times SO(2))/\BZ^{\diag}_2$, where $\BZ^{\diag}_2$ denotes the diagonal $\BZ_2$ gauging.  As we will see in Section \ref{sec:bdividespSO2}, this is a $3$d mirror for the $(A_{2\fm-1}, D_2)$ theory. We find that this mirror theory is isomorphic to the product of two copies of the SQED with $\fm$ flavors:
%\bes{ \label{SO2SO2factorize}
%\SO(2) \begin{tikzpicture}[baseline] \draw[draw,solid,light-gray,thick] (0,0.1)--(1,0.1) node[midway, above] {\scriptsize $\fm$}; \end{tikzpicture} \SO(2) ~~_{/\BZ_2} \quad \longleftrightarrow \quad \left( 1-[\fm] \right)^2
%}
\bes{ \label{SO2SO2factorize}
	D_1 \begin{tikzpicture}[baseline] \draw[draw,solid,light-gray,thick] (0,0.1)--(1,0.1) node[midway, above] {\scriptsize $\fm$}; \end{tikzpicture} D_1 ~~_{/\BZ_2} \quad \longleftrightarrow \quad \left( 1-[\fm] \right)^2
}
We match the Higgs and Coulomb branch Hilbert series of the theories on the left and right-hand sides in Appendix \ref{app:SO2SO2factorize}.
Since $D_2=A_1 \times A_1$, it is expected that $(A_{2\fm-1}, D_2)=(A_{2\fm-1}, A_1)^{\otimes 2}$. Indeed, the SQED with $\fm$ flavors is the mirror theory for $(A_{2\fm-1}, A_1)= I_{2\fm,2}$ (see e.g. \cite[(4.9)]{Giacomelli:2020ryy}).  

The second example is the following correspondence: 
\bes{ \label{threeSO2nodes}
\scalebox{0.9}{
\begin{tikzpicture}[baseline=0, font=\scriptsize]
\tikzstyle{every node}=[minimum size=0.5cm]
\node[draw, circle] (c2) at (0,1) {$D_1$};
\node[draw, circle] (f1) at (-1.2,0)  {$D_1$}; 
\node[draw, circle] (f2) at (1.2,0)  {$D_1$}; 
\draw[very thick,light-gray] (c2)--(f1) node[midway,left]  {\footnotesize $\fm$};
\draw[very thick, light-gray] (c2)--(f2) node[midway,right]   {\footnotesize $\fm$};
\draw[very thick, light-gray] (f1)--(f2) node[midway,below]  {\footnotesize $\fm$};
\end{tikzpicture}} \qquad /\BZ_2
\qquad \longleftrightarrow
\qquad
\scalebox{0.7}{
\begin{tikzpicture}[baseline=0]
\def\n{4}% how many nodes
\node[draw=none] at (-1.3,1) {\blue $\fm$};
\node[circle,minimum size=3 cm] (b) {};
\foreach\x in{1,...,\n}{
  \node[minimum size=0.75cm,draw,circle] (n-\x) at (b.{360/\n*\x}){1};
}
\foreach\x in{1,...,\n}{
  \foreach\y in{1,...,\n}{
    \ifnum\x=\y\relax\else
      \draw (n-\x) edge[very thick, blue] (n-\y);
    \fi
  }
} 
\end{tikzpicture}}
}
As we will see in Section \ref{sec:bdividespSO2}, the theory on the left is the mirror theory for $(A_{4\fm-1}, D_3)$.  Since $D_3 \cong A_3$, the theory in question is equivalent to $(A_{4\fm-1}, A_3)=I_{4\fm,4}$, and so there is another description of the mirror theory in terms of a complete graph with 4 $\U(1)$ nodes where each edge has multiplicity $\fm$ (see the discussion below (4.12) of \cite{Giacomelli:2020ryy}), which is depicted on the right-hand side.  Note that the latter has an overall $\U(1)$ that decouples. Both theories have an $\SU(\fm)^6 \times \U(1)^3$ flavor symmetry. We match the Higgs and Coulomb branch Hilbert series of the two descriptions in Appendix \ref{app:threeSO2nodes}.

The third example is the following correspondence between two descriptions of the mirror theory for $(A_1, D_{2N})$.  We will discuss this in more detail in Section~\ref{sec:D4N-2pSO4Ngcd2}.
%\bes{ \label{A1D2Nmirr1}
%~[N-1]_2 \begin{tikzpicture}[baseline] \draw[draw,solid,black,snake it] (0,0.1)--(1,0.1) node[midway, above] {}; \end{tikzpicture} \SO(2) \begin{tikzpicture}[baseline] \draw[draw,solid,light-gray,thick] (0,0.1)--(1,0.1) node[midway, above] {\scriptsize $1$}; \end{tikzpicture} \SO(2) \quad_{/\BZ_2}
%\quad \longleftrightarrow \quad
%\begin{tikzpicture}[baseline=0, font=\scriptsize]
%\tikzstyle{every node}=[minimum size=0.5cm]
%\node[draw, circle] (c) at (0,0)  {$1$}; 
%\node[draw, circle] (f1) at (1,1)  {$1$}; 
%\node[draw, circle] (f2) at (1,-1)  {$1$}; 
%\draw[draw, solid] (f1)--(c) node[midway, above] {};
%\draw[draw, solid] (f2)--(c) node[midway, below] {};
%\draw[draw, solid, blue, thick] (f1)--(f2) node[midway,right] {\blue $N-1$}; 
%\end{tikzpicture}
%}
\bes{ \label{A1D2Nmirr1}
	~[N-1]_2 \begin{tikzpicture}[baseline] \draw[draw,solid,black,snake it] (0,0.1)--(1,0.1) node[midway, above] {}; \end{tikzpicture} D_1 \begin{tikzpicture}[baseline] \draw[draw,solid,light-gray,thick] (0,0.1)--(1,0.1) node[midway, above] {\scriptsize $1$}; \end{tikzpicture} D_1 \quad_{/\BZ_2}
	\quad \longleftrightarrow \quad
	\begin{tikzpicture}[baseline=0, font=\scriptsize]
	\tikzstyle{every node}=[minimum size=0.5cm]
	\node[draw, circle] (c) at (0,0)  {$1$}; 
	\node[draw, circle] (f1) at (1,1)  {$1$}; 
	\node[draw, circle] (f2) at (1,-1)  {$1$}; 
	\draw[draw, solid] (f1)--(c) node[midway, above] {};
	\draw[draw, solid] (f2)--(c) node[midway, below] {};
	\draw[draw, solid, blue, thick] (f1)--(f2) node[midway,right] {\blue $N-1$}; 
	\end{tikzpicture}
}
where the blue edge has multiplicity $N-1$. The quiver on the right-hand side was presented in, \eg~ \cite[Section 2.4]{Benvenuti:2017bpg} and \cite[Figure 36(X)]{Dey:2020hfe}. Note that the quiver on the right-hand side has an overall $\U(1)$ that decouples.  We match the Higgs and Coulomb branch Hilbert series of the theories on both sides of \eref{A1D2Nmirr1} in Appendix \ref{app:twonodeswflv}.

\section{$D^{2N-2}_p(\SO(2N))$ with $p\geq 2N-2$ and $\GCD(2N-2,p)$ odd}
\label{sec:D2N-2SO2NGCDodd}
All theories in this class do not have any mass parameters in addition to those associated with the $\SO(2N)$ flavor symmetry.

\subsection{General result: $\GCD(2N-2,p)=2\mu-1$} \label{sec:genresultGCDodd}
We parametrize $p$ in the following way
\bes{
p = (2N-2)+(4\mu-2) \fm-(2\mu-1)~, \quad \fm \in \BZ_{\geq 1}~,
}
where we restrict to $\fm$ such that
\bes{ \label{constrA}
 \GCD(2N-2, (4\mu-2) \fm-(2\mu-1)) =2\mu-1~.
 }
The $3$d reduction gives 
\bes{ \label{reductionGCDgen}
&\text{ the $T[\SO(2N)]$ theory,} \\
&\text{along with $H_{\text{free}} = [(2\mu-1)\fm-\mu]N$ twisted hypermultiplets.}
}
The mirror theory is then
\bes{\label{mirrreductionGCDgen}
&\text{ the $T[\SO(2N)]$ theory,} \\
&\text{along with $H_{\text{free}} = [(2\mu-1)\fm-\mu]N$ free hypermultiplets.}
}

Let us test this proposal along the line of Section \ref{sec:MS}, where $F=N$ and $n_i=0$.  According to \eref{deltap} and \eref{deltaHfree}, we have $\delta \fm = \frac{N-1}{2\mu-1}$ and $\delta H_{\text{free}}= N(N-1)$.  This is in accordance with \eref{mirrreductionGCDgen}, where $\delta H_{\text{free}} = (2\mu-1)N \delta \fm = N(N-1)$.

Upon closing the full puncture, the $D^{2N-2}_p(\SO(2N))$ theory flows upon higgsing to the AD theory $(A_{(4\mu-2)\fm -2\mu},D_{N})$.   Upon decoupling the $T[\SO(2N)]$ theory from \eref{mirrreductionGCDgen}, the mirror theory for the latter is a collection of $[(2\mu-1)\fm-\mu]N$ free hypermultiplets.  We thus claim that the $(A_{(4\mu-2)\fm -2\mu},D_{N})$ theory has no Higgs branch (but it has a non-trivial Coulomb branch of dimension $[(2\mu-1)\fm-\mu]N$), and so it is a non-Higgsable interacting SCFT.

We provide some explicit examples of non-Higgsable SCFTs, together with their values of $24(c-a)$ and their ranks in Appendix \ref{app:nonHiggsableSCFTs}.

\subsubsection*{Smaller non-Higgsable SCFTs}
We remark that the non-Higgsable SCFT in question $\CT$, which is $(A_{(4\mu-2)\fm -2\mu},D_{N})$ in the current example, {\it may} `contain' a smaller non-Higgsable SCFT $\CT'$, in the sense that 
\ben
\item the rank of $\CT'$ is smaller than that of $\CT$,
\item the Coulomb branch spectrum of $\CT'$ is contained in that of $\CT$, and
\item the value of $24(c-a)$ of $\CT'$ is equal to that of $\CT$.
\een

For $\fm=1$, we conjecture that for $\CT= (A_{2\mu-2},D_N)$ with constraint \eref{constrA}, i.e.  $2\mu-1$ divides $N-1$, we have 
\bes{
\CT'=(A_1, A_{\frac{2}{2\mu-1}(N-1)-2})^{\otimes (\mu-1)}
}
whose $24(c-a) = (\mu-1) \frac{N-2\mu}{2N+(2\mu-3)}$ equal to that of $\CT$ and whose rank is equal to $(\mu-1) \left( \frac{N-1}{2\mu-1} -1 \right)$, less than $(\mu-1)N$.

Below we will consider an example of $\CT=(A_2, D_7)$. This is a non-Higgsable SCFT as discussed above; it has $24(c-a)=1/5$ and rank $7$. Nevertheless, according to the discussion below \eref{defeq2A} with $p=3$ and $N=7$, it has a one-dimensional conformal manifold. There is a weakly coupling cusp that contains an $\SO(3)$ gauging of two theories of class $\CS$.  Upon closing the puncture, we obtain the $\CT'=(A_1,A_2)$ theory.  The latter has $24(c-a)=1/5$ and rank $1$. One can also show that $\CT$ has a Coulomb branch operator of dimension $6/5$, which is the Coulomb branch operator of $\CT'$.

Currently, we do not have the full understanding of the relation between the theory $\CT'$ and $\CT$.  Under the assumption that $\CT$ has no Higgs branch, obviously one cannot obtain $\CT'$ from $\CT$ by a Higgs branch flow.  There is a possibility that such an assumption is wrong, namely, $\CT$ may contain a `Higgs branch' of which a generic point contains a collection of hypermultiplets, vector multiplets, and $\CT'$, in such a way that their contributions make $24(c-a)$ fractional and less than $1$.  However, we regard this possibility as unlikely, since the mirror theory of $\CT$ is simply a theory of free hypermultiplets (no vector multiplet); in other words, mirror symmetry does not give any indication of the presence of the aforementioned hypermultiplets at a generic point of the `Higgs branch' of $\CT$.  We leave the detailed study of $\CT'$ and its relation with $\CT$ for future work.

\subsection{Examples of $D^{12}_{15}(\SO(14))$ and $(A_2, D_7)$}
Let us consider the case of $N=7$, $\mu=2$ and $\fm= 1$, i.e.  the $D^{12}_{15}(\SO(14))$ theory.  Upon closing the full $\SO(14)$ puncture, we obtain the $(A_2, D_7)$ theory.  As described towards the end of Section \ref{sec:confmfoldAD}, with $k=2$, $q=1$, $n=2$ and $m=3$, the latter has the following description:
\bes{ \label{A2D7gauge}
(A_2, D_7) \, = \, \left[ D^2_5(\SO(3)) \longleftarrow \SO(3)  \longrightarrow  D^4_5(\SO(6)) \right]
}
The sector at $z=0$ contains $D_5(\SO(3))$ and the $\SO(3)$ gauge group, whereas the sector at $z=\infty$ contains a descendant of the $D_5(\SO(6))$ theory with the puncture $\left[3,1^3\right]$ in the $\SO(6)$ notation.

Closing the $\SO(3)$ puncture (in the same way as described in \cite[Section 3.3]{Giacomelli:2020ryy}), we obtain the $(A_1, A_2)$ theory from $D^2_5(\SO(3))=D^2_5(\SU(2))$, whereas $D^4_5(\SO(6)) = D^4_5(\SU(4))$ becomes the $(A_0, A_3)$ theory, which is trivial.  We thus obtain the $(A_1, A_2)$ theory from the $(A_2,D_7)$ theory, as previously discussed. 

From \eref{A2D7gauge}, we can obtain the mirror theory for $(A_2, D_7)$ as follows.  The mirror for of $D^2_5(\SO(3))=D^2_5(\SU(2))$ is $\SO(2)-[\USp(2)]$, with $1$ free hypermultiplet.  The relevant theory for $D^4_5(\SU(4))=D^4_5(\SO(6))$ after partially closing the $\SO(6)$ puncture is $T_{\left[3,1^3\right]}[\SO(6)]$, whose description is $\SO(2)-\USp(2)-[\SO(6)]$.  After commonly gauging the enhanced $\SU(2)$ topological symmetries of both theory associated with each $\SO(2)$ node, we obtain $[\USp(2)]-[\USp(6)]$ with $1$ free hypermultiplet, i.e.  7 free hypermultiplets in total, in accordance with \eref{reductionGCDgen}.

\subsection{Discrete gaugings and defect groups}
In some cases, the collection of free hypermultiplets indicated in \eref{mirrreductionGCDgen} {\it could} be subject to a discrete gauging.  Let us discuss this issue using an example.  We consider the $D^6_9(\SO(8))$ theory, i.e.  $\mu=2$, $N=4$ and $\fm = 1$.  Upon closing the full $\SO(8)$ puncture we obtain the $(A_2, D_4)$ theory.  This $4$d theory can be realized as 3 copies of the $(A_1, A_3)$ theory gauged by an $\SU(2)$ gauge algebra (see \cite[Figure 1]{Buican:2016arp} and \cite[(4.30)]{Closset:2020scj}).  As can be seen from the latter reference, there are two choices for the corresponding gauge symmetry, namely $\SU(2)$ or $\SU(2)/\BZ_2$.\footnote{Note that the $\SU(2)/\BZ_2$ group here corresponds to $\SO(3)_+$ in the notation of \cite{Aharony:2013hda}. As remarked below \cite[(4.30)]{Closset:2020scj}, there is also a possibility to consider $\SO(3)_-$, i.e.  the choice with a non-trivial discrete $\theta$-angle.  We shall not consider the latter possibility here.}  Upon reducing the $(A_2, D_4)$ theory to $3$d, we have a star-shaped quiver with the $\U(1)$ gauge node at each of its three legs and central node being $\SU(2)$ or $\SU(2)/\BZ_2$ (see also \cite[(4.31)]{Closset:2020scj}).  If the central node is $\SU(2)/\BZ_2$, this is precisely the mirror of the theory of four free hypermultiplets (the $T_2$ theory), in agreement with $H_{\text{free}}$ in \eref{mirrreductionGCDgen}.  On the other hand, if the central node is $\SU(2)$, this is the mirror of the $\BZ_2$ discrete gauging of the $T_2$ theory, i.e.  the mirror of the $\text{O}(1)-[\USp(8)]$ theory.  As explained below \cite[(4.30)]{Closset:2020scj}, the two aforementioned choices originate from the fact that the defect group of the $(A_2, D_4)$ theory is $\BZ_2^2$ (see also \cite[Table 1]{DelZotto:2020esg}), and so there are two versions of the $(A_2, D_4)$ theory, namely that with a $\BZ_2$ electric one-form symmetry and the other with a $\BZ_2$ magnetic one-form symmetry.  In four dimensions, the former corresponds to the choice of the $\SU(2)$ gauge group and the latter corresponds to the choice of the $\SU(2)/\BZ_2$ gauge group.

In general, the defect group is expected to indicate the presence of the one-form global symmetry and can be used to determine whether it is possible to apply a discrete gauging to the set of free hypermultiplets.  

As a consequence, when the defect group is empty, such as in the case of the $(A_{N-1}, A_{k-1})$ theory with $\GCD(N,k)=1$, the $(A_2, D_3) \cong(A_2, A_3)$ theory, and the $(A_2, D_5)$ theory \cite[Table 1]{DelZotto:2020esg}, it is expected that the one-form global symmetry is absent in such non-Higgsable SCFTs.  Upon reduction to $3$d and applying mirror symmetry, one obtains a collection of hypermultiplets, and we do {\it not} expect any discrete gauging for the latter.\footnote{We thank Stefano Cremonesi for asking us the question regarding this issue.}

\section{$D^{2 N -2}_p(\SO(2 N ))$ with $p\geq 2 N -2$ and $\GCD(2 N -2,p)$ even} \label{sec:generalGCDeven}
In this section, we discuss $3$d mirror theories for $D^{2 N -2}_p(\SO(2 N ))$ with $p\geq 2 N -2$ and $\GCD(2 N -2,p)$ even. We first discuss the general results and then provide several explicit examples. 

\subsection{General results}
Let us write
\bes{
\GCD(2 N -2,p)=2 \mu~.
}
There are two possibilities to consider, depending whether $2 N -2$ divided by $2\mu$ is an even number or an odd number.

\subsubsection{$\GCD(2 N -2,p)$ is even and $\frac{2 N -2}{\GCD(2 N -2,p)}$ is even} \label{case1GCDeven}
In this case, we write
\be
2 N = 4\mu \fN+2~.
\ee 
There is one mass parameter in addition to those associated with the $\SO(2 N )$ flavor symmetry.  Let us parametrize $p$ as
\bes{
p= 4\mu \fN+(4\mu \fm -2\mu)~, \quad \mu\in \BZ_{\geq 1}, \,\, \GCD(\fN, 2\fm-1) =1~.
}
The $3$d reduction of the $D^{2 N -2}_p(\SO(2 N ))$ theory in question flows to 
\bes{ \label{eq:HfreeGCDevendivisioneven}
H_{\text{free}} = \mu [ 2\fm ( 2 \fN \mu -1 ) - \fN (2 \mu+1) +1 ]
}
twisted hypermultiplets, together with the $T^{\vec \sigma}_{\vec \rho}[\SO(2\fn)]$ theory such that
\bes{
\begin{array}{ll}
\vec \sigma= \left[(4 \mu \fN+2x-1)^2,1^{4\mu \fN +2}\right]~, &\quad \vec \rho = \left[3^{4\mu \fN},2^{2x}\right]~,\\
2\fn =12 \mu \fN+4x ~, &\quad  x= (2\fm-1)\mu
\end{array}
}
whose quiver description is\footnote{We propose this quiver based on the observation that the mirror theory, namely \eref{GCDevendivisioneven}, has the required properties: (1) there is one topological $\U(1)$ symmetry in addition to the $\SO(2 N )$ enhanced topological symmetry arising from the tail, (2) the HB dimension, taking into account $H_{\text{free}}$, agrees with the CB dimension of the 4d theory, (3) the CB dimension is in expected relation with the value of $24(c-a)$ of the 4d theory after taking into account non-Higgsable theories, and (4) the quiver \eref{GCDevendivisioneven} satisfies the constraints from the Maruyoshi-Song flow.  Upon computing the mirror theory, we arrive at \eref{quivunderbalancedUSp2}. Note that, as a result of this process, the quiver \eref{quivunderbalancedUSp2} contains $\USp(2)$ gauge nodes that are underbalanced.  We currently do not have an interpretation of this fact. We leave it for future work.}
%\bes{ \label{quivunderbalancedUSp2}
%\begin{split}
%&[\SO(4\mu \fN+2)]-\USp(4\mu \fN)-\SO(4\mu \fN)-\USp(4\mu \fN-2)-\SO(4\mu \fN-2)- \\
%&-\USp(4)-\SO(4)-(\USp(2)-\SO(2))^{x}-\USp(2)-[\SO(2)]
%\end{split}
%}
\bes{ \label{quivunderbalancedUSp2}
	\begin{split}
		&[D_{2\mu\fN+1}]-C_{2\mu\fN}-D_{2\mu\fN}-C_{2\mu\fN-1}-D_{2\mu\fN-1}-C_2-D_2-(C_1-D_1)^{x}-C_1-[D_1]
	\end{split}
}
Note that the total number of gauge groups is $2x+1+2(2\mu \fN-1)= p-1$.  

The mirror theory consists of $H_{\text{free}}$ free hypermultiplets, together with the $T_{\vec \sigma}^{\vec \rho}[\SO(2\fn)]$ theory whose quiver description is
%\bes{ \label{GCDevendivisioneven}
%\begin{array}{lll}
%&\SO(2)-&\USp(4\mu \fN)-\SO(4\mu \fN)-\USp(4\mu \fN-2)-\SO(4\mu \fN-2)-\cdots-\USp(2)-\SO(2) \\
%&\,\,\,\, |  & \,\,\,\, | \\
%&\hspace{-0.6cm}  [\USp(2x)]  &[\SO(4\mu \fN)]
%\end{array}
%}
\bes{ \label{GCDevendivisioneven}
	\begin{array}{lll}
		&D_1-&C_{2\mu\fN}-D_{2\mu\fN}-C_{2\mu\fN-1}-D_{2\mu\fN-1}-\cdots-C_1-D_1 \\
		&\,\,\,\, |  & \,\,\,\, | \\
		&[C_{x}]  &[D_{2\mu\fN}]
	\end{array}
}
The mirror theory for the $(A_{2\mu(2\fm-1)-1},D_{2\mu \fN+1})$ theory is therefore SQED with $2x$ flavors, together with $H_{\text{free}}$ hypermultiplets.\footnote{It can be checked that the expression of $H_{\text{free}}$ given by \eref{eq:HfreeGCDevendivisioneven} is consistent with the fact that the Milnor number is equal to twice the rank $r$ of the CB of the 4d theory plus the rank $\mathrm{rk}(G_F)$ of the global symmetry \cite{Cecotti:2010fi}. Suppose that we focus on the 4d $(A_\fm, D_\fn)$ theory (or the $\SO(2N)^b[p]$ theory in general).  The Milnor number is $\fm \fn$ (or can be computed from \cite[(2.9)]{Giacomelli:2017ckh} respectively). By mirror symmetry, $r$ is equal to the dimension of the HB of the 3d mirror theory plus $H_{\text{free}}$, and $\mathrm{rk}(G_F)$ is equal to the number of $\SO(2)$ nodes in the 3d mirror theory.}

Let us test proposal \eref{GCDevendivisioneven} along the line of Section \ref{sec:MS}, where $F=2 \mu \fN$, $n_1=1$ and $n_i=0$ otherwise.  According to \eref{deltap} and \eref{deltaHfree}, we have $\delta p= 4 \mu \delta \fm = 4\mu \fN$, i.e. $\delta \fm = \fN$, and $\delta H_{\text{free}}= 2 \mu \fN(2 \mu \fN-1)$.  This is in accordance with \eref{eq:HfreeGCDevendivisioneven}, where $\delta H_{\text{free}} = 2\mu(2\mu \fN-1) \delta \fm = 2\mu \fN( 2\mu \fN -1)$.  Moreover, as stated below \eref{deltaHfree}, the increment of the number of hypermultiplets carrying charge 1 under $\U(1)$ that is isomorphic to the leftmost $\SO(2)$ is precisely $4\mu \fN = 4 \mu \delta \fm = 2\delta x$, in agreement with \eref{GCDevendivisioneven}. 

%\subsubsection*{The special case of $\fm=1$} For $\fm=1$, the non-Higgsable SCFTs are 
%\bes{
%(A_1, A_{2\fN-2})^{\otimes \mu}~,
%}
%giving rise to $\mu (\fN-1)$ free hypermultiplets as expected.  Moreover, the $3$d mirror of $(A_{2\mu-1},D_{2\mu \fN+1})$ is SQED with $2x=2\mu$ flavors, together with $\mu(\fN-1)$ free hypermultiplets.

\subsubsection{$\GCD(2 N -2,p)$ is even and $\frac{2 N -2}{\GCD(2 N -2,p)}$ is odd} \label{case2GCDeven}
In this case, we write
\bes{2 N = 4\mu \fN-2\mu+2~.}
There are $\mu+1$ mass parameters in addition to those associated with the $\SO(2 N )$ flavor symmetry. We parametrize $p$ as follows: 
\bes{
p = (4\mu \fN-2\mu)+2\mu\fm~, \quad \fm \in \BZ_{\geq 1}~, \GCD(2\fN-1,\fm)=1~.
}
The mirror theory involves 
\bes{ \label{HfreeGCDeven}
H_{\text{free}} = \mu(\fm-1)(\fN-1)
}
free hypermultiplets, together with the following quiver gauge theory: the $T[\SO(2N)]$ tail 
\be
D_1-C_1-D_2-C_2- \cdots- C_{ N -1}
\ee
connected to a complete graph of $\mu+1$ $\SO(2)$ nodes in the following way:\footnote{We find that the total number of hypermultiplets in the complete graph part (not including the tail and connections)
\bes{ \label{totalhypersgraph}
\mu F +\frac{1}{2} \mu(\mu-1)(2M) + \mu (2\fm) = \mu \fm [(2\mu-1)\fN-(\mu-2)]~.}}
\ben
\item Among the $\mu+1$ $\SO(2)$ nodes in the complete graph, $\mu$ of them are connected to the $C_{N -1}$ node in the tail by edges with multiplicity 
\be
R= \frac{2 N -2}{2\mu}=2\fN-1\coma
\ee
and the remaining $\SO(2)$ node is connected to the $C_{N -1}$ node in the tail by an edge with multiplicity $1$.  Note that the total number of flavors of the $C_{N -1}$ node is indeed $\mu R+1 =2\mu \fN -\mu+1$, as required. For convenience, we refer to the edges with multiplicity $R$ as $A_1, A_2, \ldots, A_{\mu}$ and the edge with multiplicity one as $B$.
\item Each $\SO(2)$ gauge node connected by the edges $A_1, \ldots, A_{\mu}$ has 
\be \label{FmNm1}
F=\fm(\fN-1)
\ee
flavors of hypermultiplets with carrying $2$ under $\U(1) \cong \SO(2)$.   The $\SO(2)$ node connected by the edge $B$ has no flavor charged under it.
\item Each edge in the complete graph that connects any two $\SO(2)$ gauge nodes attached to the edges $A_i$ and $A_j$ has multiplicity
\bes{
M= \fm \frac{2 N -2}{2\mu}=\fm R= \fm(2\fN-1)~,
}  
and each edge that connect any two $\SO(2)$ gauge nodes attached to the edges $B$ and $A_i$ multiplicity $\fm$.
\item The gauge symmetry of the theory is in fact
\bes{ \label{Z2quotient}
(C_1 \times D_1 \times C_2 \times D_2 \times \cdots \times C_{ N -1} \times D_1^{\mu+1})/\BZ^{\diag}_2~.
}
where the factors $C_1 \times D_1 \times C_2 \times D_2 \times \cdots \times C_{ N -1}$ come from the tail, the factor $D_1^{\mu+1}$ comes the complete graph and $\BZ^{\diag}_2$ denotes the quotient of the diagonal $\BZ_2$ symmetry (see a detailed discussion in \cite{Bourget:2020xdz}).  As pointed out in \cite{Bourget:2020xdz}, the $\BZ^{\diag}_2$ quotient affects the magnetic fluxes of each gauge factor in such a way that the half-integral values must be taken into account.  We will denote this $\BZ^{\diag}_2$ quotient by $/\BZ_2$ in the subsequent part of the paper.
\een
In general, we also conjecture that the non-Higgsable SCFTs are 
\bes{
(A_{\fm-1}, A_{2\fN-2})^{\otimes \mu}~,
}
giving rise to $H_{\text{free}}$ free hypermultiplets, as required.

The mirror theory for $(A_{2\mu\fm-1},D_{2\mu \fN-\mu+1})$ is therefore the complete graph as described above, together with $H_{\text{free}}$ free hypermultiplets.

Let us test the above procedure of constructing the mirror theory along the line of Section \ref{sec:MS}, where $F=0$, $n_1=1$ and $n_2=n_3=\cdots=n_\mu=R=2\fN-1$ otherwise.  According to \eref{deltap} and \eref{deltaHfree}, we have $\delta p= 2 \mu \delta \fm = 4\mu \fN-2\mu$, i.e. $\delta \fm = 2\fN-1$, and $\delta H_{\text{free}}= \mu \frac{1}{2}(2\fN-1)(2\fN-2) =(\fN-1)(2\fN-1)\mu$.  This is in accordance with \eref{HfreeGCDeven}, where $\delta H_{\text{free}} = \mu (\fN-1) \delta \fm =(\fN-1)(2\fN-1)\mu$.  Moreover, as stated below \eref{deltaHfree}, the increment of the number of hypermultiplets carrying charge 2 under the $\U(1)$ gauge groups, which are isomorphic to the 2nd, 3rd, $\ldots$,  $\mu$-th $\SO(2)$ gauge groups, is precisely $\frac{1}{2}(2\fN-1)(2\fN-2)=(\fN-1)(2\fN-1)=(\fN-1)\delta \fm = \delta F$, in agreement with \eref{FmNm1}.

\subsection{$D^{2 N -2}_p(\SO(2 N ))$ such that $2 N -2$ divides $p$} \label{sec:bdividespSO2}
Suppose that $p= (2  N -2) \fr$, with $\fr \geq 1$.  This is  a special case of Section \ref{case2GCDeven}, with
\bes{
\fN=1~, \quad \mu= N -1~, \quad \fm =\fr-1~.
}  

According to the proposal, the $3$d mirror for the $D^{2  N -2}_{(2  N -2) \fr}(\SO(2  N ))$ theory is described by a complete graph with $ N $ $\SO(2)$ gauge nodes with edge multiplicity $\fm =\fr-1$,\footnote{The total number of the hypermultiplets in the complete graph is $2\times\frac{1}{2} N ( N -1) \times (r-1)= N ( N -1)(\fr-1)$ in agreement with \eref{totalhypersgraph}. \label{foot:totalhyp}} such that each $\SO(2)$ node is connected to $C_{ N -1}$ node in the $T[\SO(2  N )]$ tail with an edge whose multiplicity is 1. There is a $\BZ_2$ quotient of the gauge factors, as indicated in \eref{Z2quotient}. There is no free hypermultiplet in this case, in agreement with \eref{HfreeGCDeven} with $\fN=1$.  

The Higgs branch symmetry of this theory is $[\SU(\fr-1)^2 \times \U(1)]^{\frac{1}{2}  N ( N -1)} \times \U(1)$, where each edge between two $\SO(2)$ nodes with a multiplicity $\fm$ gives rise to the symmetry $[\SU(\fm)^2 \times \U(1)]$ and there is another $\U(1)$ coming from tail.

Upon closing the full puncture, the $D^{2 N -2}_{(2 N -2) \fr}(\SO(2 N ))$ theory flows upon higgsing to the AD theory $\left(A_{(2 N -2)\fm-1},D_{ N }\right)$, with $\fm=\fr-1$.  The $3$d mirror for the latter is a complete graph with $ N $ $\SO(2)$ nodes with edge multiplicity $\fm$.  The gauge symmetry is $\SO(2)^ N /\BZ_2$. There are two interesting special cases to consider:
\bi
\item For the special case of $ N =2$, the theory in question is $(A_{2\fm-1},D_{2})$, with $\fm = \fr-1$. We have discussed this theory and its correspondence with two copies of SQED with $\fm$ flavors in \eref{SO2SO2factorize}.
\item For the special case of $ N =3$, the mirror theory for $(A_{4\fm-1}, D_3)$ was presented in \eref{threeSO2nodes}.  Its correspondence with the mirror theory for $(A_{4\fm-1}, A_3)$, namely the complete graph with 4 $\U(1)$ nodes where each edge has multiplicity $\fm$ was also discussed there.
\ei

\subsection{$D^{2 N -2}_p(\SO(2 N ))$ with $\GCD(2 N -2,p)=2$}
\label{sec:D2N-2SO2Ngcd2}
There are two cases for $D^{2 N -2}_p(\SO(2 N ))$, with $\GCD(2 N -2,p)=2$ and $ N  \geq 2$, to be considered: those are odd after divided by $2$, and those are even after divided by $2$. 
\bi
\item For $2  N -2=2(2\fN)$, i.e.  $2 N =4\fN+2$, there is one mass parameter in addition to those associated with the $\SO(2 N )$ flavor symmetry.
\item For $2  N -2=2(2\fN-1)$, i.e.  $2 N = 4\fN$, there are two mass parameters in addition to those associated with the $\SO(2 N )$ flavor symmetry.
\ei
\subsubsection{$D^{4\fN}_p(\SO(4\fN+2))$ such that $\GCD(4\fN,p)=2$}
We write
\bes{
p = 4\fN+(4\fm-2)~, \quad \fm \in \BZ_{\geq 1}
}
Note that although this parametrization includes all $p$ such that $\GCD(4\fN,p)=2$, it also include those with $\GCD(4\fN,p)\neq 2$, in which case we should exclude those values of $p$ from the following analysis.  In other words, we consider $\fm$ such that 
\bes{
\GCD(2\fN, 2 \fm-1) =1~.
}

The $3$d reduction flows to 
\bes{\label{Hfreegcd2}
H_{\text{free}} = 2 \fm (2 \fN - 1) - (3 \fN - 1) ~\text{twisted hypermultiplets}
}
together with the $T^{\vec \sigma}_{\vec \rho}[\SO(2\fn)]$ theory with
\bes{
\begin{array}{ll}
\vec \sigma= [(4 \fN+2x-1)^2,1^{4 \fN +2}]~, &\quad \vec \rho = [3^{4 \fN},2^{2x}]~,\\
2\fn =12 \fN+4x~, &\quad x= 2\fm-1~.
\end{array}
}
whose quiver description is
%\bes{
%&[\SO(4\fN+2)]-\USp(4\fN)-\SO(4\fN)-\USp(4\fN-2)-\SO(4\fN-2)\\
%&-\cdots- \USp(4)-\SO(4)-(\USp(2)-\SO(2))^{x}-\USp(2)-[\SO(2)]
%}
\bes{
	&[D_{2\fN+1}]-C_{2\fN}-D_{2\fN}-C_{2\fN-1}-D_{2\fN-1}-\cdots- C_2-D_2-(C_1-D_1)^{x}-C_1-[D_1]
}
The $[D_1]$ flavor symmetry corresponds to the fact that all of the $4$d theories in this class has one mass parameter in addition to those associated with the $\SO(4\fN+2)$ flavor symmetry.

The mirror theory consists of 
\bes{
H_{\text{free}} ~\text{free hypermultiplets}
}
and the $T_{\vec \sigma}^{\vec \rho}[\SO(2\fn)]$ theory whose quiver description is
%\bes{
%\begin{array}{lll}
%&\SO(2)-&\USp(4\fN)-\SO(4\fN)-\USp(4\fN-2)-\cdots-\USp(2)-\SO(2) \\
%&\,\,\,\, |  & \,\,\,\, | \\
%&\hspace{-0.5cm}  [\USp(2x)]  &\,\,[\SO(4\fN)]
%\end{array}
%}
\bes{
	\begin{array}{lll}
		&D_1-&C_{2\fN}-D_{2\fN}-C_{2\fN-1}-\cdots-C_1-D_1 \\
		&\,\,\,\, |  & \,\,\,\, | \\
		&\hspace{-0.0cm}  [C_{x}]  &\,\,[D_{2\fN}]
	\end{array}
}

Upon decoupling the tail, we obtain the mirror theory for $(A_{4 \fm-3}, D_{2\fN+1})$ such that $\GCD(2\fN, 2 \fm-1) =1$, namely
%\bes{
%\SO(2)-[\USp(4\fm-2)] 
%}
\bes{
	D_1-[C_{2\fm-1}] 
}
with $H_{\text{free}}$ free hypermultiplets.  In particular, setting $\fm=1$, we propose that the mirror for the $(A_1, D_{2\fN+1})$ theory is the SQED with $2$ flavors with $\fN-1$ free hypermultiplets.

%For $\fm=1$, we propose that the non-Higgsable SCFT is $(A_1, A_{2\fN-2})$, giving rise to $\fN-1$ twisted hypermultiplets upon reduction to $3$d.

For $\fN=1$ (i.e.  the case of $\SO(6)$ whose algebra is isomorphic to $\SU(4)$), the non-Higgsable SCFTs are $(A_1, A_{2\fm-2})^{\otimes 2}$ \cite[(3.27)]{Giacomelli:2020ryy}; these give rise to $2(\fm-1)$ twisted hypermultiplets, as expected.  

We tabulate some non-Higgsable SCFTs, including some cases for $\fm \geq 2$ and $\fN\geq 2$, below;
\be
\renewcommand{\arraystretch}{1.25}
\begin{tabular}{c|c|c|c|c}
$(b,p)$ & $(\fN,\fm)$ & non-Higgsable SCFT & $24(c-a)$ & rank\\
\hline
$(8,10)$ & $(2, 1)$ &  $(A_1, A_2)$ & $\frac{1}{5}$ & 1 \\
$(8,14)$ & $(2, 2)$ & {$I_{2,5} \otimes I_{3,4} = (A_1, A_4) \otimes (A_2, A_3)$} & {$\frac{2}{7}+\frac{3}{7}=\frac{5}{7}$ } & { $2+3=5$} \\
$(12, 14)$ & $(3,1)$ & $I_{2,5} = (A_1, A_4)$ & $\frac{2}{7}$ & 2 \\
$(12,22)$ & $(3, 3)$ & {\red $(A_4, D_4) \, \otimes \,  X$} & {\red $\frac{8}{11} + \frac{6}{11} = \frac{14}{11}$} & {\red $8 + \mathrm{rank}(X)$}
\end{tabular}
\ee

Let us comment on some of the above cases.  
\bi
\item We first consider the $D^{8}_{14}(\SO(10))$ theory, i.e. $(b,p)= (8,14)$ and $(\fN,\fm)=(2, 2)$.  It is convenient to close the full $\SO(10)$ puncture and analyze the $(A_5, D_5)$ theory, which admits the description
\bes{
(A_5, D_5) = \left[D_{7/2}\left(\USp'(2)\right) \,\,\longleftarrow \USp(2) \longrightarrow \,\, D_{7}(\USp(4)) \right]
}
where this was described in Section \ref{sec:confmfoldAD}, with $m=2, q=3, n=2, k=1$.  Upon closing the $\USp(2)$ puncture, we obtain the $(A_2, D_3)=(A_2, A_3)=I_{3,4}$ theory from $D_{7/2}\left(\USp'(2)\right)$ according to \eref{IIBAD}, whereas from the $D_{7}(\USp(4))$ theory we obtain the $\SO(10)^5[2]$ theory, whose Type IIB hypersurface is given by \eref{IIBSO2pm2N}, according to \eref{closeDpUSp2Nm2}. At a generic point of the Higgs branch of the $\SO(10)^5[2]$ theory, we have the $(A_1, A_4)$ theory from \eref{nHSCFTsDNpSO2Nodd} with $\fN=5$, $p=7$, $k=2$, $m=1$ and $\fn=3$, along with $2$ hypermultiplets from \eref{HfreeDNpSO2Nodd}.  For this reason, we say that the non-Higgsable SCFTs for the $D^{8}_{14}(\SO(10))$ theory and for the $(A_5, D_5)$ theory are $(A_1, A_4) \otimes (A_2, A_3)$, which have total rank equal to $5$. Upon reduction to $3$d and applying mirror symmetry, this gives rise to $5$ hypermultiplets; together with the said 2 hypermultiplets, we obtain $H_{\text{free}} = 7$ hypermultiplets, as indicated in \eref{Hfreegcd2}.
\item Next, we comment on the $D^{12}_{22}(\SO(14))$ theory, i.e. $(b,p)= (12,22)$ and $(\fN,\fm)=(3, 3)$.  It is convenient to close the full $\SO(14)$ puncture and analyze the $(A_9, D_7)$ theory, which admits the description
\bes{
(A_9, D_7) = \left[D_{11/2}\left(\USp'(4)\right) \,\,\longleftarrow \USp(4) \longrightarrow \,\, D_{11}(\USp(6)) \right]
}
where this was described in Section \ref{sec:confmfoldAD}, with $m=2, q=5, n=3, k=1$. Upon closing the $\USp(4)$ puncture, we obtain the $(A_4, D_4)$ theory from $D_{11/2}\left(\USp'(4)\right)$ according to \eref{IIBAD}, whereas from the $D_{11}(\USp(6))$ theory we obtain a theory that is described by the Type IIB hypersurface singularity given by \eref{closeDpUSp2Nm2} with $\fN=4, p=11$:
\bes{ \label{exotic1}
u^2+x^6+xy^3+yz^2=0~, \quad \Omega=\frac{dudxdydz}{dF}
}
according to \eref{closeDpUSp2Nm2}. To the best of our knowledge, the latter theory has not been studied anywhere in the literature.  We denote the by $X$ the non-Higgsable SCFT(s) for such a theory. Since we know that the values of $24(c-a)$ of the $D^{12}_{22}(\SO(14))$ theory and the $(A_4, D_4)$ theory are $14/11$ and $8/11$ respectively, the value of $24(c-a)$ of $X$ should be $6/11$.  Moreover, we expect that a generic point of the Higgs branch of theory \eref{exotic1} should contain $X$ and a collection of $h$ hypermultiplets, satisfying
\bes{
8 + \mathrm{rank}(X) + h = H_{\text{free}} =22
}
and so $\mathrm{rank}(X) + h= 14$.  We leave the detailed study of \eref{exotic1} as well as $X$ for future work.
\ei

\subsubsection{$D^{4\fN-2}_p(\SO(4\fN))$ such that $\GCD(4\fN,p)=2$}
\label{sec:D4N-2pSO4Ngcd2}
Let us write 
\bes{
p =  (4\fN-2)+2 \fm~, \quad \fm \in \BZ_{\geq 1}
}
Note that although this parametrization includes all $p$ such that $\GCD(4\fN-2,p)=2$, it also include those with $\GCD(4\fN-2,p)\neq2$, in which case we should exclude those values of $p$ from the following analysis.   In other words, we consider $\fm$ such that 
\bes{\label{condgcd2} \GCD(\fm,2\fN-1)=1~.}

We propose that the mirror theory of $D^{4\fN-2}_{4\fN-2+2\fm}(\SO(4\fN))$ with the condition \eref{condgcd2} is\footnote{The total number of hypermultiplets in the complete graph part (not included the tail and the connections) is $\fm(\fN+1)$.}
\bes{
\begin{tikzpicture}[baseline=0], font=\footnotesize]
\tikzstyle{every node}=[minimum size=0.5cm,inner sep=1.5pt]
\node[draw, circle] (c2) at (-1.5,0) {$D_1$};
\node[draw, circle] (c3) at (0,0) {$C_1$} ;
\node[draw, circle] (c4) at (1.5,0)   {$D_2$}; 
\node[draw, circle] (c5) at (3,0)  {$C_2$}; 
\node[draw=none] (c6) at (4.5,0)  {$\cdots$}; 
\node[draw, circle] (c7) at (6,0)  {$C_{2\fN-1}$}; 
\node[draw, circle] (f1) at (8,1)  {$D_1$}; 
\node[draw, circle] (f2) at (8,-1)  {$D_1$}; 
\node[draw=none] (Z2) at (10,-1)  {$/\BZ_2$}; 
\node[draw=none] (s1) at (10.2,1)  {\hspace{-0.2cm} $[\fm(\fN-1)]_2$}; 
\draw[draw, solid] (c2)--(c3)--(c4)--(c5)--(c6)--(c7);
\draw[draw, solid, red, thick] (f1)--(c7) node[midway, above] {\hspace{-0.8cm} \red $2\fN-1$};
\draw[draw, solid, black] (f2)--(c7) node[midway, below] {$1$};
\draw[draw, solid, light-gray, thick] (f1)--(f2) node[midway,right] {$\fm$}; 
\draw[draw, solid, snake it] (f1)--(s1) node[midway,right] {}; 
\end{tikzpicture}}
together with
\bes{
H_{\text{free}} = (\fN-1)(\fm-1) ~\text{free hypermultiplets}~.
}
The notation in the above quiver is as described in section \ref{sec:notation}.

The non-Higgsable SCFTs for this class of theory are
\bes{
(A_{\fm-1}, A_{2\fN-2})~.
}

Decoupling the tail, we see that the mirror for the $(A_{2\fm-1},D_{2\fN})$ theory, with the condition \eref{condgcd2}, is
%\bes{ \label{AoddD2N}
%[\fm(\fN-1)]_2 \begin{tikzpicture}[baseline] \draw[draw,solid,black,snake it] (0,0.1)--(1,0.1) node[midway, above] {}; \end{tikzpicture} \SO(2) \begin{tikzpicture}[baseline] \draw[draw,solid,light-gray,thick] (0,0.1)--(1,0.1) node[midway, above] {\footnotesize $\fm$}; \end{tikzpicture} \SO(2) \quad_{/\BZ_2}
%}
\bes{ \label{AoddD2N}
	[\fm(\fN-1)]_2 \begin{tikzpicture}[baseline] \draw[draw,solid,black,snake it] (0,0.1)--(1,0.1) node[midway, above] {}; \end{tikzpicture} D_1 \begin{tikzpicture}[baseline] \draw[draw,solid,light-gray,thick] (0,0.1)--(1,0.1) node[midway, above] {\footnotesize $\fm$}; \end{tikzpicture} D_1 \quad_{/\BZ_2}
}
together with $H_{\text{free}}$ free hypermultiplets.  Let us test this proposal in two special cases as follows:
\bi
\item For $\fm=1$, we expect that this reduces the theory depicted on the left-hand side of \eref{A1D2Nmirr1}, with no free hypermultiplets.  In fact, the $3$d mirror for the $(A_1, D_{2\fN})$ theory admits another description in terms of unitary gauge groups as depicted on the right-hand side of \eref{A1D2Nmirr1}. We have also discussed the correspondence between the two descriptions.
\item The special case of $\fN=1$ corresponds to the $(A_{2\fm-1}, D_2)$ theory.  We have discussed the mirror for this theory and its factorization in \eref{SO2SO2factorize}.
\ei

\subsection{$D^{2 N -2}_p(\SO(2 N ))$ with $\GCD(2 N -2,p)=4$}
There are two cases for $2 N -2$, with $ N  \geq 2$, to be considered: those are even after divided by 4, and those are odd after divided by 4;  
\bi
\item For $2 N -2 = 4(2\fN)$, i.e.  $2 N =8\fN+2$, there is only one mass parameter in addition to those associated with the $\SO(2 N )$ flavor symmetry.
\item For $2 N -2=4(2\fN-1)$, i.e.  $2 N =8\fN-2$, there are three mass parameters in addition to those associated with the $\SO(2 N )$ flavor symmetry.
\ei

\subsubsection{$D^{8\fN}_p(\SO(8\fN+2))$ with $\GCD(8\fN,p)=4$}
We parametrize $p$ by
\bes{
p= 8\fN+(8 \fm-4)~, \quad \fm \in \BZ_{\geq 1}
}
with the restriction
\bes{
\GCD(2\fN, 2\fm-1) = 1~.
}
The $3$d reduction of this theory gives 
\bes{
H_{\text{free}}=4 \fm (4 \fN-1)-2 (5 \fN-1)
}
twisted hypermultiplets, together with the $T^{\vec \sigma}_{\vec \rho}[\SO(2 \fn)]$ theory whose quiver description is
the quiver
%\bes{ \label{linquivD4Nmp2SO4Np2}
%&[\SO(8\fN+2)]-\USp(8\fN)-\SO(8\fN)-\USp(8\fN-2)-\SO(8\fN-2)- \\
%&-\USp(4)-\SO(4)-(\USp(2)-\SO(2))^{x}-\USp(2)-[\SO(2)]
%}
\bes{ \label{linquivD4Nmp2SO4Np2}
	&[D_{4\fN+1}]-C_{4\fN}-D_{4\fN}-C_{4\fN-1}-D_{4\fN-1}-C_2-D_2-(C_1-D_1)^{x}-C_1-[D_1]
}
where
\bes{
\begin{array}{ll}
\vec \sigma= [(8 \fN+2x-1)^2,1^{8 \fN +2}]~, &\quad \vec \rho = [3^{8 \fN},2^{2x}]~,\\
2\fn =24 \fN+4x   & \quad x= 2(2\fm-1)~.
\end{array}
}

The mirror theory for this theory is described by
%\bes{
%\begin{array}{lll}
%&\SO(2)-&\USp(8\fN)-\SO(8\fN)-\USp(8\fN-2)-\SO(8\fN-2)-\cdots-\USp(2)-\SO(2) \\
%&\,\,\,\, |  & \,\,\,\, | \\
%&\hspace{-0.6cm}  [\USp(2x)]  &[\SO(8\fN)]
%\end{array}
%}
\bes{
	\begin{array}{lll}
		&D_1-&C_{4\fN}-D_{4\fN}-C_{4\fN-1}-D_{4\fN-1}-\cdots-C_1-D_1 \\
		&\,\,\,\, |  & \,\,\,\, | \\
		&  [C_{x}]  &[D_{4\fN}]
	\end{array}
}
together with $H_{\text{free}}$ free hypermultiplets.

Upon decoupling the tail, we obtain the mirror theory for $(A_{8 \fm-5}, D_{4\fN+1})$ theory. It is described by
%\bes{
%\SO(2)-[\USp(8\fm-4)]
%}
\bes{
	D_1-[C_{4\fm-2}]
}  
with $H_{\text{free}}$ free hypermultiplets.

%For $\fm=1$, the non-Higgsable SCFTs are $(A_1, A_{2\fN-2})^{\otimes 2}$, giving rise to $H_{\text{free}} = 2(\fN-1)$ twisted hypermultiplets upon reduction to $3$d.

\subsubsection{$D^{8\fN-4}_p(\SO(8\fN-2))$ with $\GCD(8\fN-4,p)=4$}
We parametrize $p$ as follows:
\bes{
p = (8\fN-4) + 4\fm~, \fm \in \BZ_{\geq 1}
}
with the restriction
\bes{
\GCD(2\fN-1, \fm)=1~.
}

We propose the following mirror theory:
\bes{
&\begin{tikzpicture}[baseline=0, font=\footnotesize]
\tikzstyle{every node}=[minimum size=0.5cm,inner sep=1.5pt]
\node[draw, circle] (c2) at (-3.5,0) {$D_1$};
\node[draw, circle] (c3) at (-0.5,0) {$C_{4\fN-2}$};
\node[draw, circle] (c4) at (1.2,0)   {$D_{4\fN-2}$};
\node[draw=none] (c5) at (2.7,0)  {$\cdots$}; 
\node[draw, circle] (c6) at (4,0)  {$C_1$};
\node[draw, circle] (c7) at (5,0)  {$D_1$};
\node[draw, circle] (f1) at (-2,1.5)  {$D_1$}; 
\node[draw, circle] (f2) at (-2,-1.5)  {$D_1$}; 
\node[draw=none] (s1) at (-2,2.7)  {$[\fm(\fN-1)]_2$}; 
\node[draw=none] (s3) at (-5.5,0)  {$[\fm(\fN-1)]_2$}; 
\draw[draw, solid, very thick, red] (c2)--(c3) node[midway,above] {$2\fN-1$};
\draw[draw, solid] (c3)--(c4)--(c5)--(c6)--(c7);
\draw[draw, solid] (f1)--(c2);
\draw[draw, solid, snake it] (s1)--(f1);
\draw[draw, solid, snake it] (s3)--(c2);
\draw[very thick,blue] (c2)--(f1) node[above,midway]  {\hspace{-0.2cm}\footnotesize $M$};
\draw[very thick, red] (c3)--(f1) node[midway, right] {\hspace{0.1cm}$2\fN-1$};
\draw[very thick, light-gray] (c2)--(f2) node[above,pos=0.7]  {\footnotesize $\fm$};
\draw[black] (c3)--(f2) node[midway, below] {$1$};
\draw[very thick, light-gray] (f1)--(f2); 
\end{tikzpicture}  \quad /\BZ_2\\
& \text{+ $H_{\text{free}} = 2(\fN-1)(\fm-1)$ free hypermultiplets}
}
where each red line has multiplicity $4\fN-2$, each gray line has multiplicity $\fm$ and the blue line has multiplicity\footnote{The total number of hypermultiplets in the complete graph part (not included the tail and the connections) is $6\fm \fN$.}
\bes{
M = \fm (2\fN - 1)~.
}

Decoupling the tail, we obtain the mirror theory for $(A_{4\fm-1},D_{4\fN-1})$ with $\GCD(2\fN-1, \fm)=1$:
\bes{
&
\scalebox{0.9}{
\begin{tikzpicture}[baseline=0, font=\footnotesize]
\tikzstyle{every node}=[minimum size=0.5cm,inner sep=1.5pt]
\node[draw, circle] (c2) at (0,1) {$D_1$};
\node[draw, circle] (f1) at (-1.2,0)  {$D_1$}; 
\node[draw, circle] (f2) at (1.2,0)  {$D_1$}; 
\node[draw=none] (s1) at (-3.5,0)  {\footnotesize $[\fm(\fN-1)]_2$}; 
\node[draw=none] (s2) at (3.5,0)  {\footnotesize  $[\fm(\fN-1)]_2$}; 
\draw[draw, solid, snake it] (s1)--(f1);
\draw[draw, solid, snake it] (s2)--(f2);
\draw[very thick,blue] (c2)--(f1) node[above,midway]  {\hspace{-0.4cm}\footnotesize $M$};
\draw[very thick, light-gray] (c2)--(f2);
\draw[very thick, light-gray] (f1)--(f2) node[below,midway]  {\footnotesize $\fm$}; 
\end{tikzpicture}} \quad /\BZ_2
 \\ 
& \text{+ $H_{\text{free}}$ free hypermultiplets}
}

For $\fN=1$, we have $M=\fm$ and so we recover the mirror for the $(A_{4\fm-1}, D_3)$ theory, as discussed around \eref{threeSO2nodes}, as expected.

For $\fm=1$, there is no free hypermultiplet.  For $\fm=2$, the non-Higgsable SCFTs are $(A_1, A_{2\fN-2})^{\otimes 2}$, giving rise to $2\fN-2$ free hypermultiplets.  We conjecture that, in general, the non-Higgsable SCFTs are 
\bes{
(A_{\fm-1}, A_{2\fN-2})^{\otimes 2}~.}

\subsection{$D^{2 N -2}_p(\SO(2 N ))$ with $\GCD(2 N -2,p)=6$}
There are two cases for $2 N -2$, with $ N  \geq 2$, to be considered: those are even after divided by 6, and those are odd after divided by 6.  
\bi
\item For $2 N -2 = 6(2\fN)$, i.e.  $2 N =12\fN+2$, there is only one mass parameter in addition to those associated with the $\SO(2 N )$ flavor symmetry.
\item For $2 N -2=6(2\fN-1)$, i.e.  $2 N =12\fN-4$, there are four mass parameters in addition to those associated with the $\SO(2 N )$ flavor symmetry.
\ei

The first case will be discussed as a special case of the general result in Section \ref{sec:generalGCDeven} with $\mu=3$.

For the second case, namely $D^{12\fN-6}_{p}(\SO(12\fN-4))$ with $\GCD(p, 12\fN-6)=6$, we write
\bes{
p = (12\fN-6)+6\fm~, \quad \fm \in \BZ_{\geq 1}~, \GCD(2\fN-1,\fm)=1~.
}
We propose that the mirror theory of $D^{12\fN-6}_{p}(\SO(12\fN-4))$, is 
\bes{
H_{\text{free}} = 3(\fN-1)(\fm-1)~,
} 
free hypermultiplets, together with\footnote{The total number of hypermultiplets in the complete graph part (not including the tail and not including the connections) is $3\fm(5\fN-1)$.}
\bes{
\scalebox{0.9}
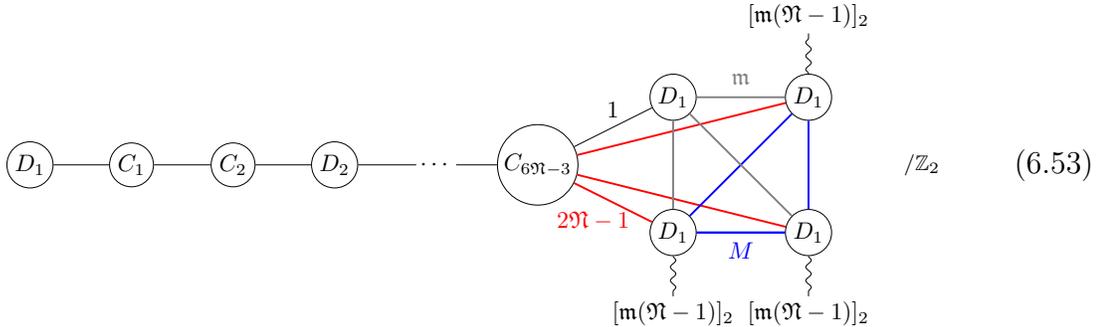
{
\begin{tikzpicture}[baseline=0, font=\footnotesize]
\tikzstyle{every node}=[minimum size=0.5cm,inner sep=1.5pt]
\node[draw, circle] (c2) at (-1.5,0) {$D_1$};
\node[draw, circle] (c3) at (0,0) {$C_1$} ;
\node[draw, circle] (c4) at (1.5,0)   {$C_2$}; 
\node[draw, circle] (c5) at (3,0)  {$D_2$}; 
\node[draw=none] (c6) at (4.5,0)  {$\cdots$}; 
\node[draw, circle] (c7) at (6,0)  {$C_{6\fN-3}$}; 
\node[draw, circle] (f1) at (8,1)  {$D_1$}; 
\node[draw, circle] (f2) at (8,-1)  {$D_1$}; 
\node[draw, circle] (f3) at (10,1)  {$D_1$}; 
\node[draw, circle] (f4) at (10,-1)  {$D_1$}; 
\node[draw=none] (s2) at (8,-2.2)  {$[\fm(\fN-1)]_2$}; 
\node[draw=none] (s3) at (10,2.2)  {$[\fm(\fN-1)]_2$}; 
\node[draw=none] (s4) at (10,-2.2)  {$[\fm(\fN-1)]_2$}; 
\draw[draw, solid] (c2)--(c3)--(c4)--(c5)--(c6)--(c7);
\draw[draw, solid, black] (f1)--(c7) node[midway, above] {$1$};
\draw[draw, solid, red, thick] (f2)--(c7) node[midway, below] {\hspace{-0.7cm} $2\fN-1$};
\draw[draw, solid, red, thick] (f3)--(c7) node[midway, above] {};
\draw[draw, solid, red, thick] (f4)--(c7) node[midway, below] {};
\draw[draw, solid, blue, thick] (f2)--(f3)--(f4)--(f2) node[midway,right] {}; 
\draw[draw, solid, light-gray, thick] (f1)--(f2) node[midway,right] {}; 
\draw[draw, solid, light-gray, thick] (f1)--(f4) node[midway,right] {}; 
\draw[draw, solid, light-gray, thick] (f1)--(f3) node[midway,above] {\color{light-gray} $\fm$}; 
\draw[draw, solid, blue, thick] (f2)--(f4) node[midway,below] {\blue $M$}; 
\draw[draw, solid, snake it] (f2)--(s2) node[midway,right] {}; 
\draw[draw, solid, snake it] (f3)--(s3) node[midway,right] {}; 
\draw[draw, solid, snake it] (f4)--(s4) node[midway,right] {}; 
\end{tikzpicture}} \quad_{/\BZ_2}
}
where $M=\fm(2\fN-1)$.  We also claim that the non-Higgsable SCFTs are 
\bes{
(A_{\fm-1}, A_{2\fN-2})^{\otimes 3}~.
}
Decoupling the tail, we obtain the following mirror theory for $(A_{6\fm-1},D_{6\fN-2})$:
\bes{
&\scalebox{0.95}{
\begin{tikzpicture}[baseline=0, font=\footnotesize]
\tikzstyle{every node}=[minimum size=0.5cm,inner sep=1.5pt]
\node[draw, circle] (f1) at (8,1)  {$D_1$}; 
\node[draw, circle] (f2) at (8,-1)  {$D_1$}; 
\node[draw, circle] (f3) at (10,1)  {$D_1$}; 
\node[draw, circle] (f4) at (10,-1)  {$D_1$}; 
\node[draw=none] (s2) at (6,-1)  {$[\fm(\fN-1)]_2$}; 
\node[draw=none] (s3) at (12,1)  {$[\fm(\fN-1)]_2$}; 
\node[draw=none] (s4) at (12,-1)  {$[\fm(\fN-1)]_2$}; 
\draw[draw, solid, blue, thick] (f2)--(f3)--(f4)--(f2) node[midway,right] {}; 
\draw[draw, solid, light-gray, thick] (f1)--(f2) node[midway,right] {}; 
\draw[draw, solid, light-gray, thick] (f1)--(f4) node[midway,right] {}; 
\draw[draw, solid, light-gray, thick] (f1)--(f3) node[midway,above] {\color{light-gray} $\fm$}; 
\draw[draw, solid, blue, thick] (f2)--(f4) node[midway,below] {\blue $M$}; 
\draw[draw, solid, snake it] (f2)--(s2) node[midway,right] {}; 
\draw[draw, solid, snake it] (f3)--(s3) node[midway,right] {}; 
\draw[draw, solid, snake it] (f4)--(s4) node[midway,right] {}; 
\end{tikzpicture}} \quad_{/\BZ_2} \\
& \text{+ $3(\fN-1)(\fm-1)$ free hypermultiplets}
}

\section{$D^{N}_p(\SO(2N))$ with $p \geq N$}
\label{sec:DNpSO2NpgrN}

In this section, we study the $D^{N}_p(\SO(2N))$ theory with $p \geq N$ and the corresponding mirror theory.  There are two subclasses to consider:
\begin{enumerate}
    \item\label{lst:subclass1} The subclass containing the theories of which $N/\GCD(N,p)$ is odd; 
    \item\label{lst:subclass2} The subclass containing the theories of which $N/\GCD(N,p)$ is even. 
\end{enumerate}
In Subclass \ref{lst:subclass1}, each theory has no additional mass parameters to those associated with the $\SO(2N)$ flavor symmetry.  Furthermore, as discussed in \eref{dimcmDNpSO2N}, the conformal manifold of the theory in this subclass is $\GCD(N,p)-1$ dimensional. In Subclass \ref{lst:subclass2}, each theory has $\GCD(N,p)$ mass deformations in addition to those associated with the $\SO(2N)$ flavor symmetry.  The conformal manifold of the theory in this subclass is $2\GCD(N,p)-1$.

\subsection{$D^N_p(\SO(2N))$ with $N/\GCD(N,p)=2\fn$ even}
We write
\bes{
&\GCD(N,p) = m~,\quad N= 2 m \fn ~, \\ 
&p = 2 m \fn + m (2k-1)~,  ~ \text{with $k \in \BZ_{\geq 1}$ and $\GCD(2\fn, 2k-1) = 1$}~.
}
All of these theories have no additional mass deformation to those associated with the $\SO(2N)$ flavor symmetry.  We expect that the $3$d reduction gives the $T[\SO(2N)]$ theory, together with
\bes{ \label{mirrreductionGCDgenB}
H_{\text{free}}= m\left[2 (2 m \fn-1) k - \fn(2m+1) +1\right]
}
twisted hypermultiplets.

this proposal along the line of Section \ref{sec:MS}, where $F=N$ and $n_i=0$.  From the discussion there, we have $\delta p =2m\delta k = N= 2m \fn$ ($i.e.  \delta k =\fn$) and $\delta H_{\text{free}}= N(N-1)$.  This is in accordance with \eref{mirrreductionGCDgenB}, where $\delta H_{\text{free}} = 2m(2m\fn-1) \delta k = 2m \fn (2m\fn-1) =N(N-1)$.

Upon closing the full $\SO(2N)$ puncture, we obtain the non-Higgsable SCFT 
\bes{ \SO(4m\fn)^{2m\fn}[m(2k-1)]~,} 
whereby the reduction to $3$d and applying mirror symmetry giving rise to $H_{\text{free}}$ free hypermultiplets, given by \eref{mirrreductionGCDgenB}.  

For $k=1$, we claim that the non-Higgsable SCFT $\CT=\SO(4m\fn)^{2m\fn}[m]$ contains the smaller non-Higgsable SCFTs
\bes{ \label{A1A2nm2}
\CT' = (A_1, A_{2(\fn-1)})^{\otimes m}
}
as discussed towards the end of Section \ref{sec:genresultGCDodd}.
We will soon provide an example of $\SO(32)^{16}[2]$ to illustrate this point. Note that \eref{A1A2nm2} has $24(c-a) = m\frac{\fn-1}{2\fn+1}$ and gives rise to $m \left(\fn -1 \right)$ twisted hypermultiplets upon reduction to $3$d.  

Moreover, for $k=1$ and $m=1$, we have the following identification:
\bes{\label{idenSO4n}
\SO(4\fn)^{2\fn}[1] = (A_1, A_{2\fn-2})~.
}
This relation can be seen from the curve.  That of the theory of the left-hand side reads $u^2+x^{2\fn-1}+xy^2+yz=0$.
Since $y$ and $z$ are massive and can be integrated out, we obtain $u^2+x^{2 \fn-1}=0$, which corresponds to the curve of the $(A_1, A_{2\fn-2})$ theory.

\subsubsection*{Examples of $D^{16}_{18}(\SO(32))$ and $\SO(32)^{16}[2]$}
We take $m=2$, $\fn=4$ and $k=1$. Let us close the full $\SO(32)$ puncture in the theory $D^{16}_{18}(\SO(32))$ and obtain the $\SO(32)^{16}[2]$ theory.  Each of such theories has a three-dimensional conformal manifold.  Subsequently, we analyze one of the weakly coupled cusps, namely, that is associated with the marginal deformation $x^3 y^3$. As described in Section \ref{sec:conformalSO2NNp}, we have the following description
\bes{ \label{descrSO32B}
SO(32)^{16}[2] = \left[ D^8_9(\SO(16)) \,\, \longleftarrow \SO(3) \longrightarrow\,\, D_9(\SO(3))\right]%~.
}
Upon closing the $\SO(3)$ puncture, the $D_9(\SO(3))=D^2_9(\SU(2))$ theory becomes the $(A_1, A_6)$ theory, and the $D^8_9(\SO(16))$ theory becomes the $\SO(16)^8[1] = (A_1, A_6)$ theory due to \eref{idenSO4n}. In summary, we obtain the $(A_1, A_6)^{\otimes 2}$ theory as claimed in \eref{A1A2nm2}.

The descriptions of the theory at other weakly coupled cusps can also be analyzed, but they are more complicated.  For example, that associated with the marginal deformation $x^7 y^2$ is
\bes{ \label{descrSO32A}
SO(32)^{16}[2] = \left[\tilde{D}_{9/2}(\USp(6)) \,\, \longleftarrow \USp(2) \longrightarrow\,\, D_{27/2}(\USp'(2)) \right]
}
The properties of the $\tilde{D}_{9/2}(\USp(6)$ theory are little known, and so we are not analyzing this theory further.  Nevertheless, as described at the end of Section \ref{partclosed}, upon closing the $\USp(2)$ puncture, the $D_{27/2}(\USp'(2))$ becomes the $(A_2, D_{13})$ theory. We will next show that the latter, in fact, contains the $(A_1, A_6)$ theory.  Although the $(A_2, D_{13})$ theory is a non-Higgsable SCFT, it has a one dimensional conformal manifold, associated with the deformation $x^4 y^2$. The theory can be described by 
\bes{
 (A_2, D_{13}) = \left[ D_9(\SO(3)) \,\, \longleftarrow \SO(3)  \longrightarrow \,\, D^8_9(\SO(10)) \right]
}
as explain in Section \ref{sec:confmfoldAD}. Upon closing the $\SO(3)$ puncture, $D_9(\SO(3))=D^2_9(\SU(2))$ becomes $(A_1, A_6)$, whereas $D^8_9(\SO(10))$ becomes trivial.

\subsection{$D^N_p(\SO(2N))$ with $N/\GCD(N,p)=2\fn-1$ odd}
We write
\bes{
&\GCD(N,p) = m~,\quad N= m (2\fn-1) ~, \\ 
&p = m (2\fn-1) + m k~,  ~ \text{with $k \in \BZ_{\geq 0}$ and $\GCD(2\fn-1, k) = 1$}~.
}
This theory has $m$ mass deformations. In general, we propose that the mirror theory contains 
\bes{ \label{HfreeDNpSO2Nodd}
H_{\text{free}} = m(\fn-1)(k-1)~,
}
free hypermultiplets, together with a quiver gauge theory that can be constructed as follows:\footnote{The total number of hypermultiplets in the complete graph part (not including the tail and the connections) is $m k [m ( 2 \fn-1) - \fn ]$.}
\ben
\item Construct a complete graph with $m$ $\SO(2)$ nodes such that every edge has multiplicity 
\be
M= k N/m = (2\fn-1)k~.
\ee
We shall use the same notation as in \eref{blueedge} and \eref{repblueedge}. 
\item There are $F=(\fn-1)k$ flavors of hypermultiplets carrying charge $2$ under each $\U(1)\cong \SO(2)$ gauge node.  We shall use the same notation as in \eref{wiggleline}.
\item Construct the $T[\SO(2N)]$ tail: $D_1-C_1-D_2-C_2- \cdots- C_{N-1}$.
\item Connect the $C_{N-1}$ gauge node in the tail to each $\SO(2)$ gauge group in the complete graph with the edges, such that each edge has multiplicity is 
\be R=N/m= 2\fn-1~.\ee  
We shall use the same notation as in \eref{rededge} and \eref{reprededge}.
\item The gauge symmetry of the theory is
\bes{ \label{overallZ2}
\left(D_1 \times C_1 \times D_2 \times C_2 \times \cdots \times C_{N-1} \times D_1^m \right)/\BZ^{\diag}_2
}
where we shall denote the diagonal $\BZ_2$ quotient $\BZ^{\diag}_2$ (see \cite{Bourget:2020xdz} for a detailed discussion) by a shorthand notation $/\BZ_2$ in the subsequent part of the paper.
\een

We propose that the non-Higgsable SCFTs are
\bes{ \label{nHSCFTsDNpSO2Nodd}
(A_{k-1},A_{2\fn-2})^{\otimes m}
}
giving rise to $H_{\text{free}}$ free hypermultiplets, as expected.

Decoupling the tail, we obtain the mirror for the 
\be 
\SO(2N)^N [p-N] = \SO(2 m(2\fn-1))^{m (2\fn-1)}[mk] 
\ee
theory as a complete graph with $m$ $\SO(2)$ nodes such that every edge has multiplicity $M= k N/m = (2\fn-1)k$ and there are $F=(\fn-1)k$ flavors of hypermultiplets carrying charge $2$ under each $\U(1)\cong \SO(2)$ gauge node.  The gauge symmetry of this mirror theory is $\SO(2)^m/\BZ_2$.

Let us test the above procedure of constructing the mirror theory along the line of Section \ref{sec:MS}, where $F=0$ and $n_1=n_2=\cdots=n_m=R=2 \fn-1$.  According to \eref{deltap} and \eref{deltaHfree}, we have $\delta p=m \delta k = m(2\fn-1)$, i.e. $\delta k = 2\fn-1$, and $\delta H_{\text{free}}= m \frac{1}{2}(2\fn-1)(2\fn-2) = m(\fn-1)(2\fn-1)$.  This is in accordance with \eref{HfreeDNpSO2Nodd}, where $\delta H_{\text{free}} = m(\fn-1)(2\fn-1)= m(\fn-1)\delta k$.  Moreover, as stated below \eref{deltaHfree}, the increment of the number of hypermultiplets carrying charge 2 under the $\U(1)$ gauge groups, which are isomorphic to the 1st, 2nd, $\ldots$,  $m$-th $\SO(2)$ gauge groups, is precisely $\frac{1}{2} (2 \fn-1)(2\fn-2) = (\fn-1)(2 \fn-1)=  (\fn-1)\delta k = \delta F$, in agreement with \eref{FmNm1}. 

\subsubsection{The case in which $N$ divides $p$} 
Let us consider the case in which 
\be m=N~, \quad \fn=1~, \quad p=N(k+1)~, \ee 
i.e., the $D^N_{N(k+1)}(\SO(2N))$ theory.  From the prescription above we have
\bes{
H_{\text{free}} =0~.
}
The mirror theory consists of a complete graph of $N$ $\SO(2)$ gauge nodes such that every edge of the graph has multiplicity $k N/m = k$ and each $\SO(2)$ node connects to the $C_{N-1}$ gauge node connects to the $T[\SO(2N)]$ tail with an edge with multiplicity $N/m=1$.  There is an overall $\BZ_2$ quotient in the gauge symmetry as indicated in \eref{overallZ2}.  Note that there is no flavor of hypermultiplets with charge $2$ under any $\U(1) \cong \SO(2)$ gauge group.  Recall from Section \ref{sec:bdividespSO2} that this is the same mirror theory as for the $D^{2N-2}_{(2N-2)(k+1)}(\SO(2N))$.  We thus claim the following identification:
\bes{
D^N_{N \kappa}(\SO(2N)) = D^{2N-2}_{(2N-2) \kappa}(\SO(2N))~, \quad \kappa \geq 1~.
}
This is analogous to \cite[(6.18)]{Giacomelli:2020ryy} for the $\SU(N)$ case; in particular, for $N=3$, this is in agreement with \cite[(6.18)]{Giacomelli:2020ryy} with $N=4$.

Decoupling the tail, we obtain the mirror theory for $\SO(2N)^{N}[Nk]$, whose description is the same as that for the $\SO(2N)^{2N-2}[(2N-2)k] = (A_{(2N-2)k-1}, D_N)$ theory, namely a complete graph of $N$ $\SO(2)$ gauge nodes such that every edge of the graph has multiplicity $k$, and the gauge symmetry is $\SO(2)^N/\BZ_2$.  We also propose the identification
\bes{
\SO(2N)^{N}[Nk] \,\, = \,\,  (A_{(2N-2)k-1}, D_N)~.
}

\subsubsection{The case of $m=1$} 
The mirror theory consists of $(\fn-1)(k-1)$ free hypermultiplets, together with the following theory
\bes{
D_1-C_1-D_2-C_2-\ldots-C_{N-1} \begin{tikzpicture}[baseline] \draw[draw,solid,red,thick] (0,0.1)--(1,0.1) node[midway, above] {\red \footnotesize $\substack{N/m \\=\, 2\fn-1}$}; \end{tikzpicture} D_1 \begin{tikzpicture}[baseline] \draw[draw,solid,black, snake it] (0,0.1)--(0.7,0.1); \end{tikzpicture} [(\fn-1)k]_2 \quad _{/\BZ_2}
}
where the red edge has multiplicity $N/m=2\fn-1$ and ${/\BZ_2}$ indicates that the gauge symmetry is as indicated in \eref{overallZ2}.

Decoupling the tail, we obtain the mirror for the $\SO(4 \fn-2)^{(2\fn-1)}[k] $ theory, whose description is as follows: $(\fn-1)(k-1)$ free hypermultiplets, together with 
\be 
D_1 \begin{tikzpicture}[baseline] \draw[draw,solid,black, snake it] (0,0.1)--(0.7,0.1); \end{tikzpicture} [(\fn-1)k]_2  \quad _{/\BZ_2}
\ee
i.e., the $\U(1)/\BZ_2$ gauge theory with $(\fn-1)k$ flavors of hypermultiplets with charge 2.  This is equivalent to the $\U(1)$ gauge theory with $(\fn-1)k$ flavors of hypermultiplets with charge 1. Hence, the mirror theory for $\SO(2N)^{(2\fn-1)}[k]$ can be rewritten as
\bes{
\left( \SO(2N)^{(2\fn-1)}[k] \right)_\text{$3$d mirr}: \quad &\U(1)-[(\fn-1)k] \quad \\
&\text{+ ~$(\fn-1)(k-1)$ free hypermultiplets}~.
}
We emphasize the \textit{importance} of the $\BZ_2$ quotient, 
discussed in \eref{overallZ2}, in reaching this conclusion.
Let us provide a test for this proposal.  We consider the case of $N=3$ and so the only possibility that is relevant to our restriction is to have $m=1$ and $\fn=2$, namely $D^3_{3+k}(\SO(6))$ theory. The above proposal for the mirror theory is
\bes{ \label{DSO6}
D^3_{3+k}(\SO(6))_{\text{$3$d mirr}}: \quad & D_1-C_1-D_2-C_2 \begin{tikzpicture}[baseline] \draw[draw,solid,red,thick] (0,0.1)--(1,0.1) node[midway, above] {\red \footnotesize $3$}; \end{tikzpicture} D_1 \begin{tikzpicture}[baseline] \draw[draw,solid,black, snake it] (0,0.1)--(0.7,0.1); \end{tikzpicture} [k]_2 \quad _{/\BZ_2} \\
& \text{+ $(k-1)$ free hypermultiplets}~.
}
Since $D^3_{3+k}(\SO(6)) = D^3_{3+k}(\SU(4))$, there is an alternative description of the mirror theory given by \cite[(6.13)]{Giacomelli:2020ryy} (with $N=4$ and $p=3+k$):  
\bes{ \label{DSU4}
D^{3}_{3+k}(\SU(4))_{\text{$3$d mirr}}: \quad\,\,
&\scalebox{0.9}{
\begin{tikzpicture}[baseline=0, font=\footnotesize]
\tikzstyle{every node}=[minimum size=0.5cm,inner sep=1.5pt]
\node[draw, circle] (c2) at (-2,0) {$1$};
\node[draw, circle] (c3) at (0,0) {\footnotesize $3$};
\node[draw, circle] (c4) at (1,0) {\footnotesize $2$};
\node[draw, circle] (c5) at (2,0) {\footnotesize $1$};
\node[draw, circle] (f1) at (-1,1)  {$1$}; 
\draw[draw, solid] (c2)--(c3);
\draw[draw, solid] (c3)--(c4)--(c5);
\draw[draw, solid] (f1)--(c2);
\draw[very thick, gray] (c2)--(f1) node[above,midway] {\hspace{-0.2cm}\footnotesize $k$};
\draw[very thick, red] (c3)--(f1) node[above,midway]  {\hspace{0.2cm}\footnotesize $3$};
\end{tikzpicture}} \\
& \text{+ $(k-1)$ free hypermultiplets}~.
}
Indeed, the Higgs branch symmetry $\SU(k) \times \SU(3) \times \U(1)$ is manifest in both descriptions.  We match the Higgs and Coulomb branch Hilbert series of the two quiver descriptions in Appendix \ref{app:SO6vsSU4}.   Upon decoupling the tail in either description, we indeed obtain the mirror theory which is the SQED with $k$ flavors, together with $k-1$ free hypermultiplets, as expected.

\subsubsection{The case of $m=2$} 
The mirror theory consists of $2(\fn-1)(k-1)$ free hypermultiplets, together with the following theory
\bes{
\scalebox{0.9}{
\begin{tikzpicture}[baseline=0, font=\footnotesize]
\tikzstyle{every node}=[minimum size=0.5cm,inner sep=1.5pt]
\node[draw, circle] (c2) at (-1.5,0) {$D_1$};
\node[draw, circle] (c3) at (0,0) {$C_1$} ;
\node[draw, circle] (c4) at (1.5,0)   {$D_2$}; 
\node[draw, circle] (c5) at (3,0)  {$C_2$}; 
\node[draw=none] (c6) at (4.5,0)  {$\cdots$}; 
\node[draw, circle] (c7) at (6,0)  {$C_{N-1}$}; 
\node[draw, circle] (f1) at (8,1)  {$D_1$}; 
\node[draw, circle] (f2) at (8,-1)  {$D_1$}; 
\node[draw=none] (s1) at (10.5,1)  {\hspace{-0.05cm}$[k(\fn-1)]_2$}; 
\node[draw=none] (s2) at (10.5,-1)  {\hspace{-0.05cm}$[k(\fn-1)]_2$}; 
\draw[draw, solid] (c2)--(c3)--(c4)--(c5)--(c6)--(c7);
\draw[draw, solid, red, thick] (f1)--(c7) node[midway, above] {\hspace{-0.8cm} \red $\substack{N/m \\ = \, 2 \fn-1}$};
\draw[draw, solid, red, thick] (f2)--(c7) node[midway, below] {\hspace{-0.6cm} \red $\substack{N/m \\ =\, 2 \fn-1}$};
\draw[draw, solid, blue, thick] (f1)--(f2) node[midway,right] {\blue $k\frac{N}{m} = k(2\fn-1)$}; 
\draw[draw, solid, snake it] (f1)--(s1) node[midway,right] {}; 
\draw[draw, solid, snake it] (f2)--(s2) node[midway,right] {}; 
\end{tikzpicture}} \quad_{/\BZ_2}
}
where each red line has multiplicity $N/m=2\fn-1$ and each blue line has multiplicity $kN/m$.  Decoupling the tail, we obtain the mirror for the $\SO(8 \fn-4)^{(4\fn-2)}[2k] $ theory, whose description is as follows:
\bes{ \label{mirrorSO8nm4}
&\scalebox{0.9}{
\begin{tikzpicture}[baseline=0, font=\footnotesize]
\tikzstyle{every node}=[minimum size=0.5cm,inner sep=1.5pt]
\node[draw, circle] (f1) at (8,1)  {$D_1$}; 
\node[draw, circle] (f2) at (8,-1)  {$D_1$}; 
\node[draw=none] (s1) at (10.5,1)  {\hspace{-0.05cm}$[k(\fn-1)]_2$}; 
\node[draw=none] (s2) at (10.5,-1)  {\hspace{-0.05cm}$[k(\fn-1)]_2$}; 
\draw[draw, solid, blue, thick] (f1)--(f2) node[midway,right] {\blue $k\frac{N}{m} = k(2\fn-1)$}; 
\draw[draw, solid, snake it] (f1)--(s1) node[midway,right] {}; 
\draw[draw, solid, snake it] (f2)--(s2) node[midway,right] {}; 
\end{tikzpicture}} \quad_{/\BZ_2} \\
& \text{+ $2(\fn-1)(k-1)$ free hypermultiplets.}
}
Let us provide a test for this proposal.  We consider the case of $N=2$ and so the only possibility that is relevant to our restriction is to have $m=2$ and $\fn=1$, namely $D^2_{2(k+1)}(\SO(4))$ theory. Since $\SO(4)=\SU(2)\times \SU(2)$, we expect that this theory factorizes into $(D^2_{2(k+1)}(\SU(2)))^{\otimes 2}$.  Indeed, this can be checked using the Higgs and Coulomb branch Hilbert series.  Upon closing the full $\SU(2)$ puncture in each factor, we obtain the $(A_{2k-1},D_2)$ theory, whose mirror theory was discussed in \eref{SO2SO2factorize}, with $\fm$ being $k$. This indeed in agreement with \eref{mirrorSO8nm4}.

\subsubsection{The case of $m=3$}
We propose the following mirror theory:
\bes{
&
\scalebox{0.8}{
\begin{tikzpicture}[baseline=0, font=\footnotesize]
\tikzstyle{every node}=[minimum size=0.5cm,inner sep=1.5pt]
\node[draw, circle] (c2) at (-3,0) {$D_1$};
\node[draw, circle] (c3) at (0,0) {$C_{N-1}$};
\node[draw, circle] (c4) at (1.5,0)   {$D_{N-1}$};
\node[draw=none] (c5) at (2.7,0)  {$\cdots$}; 
\node[draw, circle] (c6) at (3.7,0)  {$C_1$};
\node[draw, circle] (c7) at (4.7,0)  {$D_1$};
\node[draw, circle] (f1) at (-1.5,2)  {$D_1$}; 
\node[draw, circle] (f2) at (-1.5,-2)  {$D_1$}; 
\node[draw=none] (s1) at (-1.5,3.5)  {$[k(\fn-1)]_2$}; 
\node[draw=none] (s2) at (-1.5,-3.5)  {$[k(\fn-1)]_2$}; 
\node[draw=none] (s3) at (-5.5,0)  {$[k(\fn-1)]_2$}; 
\draw[draw, solid, very thick, red] (c2)--(c3); 
\draw[draw, solid] (c3)--(c4)--(c5)--(c6)--(c7);
\draw[draw, solid, thick] (f1)--(c2);
\draw[draw, solid, thick, snake it] (s1)--(f1);
\draw[draw, solid, thick, snake it] (s2)--(f2);
\draw[draw, solid, thick, snake it] (s3)--(c2);
\draw[very thick,blue] (c2)--(f1) node[midway, above]  {\hspace{-2.5cm} $k\frac{N}{m} = (2\fn-1) k$};
\draw[very thick, red] (c3)--(f1) node[midway, above] {\hspace{1.7cm} $\frac{N}{m}=2\fn-1$};
\draw[very thick, blue] (c2)--(f2) node[midway,below]  {\hspace{-0.7cm}$k\frac{N}{m}$};
\draw[very thick, red] (c3)--(f2) node[midway, below] {\hspace{0.3cm}$\frac{N}{m}$};
\draw[very thick, blue] (f1)--(f2); 
\end{tikzpicture}} \quad_{/\BZ_2} \\
& \text{+ $H_{\text{free}} = 3(\fn-1)(k-1)$ free hypermultiplets}
}
where each red line has multiplicity $N/m=2\fn-1$ and each blue line has multiplicity $kN/m=(2\fn-1)k$. Decoupling the tail, we obtain the following mirror theory for $\SO(12\fn-6)^{(6\fn-3)}[3k]$:
\bes{
&
\scalebox{0.9}{
\begin{tikzpicture}[baseline=0, font=\footnotesize]
\tikzstyle{every node}=[minimum size=0.5cm,inner sep=1.5pt]
\node[draw, circle] (c2) at (0,1) {$D_1$};
\node[draw, circle] (f1) at (-1.2,0)  {$D_1$}; 
\node[draw, circle] (f2) at (1.2,0)  {$D_1$}; 
\node[draw=none] (s1) at (-3.5,0)  {$[k(\fn-1)]_2$}; 
\node[draw=none] (s2) at (3.5,0)  {$[k(\fn-1)]_2$}; 
\node[draw=none] (s3) at (0,2.3)  {$[k(\fn-1)]_2$}; 
\draw[draw, solid, snake it] (s1)--(f1);
\draw[draw, solid, snake it] (s2)--(f2);
\draw[draw, solid, snake it] (s3)--(c2);
\draw[very thick,blue] (c2)--(f1) node at (-1.2,0.7)  {$(2\fn-1)k$};
\draw[very thick, blue] (c2)--(f2) node at (0,-0.4)  {};
\draw[very thick, blue] (f1)--(f2); 
\end{tikzpicture}}   \quad_{/\BZ_2}
 \\ 
& \text{+ $3(\fn-1)(k-1)$ free hypermultiplets}
}

\section{$D^{2N-2}_p(\SO(2N))$, with $p\leq 2N-2$}
\label{sec:D2N-2SO2Npls2N-2}

\subsection{Example: The $D^{12}_{4}(\SO(14))$ theory}
Let us analyze this theory along the line of Section \ref{sec:confmfoldDpSO2N}. We have $m=2$, $n=3$ and $q=2$. From \eref{dimcmDpSO2N}, the $D^{12}_{4}(\SO(14))$ theory has a one dimensional conformal manifold, and the corresponding marginal deformation is $x^3z^2$, which gives $k=1$, $N'=3$ and $p'=1$.  We thus see that the sector at $z=\infty$ is an $\SO(8)$ gauge theory coupled to $D^6_2(\SO(8))$. We propose the following description:
\bes{
D^{12}_{4}(\SO(14)) \, = \, \left[ D^{18}_2(\SO(20)) \, \longleftarrow \SO(8) \longrightarrow \, D^6_2(\SO(8)) \right]
}
where the $D^{18}_2(\SO(20))$ theory can be determined from the Coulomb branch spectra and the $(a,c)$ central charges of the above theories.  These are as follows;
\bes{
\renewcommand{\arraystretch}{1.25}
\begin{array}{c|c|c|c}
\text{Theory}  & \text{CB spectrum} & a & c \\
\hline
D^{12}_{4}(\SO(14)) & \{2, 3, 3, 4, 4, 5, 6, 7, 9 \} & 167/8 & 45/2 \\
\hline
D^{18}_2(\SO(20)) & \{ 3, 5, 7, 9 \} & 40/3 & 47/3 \\
\hline
D^6_2(\SO(8)) & \{ 3 \} & 41/24 & 13/6 \\
\end{array}
}
From the Coulomb branch spectra, we see that the complement of that in the first line to the second plus the third line are $\{2,4,4,6 \}$, which are precisely the Casimirs of $\SO(8)$.  Moreover, from the central charges, the differences $167/8 - (40/3+41/24) = 35/6$ and $45/2 - (47/3+13/6) = 14/3$ are respectively the $(a,c)$ central charges of the free $\SO(8)$ vector multiplet.

\subsubsection*{Reduction to $3$d and the mirror theory}
As pointed out in \cite[(7.19)]{Cecotti:2013lda}, the $D^{6}_2(\SO(8))$ theory, which coincides with the $E_6$ Minahan-Nemeschansky theory, can be realized as an IR fixed point of the $\USp(4)$ gauge theory with 5 flavors:
%\bes{\label{USp4w5flv}
%[\SO(8)]-\USp(4)-[\SO(2)]
%}
\bes{\label{USp4w5flv}
	[D_4]-C_2-[D_1]
}
Upon reduction to $3$d, \eref{USp4w5flv} gives rise to the $T_{\left[5,5\right]}[\SO(10)]$ theory, which has two quaternionic dimensional Coulomb branch.  However, the reduction of the $E_6$ MN theory to $3$d yields an SCFT with one quaternionic dimensional Coulomb branch.  In Appendix \ref{app:T55gaugedE6}, we demonstrate that upon gauging the $\SO(2)$ Coulomb branch symmetry of the $T_{\left[5,5\right]}[\SO(10)]$ theory (unfortunately, this symmetry is not manifest in the UV description \eref{USp4w5flv} upon reduction), we obtain the $3$d reduction of the $E_6$ MN theory:
\bes{ \label{T55gaugedE6}
T_{\left[5,5\right]}[\SO(10)]/\SO(2)_\CC  =  \left( D^6_2(\SO(8)) \right)_{3d}  = \left(\text{$E_6$ MN} \right)_{3d}~,
}
where the notation $/\SO(2)_\CC$ denotes the $3$d $\CN=4$ gauging of the $\SO(2)$ Coulomb branch symmetry.  Note that this $\SO(2)$ is the global symmetry associated with the D-partition $[5,5]$ of $\SO(10)$.

Since the Higgs branch of the $E_6$ MN theory is the reduced moduli space of one $E_6$ instanton on $\BC^2$ \cite{Benvenuti:2010pq}, the leftmost theory in  \eref{T55gaugedE6}, namely the $3$d $\CN=4$ $\USp(4)$ gauge theory with 5 flavors with the $\SO(2)$ topological symmetry being gauged, provides the ADHM construction of such an instanton moduli space. 

Similarly, for the $D^{18}_2(\SO(20))$ theory, it can be realized as the fixed point of the $\USp(10)$ gauge theory with 11 flavors: 
%\bes{ \label{USp10w11flv}
%[\SO(14)]-\USp(10)-[\SO(8)]
%}
\bes{ \label{USp10w11flv}
	[D_7]-C_5-[D_4]
}
The reduction of this theory to $3$d gives $T_{\left[11,11\right]}[\SO(22)]$.  Again, we propose that
\bes{ \label{T1111gauged}
T_{\left[11,11\right]}[\SO(22)]/\SO(2)_\CC =  \left( D^{18}_2(\SO(20))\right)_{3d}~,
}
where $\SO(2)_\CC$ is the Coulomb branch symmetry associated with the D-partition $[11,11]$ of $\SO(22)$.

We then propose that the reduction of the $D^{12}_{4}(\SO(14))$ theory is to consider the following $3$d theory
%\bes{
%& [\SO(14)]-\USp(10)-[\SO(8)] \,\, \longleftarrow \SO(8) \longrightarrow \,\,[\SO(8)]-\USp(4)-[\SO(2)] \\
%&= \,\, [\SO(14)]-\USp(10)-\SO(8)-\USp(4)-[\SO(2)]~,
%}
\bes{
	& [D_7]-C_5-[D_4] \,\, \longleftarrow D_4 \longrightarrow \,\,[D_4]-C_2-[D_1] =  [D_7]-C_5-D_4-C_2-[D_1]~,
}
which flows to $T^{\left[3^2,1^{14}\right]}_{\left[5^4\right]}[\SO(20)]$ in the IR, whose Coulomb branch symmetry is $\SO(4)$ associated with the $D$-partition $\left[5^4\right]$ of $\SO(20)$, and then gauge the subgroup $\SO(2) \times \SO(2)$ of this $\SO(4)$. In other words,
\bes{ \label{D64OS83d}
\left( D^{12}_{4}(\SO(14)) \right)_{3d} \,\, = \,\, \frac{T^{\left[3^2,1^{14}\right]}_{\left[5^4\right]}[\SO(20)]}{\SO(2)_{\CC_1} \times \SO(2)_{\CC_2}}~.
}
where $\SO_{\CC_1}$ and $\SO_{\CC_2}$ denote the $\SO(2)_\CC$ quotients in \eref{T55gaugedE6} and \eref{T1111gauged}.

Let us now consider the mirror theory of \eref{D64OS83d}.  The mirror theory of $T^{\left[3^2,1^{14}\right]}_{\left[5^4\right]}[\SO(20)]$ is $T^{\left[5^4\right]}_{\left[3^2,1^{14}\right]} [\SO(20)]$, whose quiver description is
\bes{
\begin{array}{ll}
D_1-C_2-D_4-&C_5-D_5-C_4-D_4-C_3-D_3-C_2-D_2-C_1-D_1\\
                        &\,\, | \\
                         & \!\![D_2]
\end{array}                       
}                                  
where we use the shorthand notations $C_n = \USp(2n)$ and $D_m = \SO(2m)$.  Under mirror symmetry, the $\SO(2)_{\CC_1} \times \SO(2)_{\CC_2}$ in the denominator of \eref{D64OS83d} becomes the $\SO(2) \times \SO(2)$ subgroup of the flavor symmetry $\left[D_2\right]=[\SO(4)]$ in the above mirror theory.  After gauging, we obtain a quiver description of the mirror theory of $\left( D^{12}_{4}(\SO(14)) \right)_{3d} $ as follows:
\bes{\label{mirrD64SO8} 
&\left( D^{12}_{4}(\SO(14)) \right)_{\text{$3$d mirr}}: \\
&
\begin{array}{ll}
                        & \!\!\U(1) \\
                        &\,\, | \\
D_1-C_2-D_4-&C_5-D_5-C_4-D_4-C_3-D_3-C_2-D_2-C_1-D_1\\
                        &\,\, | \\
                         & \!\!\U(1)
\end{array}   \quad /\BZ_2
} 
where, as before, $/\BZ_2$ denotes the diagonal $\BZ_2$ quotient.  This is a star-shaped quiver which is the mirror theory of the $3$d reduction of the following theory of class $\CS$:
\bes{ \label{classSD124SO14}
D^{12}_{4}(\SO(14)) \,\, = \,\, &\text{twisted $A_9$ theory associated with a sphere} \\
&\text{with two twisted punctures $\left[1^{11}\right]_t$, $\left[3^3, 1^2\right]_t$,} \\
&\text{and two minimal untwisted punctures $[9,1]$, $[9,1]$} \\
}
where the subscript $t$ denotes a twisted puncture, which in this case is labeled by a $B$-partition of $B_5 = \SO(11)$. Each leg of \eref{mirrD64SO8} comes from the following theories:
\bes{
T_{\left[1^{11}\right]}[\USp(10)]: &\quad [C_5]-D_5-C_4-D_4-C_3-D_3-C_2-D_2-C_1-D_1 \\
T_{\left[3^3, 1^2\right]}[\USp(10)]: &\quad D_1-C_2-D_4-[C_5] \\
T_{\left[9,1\right]}[\SU(10)]: &\quad \U(1)-[A_9] \\
T_{\left[9,1\right]}[\SU(10)]: &\quad \U(1)-[A_9]
}
where the common subgroup $C_5=\USp(10)$ of the flavor symmetry of each theory is gauged to form the central node in \eref{mirrD64SO8}.   The $(a,c)$ central charges class $\CS$ theory described in \eref{classSD124SO14} can be computed from the information given by \cite[Table 3, Section 3.5.2]{Chacaltana:2012zy} and \cite[Appendix A.4]{Chacaltana:2012ch}:
\bes{
(n_h, n_v) &= (660,637)+(576,571)+2(100,99) - (1320,1329) = (116,77)  \\
\Rightarrow \quad (a,c) &= (167/8, 45/2)~,
}
in agreement with that of the $D^{12}_{4}(\SO(14))$ theory.  It can also be checked that the Coulomb branch spectra of the two theories match perfectly.

\subsection{The $D^{8M+4}_{4}(\SO(8M+6))$ theory}
Let us analyze this theory along the line of Section \ref{sec:confmfoldDpSO2N}. We have $m=2$, $n=2M+1$ and $q=2$. From \eref{dimcmDpSO2N}, the $D^{8M+4}_{4}(\SO(8M+6))$ theory has a one dimensional conformal manifold, and the corresponding marginal deformation is $x^{2M+1}z^2$, which gives $k=1$, $N'=2M+1$ and $p'=1$.  We thus see that the sector at $z=\infty$ is an $\SO(4M+4)$ gauge theory coupled to $D^{4M+2}_2(\SO(4M+4))$. We propose the following description:
\bes{ \label{glueD4SO8Mp6}
&D^{8M+4}_{4}(\SO(8M+6)) \\ 
&=\,\, \left[ D^{12M+6}_2(\SO(12M+8)) \, \longleftarrow \SO(4M+4) \longrightarrow \, D^{4M+2}_2(\SO(4M+4))  \right] 
}
where the $D^{12M+6}_2(\SO(12M+8))$ theory can be determined from the $(a,c)$ central charges and Coulomb branch spectra of the above theories.

\subsubsection*{Reduction to $3$d and mirror theory}

The $D^{4M+2}_2(\SO(4M+4))$ theory can be realized as an IR fixed point of the $\USp(2M+2)$ gauge theory with $2M+3$ flavors:
%\bes{ \label{USpwMp2flvA}
%&[\SO(4M+4)] - \USp(2M+2)- [\SO(2)]~,
%}
\bes{ \label{USpwMp2flvA}
	&[D_{2M+2}] - C_{M+1}- [D_{1}]~.
}
Upon compactifying to $3$d, \eref{USpwMp2flvA} flows to the $T_{\left[2M+3, 2M+3\right]}[\SO(4M+6)]$ theory, with the $\U(1)_\CC$ Coulomb branch symmetry.  Upon gauging this $\U(1)_\CC$ symmetry, we obtain the $3$d reduction of the $D^{4M+2}_2(\SO(4M+4))$ theory.  The similar argument applies also for $D^{12M+6}_2(\SO(12M+8))$, whose corresponding quiver is
%\bes{ \label{USpwMp2flv}
%&[\SO(8M+6)] - \USp(6M+4)- [\SO(4M+4)]~.
%}
\bes{ \label{USpwMp2flv}
	&[D_{4M+3}] -C_{3M+2}- [D_{2M+2}]~.
}
This again flows to an SCFT with the $\U(1)$ Coulomb branch symmetry, where, upon gauging this symmetry, we obtain the $3$d reduction of the $D^{12M+6}_2(\SO(12M+8))$ theory.

To study the $3$d reduction of the $D^{8M+4}_{4}(\SO(8M+6))$ theory, we first consider the following quiver theory
%\bes{
%[\SO(8M+6)] - \USp(6M+4)- (\SO(4M+4))- \USp(2M+2)- [\SO(2)]
%}
\bes{
	[D_{4M+3}] -C_{3M+2}- (D_{2M+2})- C_{M+1}- [D_1]
}
which flows to the $T^{\left[3^2, 1^{8M+6}\right]}_{\left[(2M+3)^4\right]}[\SO(8M+12)]$ theory in the IR.  The latter has $\SO(4)$ Coulomb branch symmetry, as can be seen from the partition $\left[{(2M+3)}^4\right]$.  Upon gauging $\SO(2) \times \SO(2)$ subgroup of this $\SO(4)$ symmetry, we obtain the reduction of $D^{8M+4}_{4}(\SO(8M+6))$ to $3$d:
\bes{
\left(D^{8M+4}_{4}(\SO(8M+6)) \right)_{3d} \,\, = \,\, \left[  \frac{T^{\left[3^2, 1^{8M+6}\right]}_{\left[(2M+3)^4\right]}[\SO(8M+12)]}{ \SO(2)_{\CC_1} \times \SO(2)_{\CC_2}} \right]~.
}

Let us now consider the mirror theory.  The mirror theory of $T^{\left[3^2, 1^{8M+6}\right]}_{\left[(2M+3)^4\right]}[\SO(8M+12)]$ is $T^{\left[(2M+3)^4\right]}_{\left[3^2, 1^{8M+6}\right]}[\SO(8M+12)]$, whose quiver description is
\be
\scalebox{0.8}{$
\begin{split}
D_1-C_2-D_4-C_5 -\cdots-D_{3M+1}- &C_{3M+2} -D_{3M+2}-C_{3M+1}-D_{3M+1}- \cdots-C_2-D_2-C_1-D_1\\
                                                    &\,\, | \\
                                                    & \!\![D_2]
\end{split}$}                                                                                                     
\ee
Gauging the $\SO(2)\times \SO(2)$ subgroup of the $\left[D_2\right]$ flavor symmetry yields a quiver description for the mirror theory of $\left(D^{8M+4}_{4}(\SO(8M+6)) \right)_{3d}$:
\bes{ \label{mirrD48Mp6}
&\left(D^{8M+4}_{4}(\SO(8M+6)) \right)_{\text{3d mirr}}: \\
&\scalebox{0.75}{$
\begin{array}{ll}
\\
                                                     & \!\!\U(1) \\
                                                    &\,\, | \\
D_1-C_2-D_4-C_5 -\cdots-D_{3M+1}- &C_{3M+2} -D_{3M+2}-C_{3M+1}-D_{3M+1}- \cdots-C_2-D_2-C_1-D_1\\
                                                    &\,\, | \\
                                                    & \!\!\U(1)
\end{array}$}  \quad /\BZ_2                                                                                                     
}
This is a star-shaped quiver which is the mirror theory of the $3$d reduction of the following theory of class $\CS$:
\bes{ \label{classSD8Mp44SO8Mp6}
D^{8M+4}_{4}(\SO(8M+6))  \,\, = \,\, &\text{twisted $A_{6M+3}$ theory associated with a sphere} \\
&\text{with two twisted punctures $\left[1^{6M+5}\right]_t$, $\left[3^{2M+1}, 1^2\right]_t$,} \\
&\text{and two minimal untwisted punctures} \\
&\text{$[6M+3,1]$, $[6M+3,1]$} \\
}

\subsubsection*{The special case of $M=0$}
Due to the isomorphism between the Lie algebras of $\SU(4)$ and $\SO(6)$, the case of $M=0$ in the above discussion
 gives an alternative description of the $D^4_4(\SU(4))$ theory, which is a Lagrangian theory described by \cite{Cecotti:2013lda} (see also \cite[(4.12)]{Giacomelli:2020ryy}):
\bes{ \label{quiverD4SU4}
D^4_4(\SU(4)): \quad  [4]-\SU(3)-\SU(2)-[1]
}
In particular, it is instructive to compare this to \eref{glueD4SO8Mp6} with $M=0$, in which case $D^2_2(\SO(4)) = D^2_2(\SU(2)) \times D^2_2(\SU(2))$ is simply two copies of hypermultiplets, and the $D^6_2(\SO(8))$ theory is the $E_6$ MN theory \cite{Cecotti:2013lda}.  Theories \eref{glueD4SO8Mp6} for $M=0$ and \eref{quiverD4SU4} are related to each other by the Argyres-Seiberg duality \cite{Argyres:2007cn}, where we dualize the $\SU(3)$ node, which has six fundamental flavors transforming under it, to the $E_6$ MN theory with an $\SU(2)$ subgroup of $E_6$ being gauged and coupled to one flavor of the hypermultiplet.  Moreover, \eref{mirrD48Mp6} with $M=0$, namely
\bes{ \label{TSO6gauged}
\begin{split}
                                                     & \!\!\U(1) \\
                                                    &\,\, | \\
\left(D^{4}_{4}(\SO(6)) \right)_{\text{$3$d mirr}}: \qquad D_1- &C_{2} -D_{2}-C_{1}-D_{1} \\
                                                    &\,\, | \\
                                                    & \!\!\U(1)
\end{split} \qquad /\BZ_2
} 
gives an alternative description of the mirror theory of the reduction of $D^4_4(\SU(4))$ to $3$d:
\bes{\label{TSU4gauged}
(D^4_4(\SU(4)))_{\text{$3$d mirr}}:  \quad
&\scalebox{0.8}{
\begin{tikzpicture}[baseline=0]
\node[draw,circle] (s1) at (-1.5,3/2) {1};
\node[draw,circle] (s2) at (-1.5,1/2) {1};
\node[draw,circle] (s3) at (-1.5,-1/2) {1};
\node[draw,circle] (s4) at (-1.5,-3/2) {1};
\node[draw,circle] (3a) at (0,0) {3};
\node[draw,circle] (2a) at (1.5,0) {2};
\node[draw,circle] (1a) at (3,0) {1};
\draw (3a)--(2a)--(1a);
\draw (s1)--(3a);
\draw (s2)--(3a);
\draw (s3)--(3a);
\draw (s4)--(3a);
\end{tikzpicture}}
}
It can be checked similarly to Appendix \ref{app:SO6vsSU4} that the Coulomb and Higgs branch Hilbert series of \eref{TSO6gauged} and \eref{TSU4gauged} are in agreement with one another (see also Section 2 of \cite{Bourget:2021zyc}).

\subsection{The $D^{2\fp(2 M+1)}_{2 \fp}(\SO(4\fp M+2\fp+2))$ theory}
Similarly to the previous discussion, we propose that the $D^{2\fp(2 M+1)}_{2 \fp}(\SO(4\fp M+2\fp+2))$ theory admits the following class $\CS$ description 
\bes{ \label{classSD8Mp44SO8Mp6-2}
&D^{2\fp(2 M+1)}_{2 \fp}(\SO(4\fp M+2\fp+2)) \\
 = \,\, &\text{twisted $A_{(4 \fp-2)M+2\fp -1}$ theory associated with a sphere} \\
&\text{with two twisted punctures $\left[1^{(4 \fp-2)M+2\fp +1}\right]_t$, $\left[(2 \fp-1)^{2M+1},1^2\right]_t$,} \\
&\text{and $\fp$ minimal untwisted punctures, each labeled by} \\
&\text{$[(4 \fp-2)M+2\fp -1,1]$}~.
}
According to \eref{dimcmDpSO2N}, this theory has a $(\mathfrak{p}-1)$-dimensional conformal manifold.

Upon reduction to $3$d, this can be identified as
\bes{
\left(D^{2\fp(2 M+1)}_{2 \fp}(\SO(4\fp M+2\fp+2))  \right)_{3d} \,\, =\,\, \frac{T^{\left[(2\fp-1)^2, 1^{4 \fp M +2 \fp +2}\right]}_{\left[(2M+3)^{2\fp}\right]}[\SO(4\fp M+6\fp)]}{\SO(2)^\fp}
}
where the Cartan subalgebra $\SO(2)^\fp$ of the Coulomb branch symmetry $\SO(2\fp)$ of the $T^{\left[(2\fp-1)^2, 1^{4 \fp M +2 \fp +2}\right]}_{\left[(2M+3)^{2\fp}\right]}[\SO(4\fp M+6\fp)]$ theory is gauged.  Note that the quiver description for the latter is
%\bes{
%&T^{\left[(2\fp-1)^2, 1^{4 \fp M +2 \fp +2}\right]}_{\left[(2M+3)^{2\fp}\right]}[\SO(4\fp M+6\fp)]: \\
%&[\SO(4 \fp M+ 2\fp +2)] - \USp((4 \fp-2) M+ 2\fp ) - \SO((4 \fp-4) M+ 2\fp ) \\
%&- \USp((4 \fp-6) M+ 2\fp-2 )- \cdots -\USp(2M+2)  - [\SO(2)]
%}
\bes{
	&T^{\left[(2\fp-1)^2, 1^{4 \fp M +2 \fp +2}\right]}_{\left[(2M+3)^{2\fp}\right]}[\SO(4\fp M+6\fp)]: \\
	&[D_{2\fp M+\fp+1}] - C_{(2\fp-1)M+\fp} - D_{(2\fp-2)M+\fp} -C_{(2\fp-3)M+\fp-1}- \cdots -C_{M+1}  - [D_1]
}
where there are $2\fp-1$ gauge groups in total.
The mirror of this theory is 
\bes{
&T_{\left[(2\fp-1)^2, 1^{4 \fp M +2 \fp +2}\right]}^{\left[(2M+3)^{2\fp}\right]}[\SO(4\fp M+6\fp)]: \\
&\scalebox{0.9}{
\begin{tikzpicture}[baseline=0]
\node[draw=none] (mid) at (0,0) {$C_{x}$};
\node[draw=none] (leftleg) at (-5.3,0)  {$D_1 - C_\fp - D_{2\fp} - C_{3\fp-1}-D_{4\fp-1} - \cdots -D_{x-(\fp-1)}$};
\node[draw=none] (rightleg) at (3.8,0)  {$D_{x} - C_{x-1} - D_{x-1}\cdots -C_1-D_1 $};
\node[draw=none] (U1b) at (0,-1)  {$\left[D_{\fp}\right]$};
\draw[thick,solid] (leftleg)--(mid)--(rightleg);
\draw[thick,solid] (mid)--(U1b);
\end{tikzpicture}}
}
where we define
\bes{
x = (2\fp-1)M+\fp ~.
}
The mirror theory of the $3$d reduction of $D^{2\fp(2 M+1)}_{2 \fp}(\SO(4\fp M+2\fp+2))$ can then be obtained by gauging the Cartan subalgebra of the flavor symmetry $D_{2\fp}$ in the above quiver. As a result, we obtain
\bes{
&\left( D^{2\fp(2 M+1)}_{2 \fp}(\SO(4\fp M+2\fp+2))  \right)_{\text{$3$d mirr}}: \\
&\scalebox{0.9}{$
\begin{tikzpicture}[baseline=0]
\node[draw=none] (mid) at (0,0) {$C_{x}$};
\node[draw=none] (leftleg) at (-4.5,0)  {$D_1 - C_\fp - D_{2\fp} - C_{3\fp-1} - \cdots -D_{x-(\fp-1)}$};
\node[draw=none] (rightleg) at (3.8,0)  {$D_{x} - C_{x-1} - D_{x-1}\cdots -C_1-D_1 $};
\node[draw=none] (U1a) at (-1,-1)  {$(D_1)$};
\node[draw=none] (U1b) at (0,-1)  {\large $\cdots$};
\node[draw=none] (U1c) at (1,-1)  {$(D_1)$};
\draw[thick,solid] (leftleg)--(mid)--(rightleg);
\draw[thick,solid] (mid)--(U1a);
\draw[thick,solid] (mid)--(U1c);
\draw [thick,decorate,decoration={brace,amplitude=10pt,mirror},xshift=0.4pt,yshift=-0.4pt](-1.4,-1.3) -- (1.4,-1.3) node[black,midway,yshift=-0.6cm] {$\fp$ nodes};
\end{tikzpicture} \qquad /\BZ_2 $}
}
For the special case of $M=0$, the above mirror theory reduces to 
\bes{
\left( D^{2\fp}_{2 \fp}(\SO(2\fp+2))  \right)_{\text{$3$d mirr}}:\quad 
\begin{tikzpicture}[baseline=0]
\node[draw=none] (mid) at (0,0) {$C_{\fp}$};
\node[draw=none] (rightleg) at (3.8,0)  {$D_{\fp} - C_{\fp-1} - D_{\fp-1}\cdots -C_1-D_1 $};
\node[draw=none] (U1a) at (-1,-1)  {$(D_1)$};
\node[draw=none] (U1b) at (0,-1)  {\large $\cdots$};
\node[draw=none] (U1c) at (1,-1)  {$(D_1)$};
\draw[thick,solid] (mid)--(rightleg);
\draw[thick,solid] (mid)--(U1a);
\draw[thick,solid] (mid)--(U1c);
\draw [thick,decorate,decoration={brace,amplitude=10pt,mirror},xshift=0.4pt,yshift=-0.4pt](-1.4,-1.3) -- (1.4,-1.3) node[black,midway,yshift=-0.6cm] {$\fp+1$ nodes};
\end{tikzpicture} \qquad /\BZ_2
}

\subsection{Comments on the $D^{2Mp}_p(\SO(2Mp+2))$ theory}
It was pointed out in \cite[Appendix C.2]{Cecotti:2013lda} that $D^{2Mp}_p(\SO(2Mp+2))$ is in fact a Lagrangian theory, whose quiver description is
\bes{ \label{quivb2Mp}
\scalebox{0.9}{$
\begin{array}{ll}
~[D_{Mp+1}] - C_{M(p-1)}  - D_{M(p-2)+1}- C_{M(p-3)} - D_{M(p-4)+1}-\cdots -D_{2M+1}-C_M-[D_1]~,  &\quad \text{$p$ even}\\
~[D_{Mp+1}] - C_{M(p-1)}  - D_{M(p-2)+1}- C_{M(p-3)} - D_{M(p-4)+1}- \cdots - C_{2M} - D_{M+1}~, &\quad \text{$p$ odd} 
\end{array}
$}
}
In these $4$d $\CN=2$ theories, each $C$ and $D$ gauge group has zero beta-function,\footnote{The beta-functions of the $C_N$ gauge group with $2N+2$ flavors of fundamental hypermultiplets and the $D_N$ gauge group with $2N-2$ flavors of vector hypermultiplets are zero.} i.e.  all gauge groups are conformal.  However, upon reduction to $3$d, if we {\it assume} that we obtain the same quiver gauge theory with $3$d $\CN=4$ supersymmetry, then each conformal $C$-gauge group is overbalanced, and each conformal $D$-gauge group is underbalanced.\footnote{The conditions for a $C_N$ gauge group with $F_{C_N}$ flavors of fundamental hypermultiplets and a $D_N$ gauge group with $F_{D_N}$ flavors of vector hypermultiplets to be balanced are, respectively, $F_{C_N}=2N+1$ and $F_{D_N}=2N-1$.  In each case, if the number of flavors are fewer (resp. greater) than the said $F_{C_N}$ or $F_{D_N}$, then the corresponding gauge group is said to be underbalanced (resp. overbalanced) \cite{Gaiotto:2008ak}.}  The presence of the latter renders the quiver gauge theory in question a ``bad theory'' in the sense of \cite{Gaiotto:2008ak}.   

For example, the $D^{8M}_{4}(\SO(8M+2))$ theory has the following Lagrangian description: 
%\bes{
%[\SO(8M+2)]-\USp(6M)-\SO(4M+2)-\USp(2M)-[\SO(2)]
%}
\bes{
	[D_{4M+1}]-C_{3M}-D_{2M+1}-C_{M}-[D_{1}]
}
As a $3$d $\CN=4$ gauge theory, this quiver is a bad theory, due to the presence of the underbalanced $\SO(4M+2)$ node.  Nevertheless, it can be identified with the $T^{\vec \sigma}_{\vec \rho}[\SO(8M+8)]$ theory, where $\vec \sigma={\left[3^2,1^{8M+2}\right]}$ and $\vec \rho = {\left[2M+3, (2M+2)^2, 2M+1\right]}$   The mirror theory of the latter, namely $T_{\vec \sigma}^{\vec \rho}[\SO(8M+8)]$, admits the following Lagrangian description
\bes{
&\scalebox{0.9}{$
\begin{array}{llll}
D_1 - C_2 - D_4 - C_5 - \cdots D_{3M-2}- &C_{3M-1}- &B_{3M}- &C_{3M}- D_{3M} - \cdots - C_2-D_2 - C_1-D_1\\
                                                  &\,\, |          & \,\,|           & \,\,| \\
                                                  &\!\![O(1)]     &\!\![C_1]      &\!\![O(1)] \\
\end{array}$}
}
We propose that this is a mirror for the $D^{8M}_{4}(\SO(8M+2))$ theory.

Another example is the $D^{10M}_{5}(\SO(10M+2))$ theory, whose quiver description is
%\bes{
%[\SO(10M+2)]-\USp(8M)-\SO(6M+2)-\USp(4M)-\SO(2M+2)
%}
\bes{
	[D_{5M+1}]-C_{4M}-D_{3M+1}-C_{2M}-D_{M+1}
}
As a $3$d $\CN=4$ gauge theory, this can be identified with the $T_{\left[2M+1, (2M)^4,1\right]}[\SO(10M+2)]$ theory.
The mirror theory $T^{\left[2M+1, (2M)^4,1\right]}[\SO(10M+2)]$ admits the following quiver description:
\bes{
&\scalebox{0.9}{$
\begin{array}{llll}
~&C_2 - B_4 - C_6 -B_8 - \cdots C_{4M-2}- &B_{4M}- &C_{4M}- D_{4M}- C_{4M-1}-D_{4M-1} - \cdots  - C_1-D_1\\
~&\,\, |          & \,\,|           & \,\,| \\
~&\!\![O(1)]     &\!\![C_2]      &\!\![O(1)] \\
\end{array}$}
}
We propose that this is a mirror for the $D^{10M}_{5}(\SO(10M+2))$ theory.

The special case of $D^8_2(\SO(10))$ is also worth discussing.  This $4$d theory admits the Lagrangian description in terms of the $\USp(4)$ gauge theory with 6 hypermultiplets in the fundamental representation \cite{Cecotti:2013lda}.  In terms of a $3$d $\CN=4$ gauge theory, this flows to the $T_{\left[7,5\right]}[\SO(12)]$ theory.  The mirror theory, namely, $T^{\left[7,5\right]}[\SO(12)]$, admits the following quiver description
\bes{ \label{mirrD2SO10}
&
\begin{array}{llll}
D_1  -C_1-D_2- &C_{2}- &B_{2}- &C_{2}- D_{2}- C_1-D_1\\
                          &\,\, |          &           & \,\,| \\
                        &\!\![O(1)]     &     &\!\![O(1)] \\
\end{array}
}
This is precisely the $3$d mirror theory \cite{Benini:2010uu} for the class $\CS$ theory of the twisted $D_3$ type associated with a sphere with two untwisted punctures $\left[1^6\right]$, $\left[1^6\right]$ and two twisted punctures $[4]_t$, $[4]_t$.  Such a theory of class $\CS$ indeed describes the $4$d $\CN=2$ $\USp(4)$ gauge theory with 6 fundamental flavors\footnote{Note that this theory is also Argyres-Seiberg dual \cite{Argyres:2007cn} to the $E_7$ MN theory coupled to the $\SU(2)$ super-Yang-Mills.} \cite{Chacaltana:2013oka}.  We thus conclude that \eref{mirrD2SO10} is a mirror theory for the $D^8_2(\SO(10))$ theory.  Moreover, due to the isomorphism between $D_3$ and $A_3$, the aforementioned class $\CS$ theory can also be described as that of the twisted $A_3$ type \cite{Chacaltana:2012ch} associated with a sphere with two untwisted punctures $\left[1^4\right]$, $\left[1^4\right]$ and two twisted punctures $[5]_t$, $[5]_t$.  We thus propose another description of the mirror theory for the $D^8_2(\SO(10))$ theory as follows:
\bes{ \label{mirrD2SO10B}
&
\begin{array}{llll}
\U(1) -\U(2)-\U(3) -&\USp(4)- &\U(3)- &\U(2)- \U(1) \quad /\BZ_2
\end{array}
}

\acknowledgments
We thank Matteo Sacchi for the collaboration at the beginning of this project and Stefano Cremonesi for asking the question regarding discrete gaugings. F.C. is supported by STFC consolidated grant ST/T000708/1. The work of S.G. is supported by the ERC Consolidator Grant 682608 Higgs bundles: Supersymmetric Gauge Theories and Geometry (HIGGSBNDL). N.M. thanks Stefano Lionetti for his hospitality at \href{https://www.zetalab.com/}{Zetalab} during the completion of the project. A. M. received funding from ``la Caixa" Foundation (ID 100010434) with fellowship code LCF/BQ/IN18/11660045 and from the European Union’s Horizon 2020 research and innovation programme under the Marie Sk\l odowska-Curie grant agreement No. 713673. 

\appendix
\section{Hilbert series}
\label{sec:HilbSer}

In this appendix, we compute the Hilbert series of various theories discussed in the main text.  In particular, we match the Higgs and Coulomb branch Hilbert series for the dual theories that admit different quiver descriptions, as well as discuss certain properties of the moduli space.  Regarding the Coulomb branch Hilbert series computations, the magnetic lattices for orthosymplectic quivers were spelled out explicitly in \cite{Bourget:2020xdz}.

\subsection{Relation \eref{SO2SO2factorize}} \label{app:SO2SO2factorize}
\subsubsection*{The Higgs branch Hilbert series}
The Higgs branch Hilbert series of the theory on the left-hand side of \eref{SO2SO2factorize} can be written as follows, following the convention \eref{repblueedge}:
\bes{
&H[\text{HB}\eref{SO2SO2factorize}](t; \vec x_1, \vec x_2) \\
&= \oint_{|z_1|=1} \frac{d z_1}{2 \pi i z_1}   \oint_{|z_2|=1} \frac{d z_2}{2 \pi i z_2}  \times \\
& \quad \times\PE \left[\chi^{\SU(\fm)}_{\left[0,\ldots,0,1\right]} (\vec x_1) q^{-1} z_1 z_2  t +\chi^{\SU(\fm)}_{\left[1,0,\ldots,0\right]} (\vec x_1) q z_1^{-1} z_2^{-1}  t +\right. \\
& \left.\qquad + \chi^{\SU(\fm)}_{\left[1,0,\ldots,0\right]} (\vec x_2) q z_1z_2^{-1} t + \chi^{\SU(\fm)}_{\left[0,\ldots,0,1\right]} (\vec x_2) q^{-1} z_1^{-1}z_2 t   -2t^2  \right]~,
}
where $z_1$ and $z_2$ are gauge fugacities for each $\U(1)$ gauge factor, and $\vec x_1$, $\vec x_2$, $q$ are the fugacities for each factor of the $\SU(\fm) \times \SU(\fm) \times \U(1)$ flavor symmetry respectively.  We will see that the fugacity $q$ can be absorbed into the gauge fugacities by a redefinition. Let us rewrite the gauge fugacities as follows:
\bes{ \label{defuv}
u = q^{-1} z_1 z_2~, \quad v = q^{-1} z_1^{-1} z_2~.
}
Then, the above Hilbert series can be rewritten as
\bes{
&H[\text{HB}\eref{SO2SO2factorize}](t; \vec x_1; \vec x_2) \\
&= \oint_{|u|=1} \frac{d u}{2 \pi i u}   \oint_{|v|=1} \frac{d v}{2 \pi i v}  \times \\
& \quad\times\PE \left[\chi^{\SU(\fm)}_{\left[1,0,\ldots,0\right]} (\vec x_1) u^{-1} t +\chi^{\SU(\fm)}_{\left[0,\ldots,0,1\right]} (\vec x_1) u  t -t^2 +\right.\\
& \left.\qquad \quad + \chi^{\SU(\fm)}_{\left[1,0,\ldots,0\right]} (\vec x_2) v^{-1} t +\chi^{\SU(\fm)}_{\left[0,\ldots,0,1\right]} (\vec x_2) v  t -t^2  \right] \\
&=\prod_{j=1}^2\left( \oint_{|u|=1} \frac{d u}{2 \pi i u} \PE \left[\chi^{\SU(\fm)}_{\left[1,0,\ldots,0\right]} (\vec x_j) u^{-1} t +\chi^{\SU(\fm)}_{\left[0,\ldots,0,1\right]} (\vec x_i) u  t -t^2 \right] \right) \\
&= \prod_{j=1}^2 \left( \sum_{r=0}^\infty \chi^{\SU(\fm)}_{\left[r,0,\cdots,0,r\right]} (\vec x_j) t^{2r} \right)~.
}
The right-hand side of the second equality is indeed the product of two HB Hilbert series of the SQED with $\fm$ hypermultiplets with charge 1. This matches the Higgs branches of the theories on both sides of \eref{SO2SO2factorize}.  The right-hand side of the third equality is the product of two Hilbert series of the reduced moduli space of one $\SU(\fm)$ instanton on $\BC^2$ \cite{Benvenuti:2010pq}, which is the closure of the minimal nilpotent orbit of $\SU(\fm)$.  

\subsubsection*{The Coulomb branch Hilbert series}

On the other hand, the Coulomb branch (CB) Hilbert series of the theory on the left-hand side of \eref{SO2SO2factorize} can be written as \cite{Cremonesi:2013lqa, Bourget:2020xdz}\footnote{Note that the fugacity $t$ in \cite{Cremonesi:2013lqa} corresponds to $t^2$ in this paper.} (see also \cite{Razamat:2014pta, Fazzi:2018rkr}):
\bes{
H[\text{CB}\eref{SO2SO2factorize}](t; \omega, x_1, x_2) =\sum_{\sigma=0}^1 \sum_{r_1 \in \BZ+\frac{\sigma}{2}}  \sum_{r_2 \in \BZ+\frac{\sigma}{2} }  \omega^{\sigma}~  t^{2 \Delta(r_1,r_2)} (1-t^2)^{-2} x_1^{r_1} x_2^{r_2}~.
}
where $\Delta(r,s)$ is the dimension of the monopole operator with the Abelian gauge flux $(r, s)$ whose expression is given by
\bes{
\Delta(r_1,r_2) = \frac{\fm}{4} \sum_{\sigma_1, \sigma_2 =0}^1 | (-1)^{\sigma_1} r_1 + (-1)^{\sigma_2} r_2| = \frac{\fm}{2} |r_1+r_2| +  \frac{\fm}{2} |r_1-r_2| ~  ;
}
$x_1$ and $x_2$ are fugacities for the topological symmetries for each $\SO(2)$ gauge group; $\omega$ is the discrete fugacity for the  topological symmetry for the $\BZ_2$ quotient in $(\SO(2)\times \SO(2))/\BZ_2$ which satisfies
\bes{
\omega^2=1~;
}
and $(1-t^2)^{-1}$ is the dressing factor associated with each $\SO(2)$ gauge group.  We can rewrite
\bes{ \label{defu1u2y1y2}
u_1=r_1+r_2~, \quad u_2= -r_1+r_2~, \quad y_1 = (x_1 x_2)^{1/2}~, \quad y_2 = (x_1^{-1} x_2)^{1/2}
}
and obtain
\bes{
H[\text{CB}\eref{SO2SO2factorize}](t;\, \omega, \, y_1 y_2, \,y_1^{-1} y_2) &= \prod_{i=1}^2 \left[ \sum_{u_i=-\infty}^\infty  t^{2 \frac{\fm}{2} |u_i|} (1-t^2)^{-1}  y_i^{u_i} \,\omega^{u_i} \right]\\
&= \prod_{i=1}^2 \PE\left[ t^2 + \omega (y_i + y_i^{-1})t^\fm - t^{2\fm} \right]~.
}
The right-hand side of the second equality is indeed the CB Hilbert series of two copies of the SQED with $\fm$ hypermultiplets with charge 1. This matches the Coulomb branches of the theories on both sides of \eref{SO2SO2factorize}. The right-hand side of the third equality is the Hilbert series for the product of two copies of $\BC^2/\BZ_\fm$.

\subsection{Relation \eref{threeSO2nodes}} \label{app:threeSO2nodes}
\subsubsection*{The Higgs branch Hilbert series}
Following the convention \eref{repblueedge}, we see that the HB Hilbert series of the theory on the left-hand side of \eref{threeSO2nodes} is given by
\bes{
H[\text{HB}\,\, &\eref{threeSO2nodes}](t; \vec x_1, \vec x_2, \vec x_3, \vec y_1, \vec y_2, \vec y_3, q_1, q_2, q_3)\\
&= \left(  \prod_{j=1}^3 \oint_{|z_j|=1} \frac{d z_j}{2 \pi i z_j} \right)  \times \PE \left[ -3 t^2 \right] \times \\
& \times\PE \left[ \sum_{i=1}^3 \left( \chi^{\SU(\fm)}_{\left[0,\cdots,0,1\right]} (\vec x_i) q^{-1}_i z_i z_{i+1} t+  \chi^{\SU(\fm)}_{\left[1,0,\cdots,0\right]} (\vec x_i) q_i (z_i z_{i+1})^{-1} t \right) \right] \\
&\times\PE \left[ \sum_{i=1}^3 \left( \chi^{\SU(\fm)}_{\left[1,0,\cdots,0\right]} (\vec y_i) q_i z_i z_{i+1}^{-1} t+  \chi^{\SU(\fm)}_{\left[0,\cdots,0,1\right]} (\vec y_i) q_i^{-1} z_i^{-1} z_{i+1} t \right) \right]
}
where each $\vec x_{1,2,3}$ and $\vec y_{1,2,3}$ denotes the fugacities of each $\SU(\fm)$ factor of the $\SU(\fm)^6$ flavor symmetry and $z_{1,2,3}$ are the fugacities for each $\U(1)$ gauge symmetry.  Here we identify $z_4 \equiv z_1$.  Similarly to the previous discussion, we define
\bes{
u_i = z_i z_{i+1}~, \quad v_i =  z_i^{-1} z_{i+1}~,  \quad i=1,2,3~.
}
There are, however, the relations:
\bes{
u_1 u_2^{-1} =  v_3~, \quad u_2 u_3^{-1} = v_1~, \quad u_3 u_1^{-1} =  v_2~.
}
We can therefore rewrite the above Hilbert series as
\bes{
&~H[\text{HB}\,\,\eref{threeSO2nodes}](t; \vec x_1, \vec x_2, \vec x_3, \vec y_1, \vec y_2, \vec y_3, q_1, q_2, q_3) \\
&= \left(  \prod_{j=1}^3 \oint_{|u_j|=1} \frac{d u_j}{2 \pi i u_j} \right)  \times \PE \left[ -3 t^2 \right] \times \\
& \quad \times\PE \left[\sum_{i=1}^3 \left( \chi^{\SU(\fm)}_{\left[1,0,\ldots,0\right]} (\vec x_i) q_i u_i^{-1} t +\chi^{\SU(\fm)}_{\left[0,\ldots,0,1\right]} (\vec x_i) q_i^{-1} u_i  t  \right)+ \right.\\
& \left.\qquad + \sum_{i=1}^3 \left( \chi^{\SU(\fm)}_{\left[1,0,\ldots,0\right]} (\vec y_i) q_i u_{i+1} u_{i+2}^{-1} t +\chi^{\SU(\fm)}_{\left[0,\ldots,0,1\right]} (\vec y_i) q_i^{-1} u^{-1}_{i+1} u_{i+2}  t \right)  \right] \\
&= \left(  \prod_{j=1}^3 \oint_{|w_j|=1} \frac{d w_j}{2 \pi i w_j} \right)  \times \PE \left[ -3 t^2 \right] \times \\
& \quad \times\PE \left[\sum_{i=1}^3 \left( \chi^{\SU(\fm)}_{\left[1,0,\ldots,0\right]} (\vec x_i) w_i^{-1} t +\chi^{\SU(\fm)}_{\left[0,\ldots,0,1\right]} (\vec x_i) w_i  t  \right)+ \right.\\
& \left.\qquad + \sum_{i=1}^3 \left( \chi^{\SU(\fm)}_{\left[1,0,\ldots,0\right]} (\vec y_i) \frac{q_{i-1} q_i}{q_{i+1}} w_i w_{i+1}^{-1} t +\chi^{\SU(\fm)}_{\left[0,\ldots,0,1\right]} (\vec y_i) \frac{q_{i+1}}{q_{i-1} q_i} w^{-1}_i w_{i+1}  t \right)  \right] 
}
where we identify $u_4 \equiv u_1$, $u_0 \equiv u_3$ and define $w_i =  q^{-1}_i u_i$.
This is the HB Hilbert series for the following quiver gauge theory
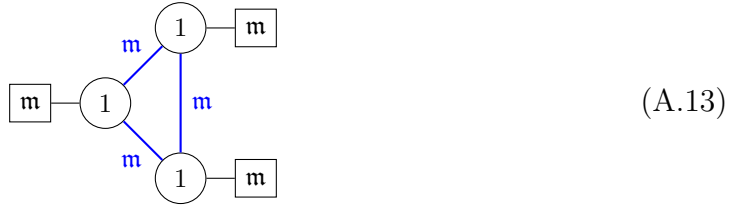
\bes{ \label{threeU1nodeswallflv}
\begin{tikzpicture}[baseline=0, font=\footnotesize]
\tikzstyle{every node}=[minimum size=0.5cm]
\node[draw, circle] (c) at (0,0)  {$1$}; 
\node[draw, circle] (f1) at (1,1)  {$1$}; 
\node[draw, circle] (f2) at (1,-1)  {$1$}; 
\node[draw, rectangle] (s0) at (-1,0)  {$\fm$}; 
\node[draw, rectangle] (s1) at (2,1)  {$\fm$}; 
\node[draw, rectangle] (s2) at (2,-1)  {$\fm$}; 
\draw[draw, solid, blue, thick] (f1)--(c) node[midway, above] {\hspace{-0.3cm}\blue $\fm$};
\draw[draw, solid, blue, thick] (f2)--(c) node[midway, below] {\hspace{-0.3cm}\blue $\fm$};
\draw[draw, solid, blue, thick] (f1)--(f2) node[midway,right] {\blue $\fm$}; 
\draw[draw, solid] (c)--(s0); 
\draw[draw, solid] (f1)--(s1); 
\draw[draw, solid] (f2)--(s2); 
\end{tikzpicture}
}
This is actually equivalent to the theory on the right-hand side of \eref{threeSO2nodes}, namely a complete graph of 4 $\U(1)$ gauge nodes where each edge has multiplicity $\fm$, upon decoupling an overall $\U(1)$ in the latter theory.  We indeed have an $\SU(\fm)^6 \times \U(1)^3$ flavor symmetry, as can also be seen from the $t^2$ terms in the power expansion of the Hilbert series\footnote{For reference, we provide a partially unrefined Hilbert series for $\fm=3$, with all elements of $\vec x_{1,2,3}$ and $\vec y_{1,2,3}$ set to $1$, as follows:
\bes{ \nn
\scalebox{0.8}{$
\begin{split}
1+51t^2+\left(27 q_1 q_2 q_3+\frac{27 q_1 q_3}{q_2}+\frac{27 q_1 q_2}{q_3}+\frac{27 q_2 q_3}{q_1}+\frac{27 q_1}{q_2 q_3}+\frac{27 q_3}{q_1 q_2}+\frac{27 q_2}{q_1 q_3} +\frac{27}{q_1 q_2 q_3} \right) t^3+\ldots,
\end{split}$}
}
where we remark that $51$ is the dimension of $\SU(3)^6 \times \U(1)^3$, which is the flavor symmetry of the theory as claimed.
}, namely
\bes{
3+\sum_{i=1}^3 \left[ \chi^{\SU(\fm)}_{[1,0,\cdots,0,1]}(\vec x_i) + \chi^{\SU(\fm)}_{[1,0,\cdots,0,1]}(\vec y_i) \right]~.
}

\subsubsection*{The Coulomb branch Hilbert series}

On the other hand, the Coulomb branch (CB) Hilbert series of the theory on the left-hand side of \eref{threeSO2nodes} can be written as 
\bes{ \label{HCBthreeSO2nodes}
H[\text{CB}\eref{threeSO2nodes}](t; \omega, x_1, x_2, x_3)
&=\sum_{\sigma=0}^1 \,\sum_{r_1 \in \BZ+\frac{\sigma}{2}} \, \sum_{r_2 \in \BZ+\frac{\sigma}{2} } \,  \sum_{r_3 \in \BZ+\frac{\sigma}{2} }  \omega^{\sigma}~  t^{2 \Delta(r_1,r_2,r_3)} (1-t^2)^{-3} \prod_{i=1}^3 x_i^{r_i}
}
where $\omega$ is the discrete fugacity for the topological symmetry associated with the $\BZ_2$ quotient satisfying $\omega^2=1$ and 
\bes{
\Delta(r_1,r_2,r_3)= \frac{\fm}{2} \sum_{i=1}^3 \left( |r_i-r_{i+1}|+  |r_i+r_{i+1}| \right)
}
with $r_4 \equiv r_1$.  We define, for $i=1,2,3$,
\bes{
u_i= r_i +r_{i+1}~, \quad v_i= -r_i +r_{i+1}~,\quad y_i = (x_i x_{i+1})^{1/4}~,\quad z_i = (x^{-1}_i x_{i+1})^{1/4}~.
}
with $x_4 \equiv x_1$.  There are the following relations:
\bes{
u_1 -u_2 = v_3~, \quad u_2 -u_3 = v_1~, \quad u_3 - u_1 = v_2 \\
y_1 y_2^{-1} = z_3~, \quad y_2 y_3^{-1} = z_1~, \quad y_3 y_1^{-1} = z_2
}
The dimension of monopole operators can be rewritten as
\bes{
\Delta(r_1,r_2,r_3) &=  \frac{\fm}{2} \left( |u_1-u_2|+|u_2-u_3|+|u_3-u_1| \right)+ \frac{\fm}{2} \sum_{i=1}^3 |u_i| \\
&=: \tilde{\Delta}(u_1,u_2,u_3)
}
The factor $\prod_{i=1}^3 x_i^{r_i}$ in \eref{HCBthreeSO2nodes} can be written as
\bes{
\prod_{i=1}^3 x_i^{r_i} = \prod_{i=1}^3 (y_i^{u_i} z_i^{v_i}) = \prod_{i=1}^3 \left(\frac{y_i^4}{y_1 y_2y_3} \right)^{u_i}~.
}
We can rewrite the above CB Hilbert series as
\bes{
H[\text{CB}\eref{threeSO2nodes}](t; \, \omega, \,x_1 , \,x_2 ,\, x_3)
&=\sum_{u_1 \in \BZ}\, \sum_{u_2 \in \BZ}\, \sum_{u_3 \in \BZ}~  t^{2 \tilde{\Delta}(u_1,u_2,u_3)} (1-t^2)^{-3} \prod_{i=1}^3 \left(\omega \frac{y_i^4}{y_1 y_2y_3} \right)^{u_i}~.
} 
We thus obtained the CB Hilbert series of \eref{threeU1nodeswallflv} such that the topological symmetry for the $i$-th node is $\omega \, y_i^4 (y_1 y_2 y_3)^{-1}$~. 

\subsection{Relation \eref{A1D2Nmirr1}} \label{app:twonodeswflv}
\subsubsection*{The Higgs branch Hilbert series}
For the sake of conciseness, let us write
\bes{
P=N-1~.
}
The Higgs branch Hilbert series of the theory of the left-hand side of \eref{A1D2Nmirr1} is
\bes{
&H[\text{HB}\eref{A1D2Nmirr1}](t; \vec x; q)  \\
&= \oint_{|z_1|=1} \frac{d z_1}{2 \pi i z_1}   \oint_{|z_2|=1} \frac{d z_2}{2 \pi i z_2}  \times \PE \left[ q^{-1} (z_1 z_2 + z_1^{-1} z_2) t+ \right.\\
& \left.\quad +  q (z_1 z_2^{-1} + z_1^{-1} z_2^{-1}) t + \chi^{\SU(P)}_{\left[0,\ldots,0,1\right]} (\vec x) z^2_2 t  + \chi^{\SU(P)}_{\left[1,0,\ldots,0\right]} (\vec x) z^{-2}_2 t  -2t^2  \right]
}
where $\vec x$ denotes the fugacities of the $\SU(P)$ flavor symmetry and $q$ denotes the fugacity for $\U(1)$ flavour symmetry arising from the edge between two $\SO(2)$ gauge groups.  We change the integration variables as in \eref{defuv} and obtain
\bes{
H[\text{HB}&\eref{A1D2Nmirr1}](t; \vec x,q) = \oint_{|u|=1} \frac{d u}{2 \pi i u}   \oint_{|v|=1} \frac{d v}{2 \pi i v}  \times\PE \left[(u+u^{-1}+v+v^{-1}) t + \right. \\
& \qquad \left. +\chi^{\SU(P)}_{\left[0,\ldots,0,1\right]} (\vec x) q^2 u v^{-1} t  + \chi^{\SU(P)}_{\left[1,0,\ldots,0\right]} (\vec x) q^{-2} u^{-1} v t  -2t^2 \right]\fstop
}
This is indeed the HB Hilbert series for the theory on the right-hand side of \eref{A1D2Nmirr1}, upon decoupling an overall $\U(1)$, where $u$ and $v$ are the gauge fugacities for the $\U(1)$ nodes connected by the blue line.  The flavor symmetry is indeed $\SU(P)\times \U(1)$, as can be seen from order $t^2$ in the power series of the above Hilbert series.

\subsubsection*{The Coulomb branch Hilbert series}

The CB Hilbert series of the theory on the left-hand side of \eref{A1D2Nmirr1} is given by
\bes{
H[\text{CB}\eref{A1D2Nmirr1}](t; \omega, x_1, x_2) =\sum_{\sigma=0}^1 \sum_{r_1 \in \BZ+\frac{\sigma}{2}}  \sum_{r_2 \in \BZ+\frac{\sigma}{2} }  \omega^{\sigma}~  t^{2 \Delta(r_1,r_2)} (1-t^2)^{-2} x_1^{r_1} x_2^{r_2}~.
}
where $\omega$ is the discrete fugacity for the topological symmetry associated with the $\BZ_2$ quotient, satisfying $\omega^2=1$, and
\bes{
\Delta(r_1,r_2) = \frac{1}{2} |r_1+r_2| +  \frac{1}{2} |r_1-r_2| + \frac{P}{2}|2r_1| ~  ;
}
We change the variables as in \eref{defu1u2y1y2} and obtain
\bes{
&H[\text{CB}\eref{A1D2Nmirr1}](t; \omega, y_1 y_2, \,y_1^{-1} y_2) \\
&=\sum_{u_1 \in \BZ}\, \sum_{u_2 \in \BZ}\,   t^{|u_1|+|u_2|+P|u_1-u_2| }  \frac{1}{(1-t^2)^2} \prod_{i=1}^2 (\omega y_i)^{u_i}~.
}
This is indeed the CB Hilbert series for the theory on the right-hand side of \eref{A1D2Nmirr1}, upon decoupling an overall $\U(1)$, where $(u_1,u_2)$ and $(\omega y_1, \omega y_2)$ are, respectively, the gauge fluxes and the fugacities for the topological symmetries for the $\U(1)$ nodes connected by the blue line.

\subsection{Theories \eref{DSO6} and \eref{DSU4}} \label{app:SO6vsSU4}
\subsubsection*{The Higgs branch Hilbert series}
Theories \eref{DSO6} and \eref{DSU4} contain a $T[\SO(6)]$ tail and a $T[\SU(4)]$ tail, respectively.  Both of the latter theories have the Higgs and Coulomb branches isomorphic to the nilpotent cone of the $A_3$ algebra, whose Hilbert series is (see e.g.  \cite[(3.4)]{Hanany:2011db})
\bes{
H[\CN_{A_3}] (t; \vec x) = \PE \left[ \chi^{\SU(4)}_{\left[1,0,1\right]}(\vec x) t^2 \right] \prod_{p=2}^4 (1-t^{2p})~.
}
where $\vec x$ denotes the fugacities for the $\SU(4)$ symmetry.
To compute the Higgs branch Hilbert series, we decompose the adjoint representation $[1,0,1]$ of $\SU(4)$ to representations of the $\SU(3) \times \U(1)$ subgroup as follows:
\bes{
[1,0,1] \, \longrightarrow \, [1,1; 0] \oplus [1,0; +4] \oplus [0,1; -4] \oplus [0,0; 0]~.
} 
Let us denote by $\vec y$ and $z$ the fugacities for $\SU(3)$ and $\U(1)$ respectively.  Then, the Higgs branch Hilbert series of \eref{DSO6} can be written as
\bes{
&H[ \text{HB} \eref{DSO6} ] (t; \vec y, \vec f)\\ 
&= \left[ \prod_{p=2}^4 (1-t^{2p}) \right] \PE \left[ (1+\chi^{\SU(3)}_{\left[1,1\right]}(\vec y)) t^2 \right] \PE[-t^2]  \, \oint_{|z|=1} \frac{dz}{2\pi i z} \times \\
&\quad \PE\left[ \chi^{\SU(3)}_{\left[1,0\right]}(\vec y)  z^4 t^2 +  \chi^{\SU(3)}_{\left[0,1\right]}(\vec y)  z^{-4} t^2+ {\blue \chi^{\SU(k)}_{\left[1,0,\cdots,0\right]} (\vec f) z^{-2} t +\chi^{\SU(k)}_{\left[0,\cdots,0,1\right]} (\vec f) z^{2} t} \right] \\
&= \left[ \prod_{p=2}^4 (1-t^{2p}) \right] \PE \left[ \chi^{\SU(3)}_{\left[1,1\right]}(\vec y) t^2 \right] \, \oint_{|z|=1} \frac{dz}{2\pi i z} \\
&\quad\PE\left[ \chi^{\SU(3)}_{\left[1,0\right]}(\vec y)  z^4 t^2 +  \chi^{\SU(3)}_{\left[0,1\right]}(\vec y)  z^{-4} t^2+ {\blue \chi^{\SU(k)}_{\left[1,0,\cdots,0\right]} (\vec f) z^{-2} t +\chi^{\SU(k)}_{\left[0,\cdots,0,1\right]} (\vec f) z^{2} t} \right]
}
In order to compute the HB Hilbert series with that of \eref{DSU4}, we proceed similarly, noting that there is an overall $\U(1)$ that decouples.  Specifically, this can be realized as follows.  Suppose we denote by $z_1$ and $z_2$ the gauge fugacities of the two $\U(1)$ nodes in the triangle.  The hypermultiplets associated with the blue line transform as $z_1 z_2^{-1}$ and $z_1^{-1} z_2$, whereas there is no matter field that transforms as $z_1 z_2$ or $(z_1 z_2)^{-1}$. The latter correspond to the combination of an overall $\U(1)$ that decouples, and so we may set $z_1 z_2=1$, i.e.  $z_1 = z_2^{-1} \equiv z$.  Hence, the hypermultiplets associated with the blue line transform as $z^2$ and $z^{-2}$, corresponding to the terms indicated in blue in the above expression.

\subsubsection*{The Coulomb branch Hilbert series}

The CB Hilbert series of \eref{DSO6} can be written as
\bes{ \label{CBHSDSO6}
&H[\text{CB}\eref{DSO6}](t; \omega, x_1, x_2) \\
&= \sum_{\sigma=0}^1 \, \sum_{d_1 \in \BZ+ \frac{\sigma}{2}}\, \sum_{c_1 \in \BZ_{\geq0}+ \frac{\sigma}{2}} \, \sum_{ \substack{(d_2)_1 \geq |(d_2)_2| \\ \vec{d}_2 \in \left( \BZ + \frac{\sigma}{2} \right)^2}}\, \sum_{ \substack{(c_2)_1 \geq (c_2)_2 \\ \vec{c}_2 \in \left( \BZ_{\geq0} + \frac{\sigma}{2} \right)^2}} \, \sum_{d'_1 \in  \BZ+ \frac{\sigma}{2}}   \\
& \qquad \omega^{\sigma} t^{2 \Delta(d_1, c_1, \vec d_2, \vec c_2, d'_1)} (1-t^2)^{-2} P_{C_1}(t; c_1) P_{D_2}(t; \vec d_2) P_{C_2}(t;\vec c_2) x_1^{d_1} x_2^{d'_1}
}
where we use the small case letters to denote the gauge fugacities of the corresponding groups; the dimension of the monopole operator is
\bes{
\Delta(d_1, c_1, \vec d_2, \vec c_2, d'_1) &= \frac{1}{2}|d_1 \pm c_1|+  \frac{1}{2}\sum_{i=1}^2 |c_1 \pm (d_2)_i| \\
&+  \frac{1}{2}\sum_{i,j=1}^2 |(d_2)_i \pm (c_2)_j| +  \frac{3}{2}\sum_{i=1}^2 |(c_2)_i \pm d'_1|+\frac{k}{2}|2d'_1| \\
&- |2c_1| - |(d_2)_1 \pm (d_2)_2| - \sum_{i=1}^2 |2(c_2)_i |- |(c_2)_1 \pm (c_2)_2|
}
with $|x \pm y| = |x+y|+|x-y|$;
$x_1$ and $x_2$ are the fugacities for the topological symmetries associated with the left $D_1$ and right $D_1$ respectively; $\omega$ is the discrete fugacity for the topological symmetry associated with the $\BZ_2$ quotient, satisfying $\omega^2=1$; $(1-t^2)^{-1}$ is the dressing factor for each $D_1$ gauge group, whereas $P_G$ denotes the dressing factor for the group $G$ given by \cite[Appendix A]{Cremonesi:2013lqa} (with $t$ in that reference being $t^2$ in this article).  In particular,  the explicit expression for $P_{C_2}(t;\vec c_2)$ was provided in \cite[(A.18)]{Cremonesi:2013lqa}.  It should be noted that the Coulomb branch symmetry $\SU(4) \times \U(1)$ is not manifest in the Hilbert series \eref{CBHSDSO6}; this is analogous to the discussion in \cite{Kapustin:1998fa}.

On the other hand, the CB Hilbert series of \eref{DSU4} can be written as
\bes{
\label{CBHSDSU4}
&H[\text{CB}\eref{DSU4}](t;  y_1, y_2, y_3, r) \\
&= \sum_{u_1 \in \BZ} \, \sum_{(u_2)_1 \geq (u_2)_2 > -\infty}\, \sum_{(u_3)_1 \geq (u_3)_2 \geq (u_3)_3 > -\infty} \, \sum_{u'_1 \in \BZ}  t^{2\Delta(u_1, \vec{u}_2, \vec{u}_3, u'_1)} (1-t^2)^{-2} \times \\
& \quad  P_{U(2)}(t; \vec{u}_2) P_{U(3)}(t; \vec{u}_3) y_1^{u_1} y_2^{\sum_{i=1}^2 (u_2)_i} y_3^{\sum_{i=1}^3 (u_3)_i} r^{u'_1}
}
where $(y_1, y_2, y_3)$ are fugacities for the $\SU(4)$ Coulomb branch symmetry and $r$ is that for the $\U(1)$ Coulomb branch symmetry.  The dimension of the monopole operators is
\bes{
\Delta(u_1, \vec{u}_2, \vec{u}_3, u'_1) &= \frac{1}{2}\sum_{i=1}^2 |u_1-(u_2)_i|+\frac{1}{2}\sum_{i=1}^2 \sum_{j=1}^3  |(u_2)_i-(u_3)_j|\\
&\quad +\frac{1}{2} \sum_{j=1}^3  |(u_3)_j-u'_1| + \frac{3}{2} \sum_{j=1}^3  |(u_3)_j| + \frac{k}{2}|u'_1| \\
&\quad - |(u_2)_1-(u_2)_2|  -\sum_{1\leq i< j \leq 3} |(u_3)_i-(u_3)_j|~.
}
The dressing factor $P_{U(2)}$ and $P_{U(3)}$ are given by \cite[(A.4)]{Cremonesi:2013lqa} (with $t$ in that reference being $t^2$ in this article).  The Coulomb branch symmetry $\SU(4)\times \U(1)$ is manifest in this expression.

Let us report the unrefined CB Hilbert series, namely $\omega=x_i =1$ in \eref{CBHSDSO6} and $y_i=r=1$ in \eref{CBHSDSU4}, for some $k$.  Such Hilbert series obtained from \eref{CBHSDSO6} and \eref{CBHSDSU4} are in agreement with each other.
\bes{
\begin{array}{lll}
k=1: \qquad &1 + 16 t^2 + 143 t^4+O(t^6) &= \PE\left[ 16 t^2 + 7 t^4 +O(t^6)  \right] \\
k=2: \qquad & 1 + 16 t^2 + 135 t^4 + 8 t^5 +O(t^6)  &= \PE\left[ 16 t^2 - t^4 + 8 t^5+O(t^6)  \right] \\
k=3: \qquad & 1 + 16 t^2 + 135 t^4 +O(t^6)  &= \PE \left[16 t^2 - t^4 +O(t^6)  \right]~.
\end{array}
}

\subsection{Relation \eref{T55gaugedE6}} \label{app:T55gaugedE6}
In this appendix, we demonstrate the relation \eref{T55gaugedE6}.   The $T_{\left[5,5\right]}[\SO(10)]/\SO(2)_\CC$ theory in question has a mirror dual in terms of $T^{\left[5,5\right]}[\SO(10)]/\SO(2)_\CH$, where $/\SO(2)_\CH$ denotes the gauging of the $\SO(2)$ flavor symmetry of $T^{\left[5,5\right]}[\SO(10)]$.  The $T^{\left[5,5\right]}[\SO(10)]/\SO(2)_\CH$ theory admits the following star-shaped quiver description
\bes{
\begin{array}{ll}
D_1-C_1-D_2-&C_2-D_2-C_1-D_1 \\
& \,\, | \\
&D_1
\end{array} \qquad /\BZ_2 
}
This is precisely the $3$d mirror theory \cite{Benini:2010uu} for the class $\CS$ theory of the twisted $A_3$ type associated with a sphere with two twisted punctures $\left[1^5\right]_t$, $\left[1^5\right]_t$ and one untwisted puncture $[3,1]$.  According to \cite{Chacaltana:2012ch}, this can be identified with the $E_6$ MN theory.  This establishes the relation \eref{T55gaugedE6}.

\subsubsection*{The Coulomb branch Hilbert series}
We can also compute the CB Hilbert series of the theories in \eref{T55gaugedE6}.  In order to make the Coulomb branch symmetry of the $T_{\left[5,5\right]}[\SO(10)]$ theory manifest, we use the Hall-Littlewood formula \cite{Cremonesi:2014kwa, Cremonesi:2014vla}\footnote{We remark that the fugacity $t$ in these references correspond to $t^2$ in this paper.} to compute the CB Hilbert series of such a theory as follows:
\bes{
&H[\text{CB of}\,\, T_{\left[5,5\right]}[\SO(10)]](t; x; \vec n) \\
&= t^{8 n_1+6n_2+4n_3+2n_4} (1-t^2)^5 K^{\SO(10)}_{\left[5,5\right]}(x; t) \Psi^{\vec n}_{\SO(10)}(\vec a (t;x);t)
}
where $x$ is the fugacity for the $\SO(2)_\CC$ Coulomb branch symmetry, $\vec n = (n_1, n_2, \ldots, n_5)$, $\Psi^{\vec n}_{\SO(10)}(\vec a (t;x);t)$ is the Hall-Littlewood polynomial given in \cite[(B.10)]{Cremonesi:2014kwa}, and
\bes{
\vec a(t;x) &= (t^4 x, t^2 x, x, x t^{-2}, x t^{-4}) \\ 
K^{\SO(10)}_{\left[5,5\right]}(x; t) &= \PE\left[ t^{10} + t^8 (x^2+1+x^{-2}) + t^6 + t^4 (x^2+1+x^{-2})  +t^2 \right]
} 
The Hilbert series of the $T_{\left[5,5\right]}[\SO(10)]$ theory, with the $\SO(2)_\CC$ Coulomb branch symmetry being gauged, can be computed as follows:
\bes{
&H[\text{CB of}\,\, T_{\left[5,5\right]}[\SO(10)]/\SO(2)_\CC](t) \\
&\oint_{|x|=1} \frac{dx}{2 \pi i x} (1-t^2) H[\text{CB of}\,\, T_{\left[5,5\right]}[\SO(10)]](t; x; \vec 0)  \\
&= \PE\left[ t^6 + t^8 + t^{12} - t^{24}\right]~.
}
This is precisely the Hilbert series of $\BC^2/E_6$, which is the Coulomb branch of the $E_6$ MN theory.

\section{Results regarding ~\cref{closeDpUSp2Nm2,closeDpUSp2Nm2-2}}
\label{app:Featureshypesing}

The geometries in~\cref{closeDpUSp2Nm2,closeDpUSp2Nm2-2} have been identified, respectively, as the hypersurface singularities of type VII and X, following the notation of~\cite{Xie:2015rpa}. In particular, we can compute their Milnor numbers to be
\begin{equation}
\mu_{\text{\eqref{closeDpUSp2Nm2}}}=N(3+p)-(2N^2+1) \coma \mu_{\text{\eqref{closeDpUSp2Nm2-2}}}=2\left(p+\frac{1}{2}-N\right)(N-1)\fstop
\end{equation}
We can also discuss the number of marginal operators for theory \eqref{closeDpUSp2Nm2}. Such theory has a number of marginal operators whose patterns depend on if $N$ is even or odd.
\begin{itemize}
    \item If $N=2\left[k(2j+1)+(j+1)\right]$, for any $i,j\geq 0$ and $k\geq j$, there are $2j$ marginal operators whenever $p=(4k+i+3)(2j+1)$, otherwise the number of marginal operators is $0$.
    \item If $N$ is odd, for any $i,j,k\geq 0$, there are
    \begin{enumerate}
        \item $1$ marginal operators when $p+3-2N$ is odd. 
        \item $N-1$ marginal operators when $p=(N-1)(2j+3)$.
        \item $N-2$ marginal operators when $p=2(N-1)(2+j)$.
        \item $2j+2$ marginal operators when $N=(4j+6)(k+1)+1$ and $p=(3+2j)(5+2i+4k)$. Whenever different values of $i,j,k$ gives the same value of $N$ and $p$, the number of marginal operators is the one that have the largest $j$.
        \item $2j+3$ marginal operators when $N=(2j+4)k+4j+9$ and $p=2(5+2i+2k)(2+j)$. Whenever different values of $i,j,k$ gives the same value of $N$ and $p$, the number of marginal operators is the one that have the largest $j$. 
        \item Whenever $N$ and $p$ of the previous list are the same, there are a number of marginal operators corresponding to the largest value of marginal operators obtained by the previous conditions.
    \end{enumerate}
\end{itemize}
\FloatBarrier

\section{Examples of non-Higgsable SCFTs} 
\label{app:nonHiggsableSCFTs}

Here we list $(G,G')$ theories that have $0\leq 24(c-a)< 1$. They have been computed using the program provided in~\cite{Carta:2020plx}. We are dividing the theories in three main tables:
\begin{enumerate}
    \item $(A_n,A_m)$ with $1\leq n,m\leq 100$ in Table~\ref{tab:AnAmNHC}.
    \item $(A_n,D_m)$ with $1\leq n,m\leq 100$ in Table~\ref{tab:AnDmNHC}.
    \item $(E_n,A_m)$ and $(E_n,D_m)$, with $n=6,7,8$ and $1\leq m\leq 100$ in Table~\ref{tab:GGpNHC}.
\end{enumerate}
All the theories that are not in the tables have $24(c-a)\geq 1$. For instance, there are no $(D_n,D_m)$ or $(E,E)$ theories. All the non-Higgsable SCFTs are expected not to have mass parameters since they have no Higgs branch. Indeed, theories in~\cref{tab:AnAmNHC,tab:AnDmNHC,tab:GGpNHC} have rank equal to 
\begin{equation}
\rank\left(G_n,G_m\right)=\frac{nm}{2}\coma \text{for $G=A,D,E$.}
\label{eq:rankNHC}
\end{equation}
The rank in Eq.~\eqref{eq:rankNHC} is exactly equal to half the Milnor number, computed in~\cite{Xie:2015rpa}, of the associated hypersurface singularity, as expected by theories without mass parameters.
%\AM{Comments on $24(c-a)$ that is approaching $1$ asymptocally? I don't have a fit for that.}\\

It is natural to conjecture that the rank of such theories equals the sum of all the non-vanishing genus-zero Gopakumar-Vafa (GV) invariants~\cite{Gopakumar:1998ii,Gopakumar:1998jq,Gopakumar:1998ki} of the corresponding geometry. The logic for this is the following. The magnetic quiver of a $4$d theory engineered by IIB on a given CY $X$ is related to the electric quiver of the $5$d theory engineered by A-theory on the same CY $X$ by an operation consisting in gauging the topological symmetry (see Figure 1 in~\cite{Closset:2020afy}). However, the mirrors we consider here are just free hypers, so there is no topological symmetry that can be gauged. Therefore, they directly correspond to the electric quivers (i.e., the $3$d dimensional reduction) of the $5$d theory obtained by compactifying M-theory on $X$. GV invariants at genus zero count hypers of the $5$d theory, which in this case are free hypers, and are clearly in one-to-one correspondence with those of the considered 3d mirror. Their number equals the rank of the original $4$d theory. Some of those invariants were computed independently in ~\cite{Collinucci:2021wty,Collinucci:2021ofd}. We checked that our prediction matches the results of \cite{Collinucci:2021wty,Collinucci:2021ofd} in all cases in which the same CY was considered in both papers.

\begin{center}
\renewcommand{\arraystretch}{1.25}
\begin{longtable}{c|c|c||c|c|c||c|c|c}
\caption{$(A_n,A_m)$ with $1\leq n,m\leq 100$}\\
$(A_n,A_m)$  & $24(c-a)$ & $\text{rank}$ & $(A_n,A_m)$  & $24(c-a)$ & $\text{rank}$ & $(A_n,A_m)$  & $24(c-a)$ & $\text{rank}$ \\
    \hline
\endhead
\label{tab:AnAmNHC}
$(A_1,A_2)$     & $\frac{1}{5}$    & $1  $ &  $(A_1,A_{82})$  & $\frac{41}{85}$  & $41 $ &  $(A_2,A_{48})$  & $\frac{12}{13}$  & $48 $ \\
$(A_1,A_4)$     & $\frac{2}{7}$    & $2  $ &  $(A_1,A_{84})$  & $\frac{14}{29}$  & $42 $ &  $(A_2,A_{49})$  & $\frac{49}{53}$  & $49 $ \\
$(A_1,A_6)$     & $\frac{1}{3}$    & $3  $ &  $(A_1,A_{86})$  & $\frac{43}{89}$  & $43 $ &  $(A_2,A_{51})$  & $\frac{51}{55}$  & $51 $ \\
$(A_1,A_8)$     & $\frac{4}{11}$   & $4  $ &  $(A_1,A_{88})$  & $\frac{44}{91}$  & $44 $ &  $(A_2,A_{52})$  & $\frac{13}{14}$  & $52 $ \\
$(A_1,A_{10})$  & $\frac{5}{13}$   & $5  $ &  $(A_1,A_{90})$  & $\frac{15}{31}$  & $45 $ &  $(A_2,A_{54})$  & $\frac{27}{29}$  & $54 $ \\
$(A_1,A_{12})$  & $\frac{2}{5}$    & $6  $ &  $(A_1,A_{92})$  & $\frac{46}{95}$  & $46 $ &  $(A_2,A_{55})$  & $\frac{55}{59}$  & $55 $ \\
$(A_1,A_{14})$  & $\frac{7}{17}$   & $7  $ &  $(A_1,A_{94})$  & $\frac{47}{97}$  & $47 $ &  $(A_2,A_{57})$  & $\frac{57}{61}$  & $57 $ \\
$(A_1,A_{16})$  & $\frac{8}{19}$   & $8  $ &  $(A_1,A_{96})$  & $\frac{16}{33}$  & $48 $ &  $(A_2,A_{58})$  & $\frac{29}{31}$  & $58 $ \\
$(A_1,A_{18})$  & $\frac{3}{7}$    & $9  $ &  $(A_1,A_{98})$  & $\frac{49}{101}$ & $49 $ &  $(A_2,A_{60})$  & $\frac{15}{16}$  & $60 $ \\
$(A_1,A_{20})$  & $\frac{10}{23}$  & $10 $ &  $(A_1,A_{100})$ & $\frac{50}{103}$ & $50 $ &  $(A_2,A_{61})$  & $\frac{61}{65}$  & $61 $ \\
$(A_1,A_{22})$  & $\frac{11}{25}$  & $11 $ &  $(A_2,A_3)$     & $\frac{3}{7}$    & $3  $ &  $(A_2,A_{63})$  & $\frac{63}{67}$  & $63 $ \\
$(A_1,A_{24})$  & $\frac{4}{9}$    & $12 $ &  $(A_2,A_4)$     & $\frac{1}{2}$    & $4  $ &  $(A_2,A_{64})$  & $\frac{16}{17}$  & $64 $ \\
$(A_1,A_{26})$  & $\frac{13}{29}$  & $13 $ &  $(A_2,A_6)$     & $\frac{3}{5}$    & $6  $ &  $(A_2,A_{66})$  & $\frac{33}{35}$  & $66 $ \\
$(A_1,A_{28})$  & $\frac{14}{31}$  & $14 $ &  $(A_2,A_7)$     & $\frac{7}{11}$   & $7  $ &  $(A_2,A_{67})$  & $\frac{67}{71}$  & $67 $ \\
$(A_1,A_{30})$  & $\frac{5}{11}$   & $15 $ &  $(A_2,A_9)$     & $\frac{9}{13}$   & $9  $ &  $(A_2,A_{69})$  & $\frac{69}{73}$  & $69 $ \\
$(A_1,A_{32})$  & $\frac{16}{35}$  & $16 $ &  $(A_2,A_{10})$  & $\frac{5}{7}$    & $10 $ &  $(A_2,A_{70})$  & $\frac{35}{37}$  & $70 $ \\
$(A_1,A_{34})$  & $\frac{17}{37}$  & $17 $ &  $(A_2,A_{12})$  & $\frac{3}{4}$    & $12 $ &  $(A_2,A_{72})$  & $\frac{18}{19}$  & $72 $ \\
$(A_1,A_{36})$  & $\frac{6}{13}$   & $18 $ &  $(A_2,A_{13})$  & $\frac{13}{17}$  & $13 $ &  $(A_2,A_{73})$  & $\frac{73}{77}$  & $73 $ \\
$(A_1,A_{38})$  & $\frac{19}{41}$  & $19 $ &  $(A_2,A_{15})$  & $\frac{15}{19}$  & $15 $ &  $(A_2,A_{75})$  & $\frac{75}{79}$  & $75 $ \\
$(A_1,A_{40})$  & $\frac{20}{43}$  & $20 $ &  $(A_2,A_{16})$  & $\frac{4}{5}$    & $16 $ &  $(A_2,A_{76})$  & $\frac{19}{20}$  & $76 $ \\
$(A_1,A_{42})$  & $\frac{7}{15}$   & $21 $ &  $(A_2,A_{18})$  & $\frac{9}{11}$   & $18 $ &  $(A_2,A_{78})$  & $\frac{39}{41}$  & $78 $ \\
$(A_1,A_{44})$  & $\frac{22}{47}$  & $22 $ &  $(A_2,A_{19})$  & $\frac{19}{23}$  & $19 $ &  $(A_2,A_{79})$  & $\frac{79}{83}$  & $79 $ \\
$(A_1,A_{46})$  & $\frac{23}{49}$  & $23 $ &  $(A_2,A_{21})$  & $\frac{21}{25}$  & $21 $ &  $(A_2,A_{81})$  & $\frac{81}{85}$  & $81 $ \\
$(A_1,A_{48})$  & $\frac{8}{17}$   & $24 $ &  $(A_2,A_{22})$  & $\frac{11}{13}$  & $22 $ &  $(A_2,A_{82})$  & $\frac{41}{43}$  & $82 $ \\
$(A_1,A_{50})$  & $\frac{25}{53}$  & $25 $ &  $(A_2,A_{24})$  & $\frac{6}{7}$    & $24 $ &  $(A_2,A_{84})$  & $\frac{21}{22}$  & $84 $ \\
$(A_1,A_{52})$  & $\frac{26}{55}$  & $26 $ &  $(A_2,A_{25})$  & $\frac{25}{29}$  & $25 $ &  $(A_2,A_{85})$  & $\frac{85}{89}$  & $85 $ \\
$(A_1,A_{54})$  & $\frac{9}{19}$   & $27 $ &  $(A_2,A_{27})$  & $\frac{27}{31}$  & $27 $ &  $(A_2,A_{87})$  & $\frac{87}{91}$  & $87 $ \\
$(A_1,A_{56})$  & $\frac{28}{59}$  & $28 $ &  $(A_2,A_{28})$  & $\frac{7}{8}$    & $28 $ &  $(A_2,A_{88})$  & $\frac{22}{23}$  & $88 $ \\
$(A_1,A_{58})$  & $\frac{29}{61}$  & $29 $ &  $(A_2,A_{30})$  & $\frac{15}{17}$  & $30 $ &  $(A_2,A_{90})$  & $\frac{45}{47}$  & $90 $ \\
$(A_1,A_{60})$  & $\frac{10}{21}$  & $30 $ &  $(A_2,A_{31})$  & $\frac{31}{35}$  & $31 $ &  $(A_2,A_{91})$  & $\frac{91}{95}$  & $91 $ \\
$(A_1,A_{62})$  & $\frac{31}{65}$  & $31 $ &  $(A_2,A_{33})$  & $\frac{33}{37}$  & $33 $ &  $(A_2,A_{93})$  & $\frac{93}{97}$  & $93 $ \\
$(A_1,A_{64})$  & $\frac{32}{67}$  & $32 $ &  $(A_2,A_{34})$  & $\frac{17}{19}$  & $34 $ &  $(A_2,A_{94})$  & $\frac{47}{49}$  & $94 $ \\
$(A_1,A_{66})$  & $\frac{11}{23}$  & $33 $ &  $(A_2,A_{36})$  & $\frac{9}{10}$   & $36 $ &  $(A_2,A_{96})$  & $\frac{24}{25}$  & $96 $ \\
$(A_1,A_{68})$  & $\frac{34}{71}$  & $34 $ &  $(A_2,A_{37})$  & $\frac{37}{41}$  & $37 $ &  $(A_2,A_{97})$  & $\frac{97}{101}$ & $97 $ \\
$(A_1,A_{70})$  & $\frac{35}{73}$  & $35 $ &  $(A_2,A_{39})$  & $\frac{39}{43}$  & $39 $ &  $(A_2,A_{99})$  & $\frac{99}{103}$ & $99 $ \\
$(A_1,A_{72})$  & $\frac{12}{25}$  & $36 $ &  $(A_2,A_{40})$  & $\frac{10}{11}$  & $40 $ &  $(A_2,A_{100})$ & $\frac{25}{26}$  & $100$ \\
$(A_1,A_{74})$  & $\frac{37}{77}$  & $37 $ &  $(A_2,A_{42})$  & $\frac{21}{23}$  & $42 $ &  $(A_3,A_4)$     & $\frac{2}{3}$    & $6  $ \\
$(A_1,A_{76})$  & $\frac{38}{79}$  & $38 $ &  $(A_2,A_{43})$  & $\frac{43}{47}$  & $43 $ &  $(A_3,A_6)$     & $\frac{9}{11}$   & $9  $ \\
$(A_1,A_{78})$  & $\frac{13}{27}$  & $39 $ &  $(A_2,A_{45})$  & $\frac{45}{49}$  & $45 $ &  $(A_3,A_8)$     & $\frac{12}{13}$  & $12 $ \\
$(A_1,A_{80})$  & $\frac{40}{83}$  & $40 $ &  $(A_2,A_{46})$  & $\frac{23}{25}$  & $46 $ &  $(A_4,A_5)$     & $\frac{10}{11}$  & $10 $ \\
\end{longtable}
\end{center}

\begin{center}
\renewcommand{\arraystretch}{1.25}
\begin{longtable}{c|c|c||c|c|c||c|c|c}
\caption{$(A_n,D_m)$ with $1\leq n,m\leq 100$}\\
$(A_n,D_m)$  & $24(c-a)$ & $\text{rank}$ & $(A_n,D_m)$  & $24(c-a)$ & $\text{rank}$ & $(A_n,D_m)$  & $24(c-a)$ & $\text{rank}$ \\
    \hline
\endhead
\label{tab:AnDmNHC}
$(A_2,D_3)$     & $\frac{3}{7}$    & $3  $ &  $(A_2,D_{87})$  & $\frac{87}{175}$  & $87 $  & $(A_4,D_{73})$     & $\frac{146}{149}$ & $146 $ \\
$(A_2,D_4)$     & $0$              & $4  $ &  $(A_2,D_{88})$  & $\frac{28}{59}$   & $88 $  & $(A_4,D_{74})$     & $\frac{148}{151}$ & $148 $ \\
$(A_2,D_5)$     & $\frac{5}{11}$   & $5  $ &  $(A_2,D_{89})$  & $\frac{89}{179}$  & $89 $  & $(A_4,D_{75})$     & $\frac{50}{51}$   & $150 $ \\
$(A_2,D_6)$     & $\frac{6}{13}$   & $6  $ &  $(A_2,D_{90})$  & $\frac{90}{181}$  & $90 $  & $(A_4,D_{76})$     & $\frac{28}{31}$   & $152 $ \\
$(A_2,D_7)$     & $\frac{1}{5}$    & $7  $ &  $(A_2,D_{91})$  & $\frac{29}{61}$   & $91 $  & $(A_4,D_{77})$     & $\frac{154}{157}$ & $154 $ \\
$(A_2,D_8)$     & $\frac{8}{17}$   & $8  $ &  $(A_2,D_{92})$  & $\frac{92}{185}$  & $92 $  & $(A_4,D_{78})$     & $\frac{52}{53}$   & $156 $ \\
$(A_2,D_9)$     & $\frac{9}{19}$   & $9  $ &  $(A_2,D_{93})$  & $\frac{93}{187}$  & $93 $  & $(A_4,D_{79})$     & $\frac{158}{161}$ & $158 $ \\
$(A_2,D_{10})$  & $\frac{2}{7}$    & $10 $ &  $(A_2,D_{94})$  & $\frac{10}{21}$   & $94 $  & $(A_4,D_{80})$     & $\frac{160}{163}$ & $160 $ \\
$(A_2,D_{11})$  & $\frac{11}{23}$  & $11 $ &  $(A_2,D_{95})$  & $\frac{95}{191}$  & $95 $  & $(A_4,D_{81})$     & $\frac{10}{11}$   & $162 $ \\
$(A_2,D_{12})$  & $\frac{12}{25}$  & $12 $ &  $(A_2,D_{96})$  & $\frac{96}{193}$  & $96 $  & $(A_4,D_{82})$     & $\frac{164}{167}$ & $164 $ \\
$(A_2,D_{13})$  & $\frac{1}{3}$    & $13 $ &  $(A_2,D_{97})$  & $\frac{31}{65}$   & $97 $  & $(A_4,D_{83})$     & $\frac{166}{169}$ & $166 $ \\
$(A_2,D_{14})$  & $\frac{14}{29}$  & $14 $ &  $(A_2,D_{98})$  & $\frac{98}{197}$  & $98 $  & $(A_4,D_{84})$     & $\frac{56}{57}$   & $168 $ \\
$(A_2,D_{15})$  & $\frac{15}{31}$  & $15 $ &  $(A_2,D_{99})$  & $\frac{99}{199}$  & $99 $  & $(A_4,D_{85})$     & $\frac{170}{173}$ & $170 $ \\
$(A_2,D_{16})$  & $\frac{4}{11}$   & $16 $ &  $(A_2,D_{100})$ & $\frac{32}{67}$   & $100$  & $(A_4,D_{86})$     & $\frac{32}{35}$   & $172 $ \\
$(A_2,D_{17})$  & $\frac{17}{35}$  & $17 $ &  $(A_4,D_3)$     & $\frac{2}{3}$     & $6  $  & $(A_4,D_{87})$     & $\frac{58}{59}$   & $174 $ \\
$(A_2,D_{18})$  & $\frac{18}{37}$  & $18 $ &  $(A_4,D_4)$     & $\frac{8}{11}$    & $8  $  & $(A_4,D_{88})$     & $\frac{176}{179}$ & $176 $ \\
$(A_2,D_{19})$  & $\frac{5}{13}$   & $19 $ &  $(A_4,D_5)$     & $\frac{10}{13}$   & $10 $  & $(A_4,D_{89})$     & $\frac{178}{181}$ & $178 $ \\
$(A_2,D_{20})$  & $\frac{20}{41}$  & $20 $ &  $(A_4,D_6)$     & $0$               & $12 $  & $(A_4,D_{90})$     & $\frac{60}{61}$   & $180 $ \\
$(A_2,D_{21})$  & $\frac{21}{43}$  & $21 $ &  $(A_4,D_7)$     & $\frac{14}{17}$   & $14 $  & $(A_4,D_{91})$     & $\frac{34}{37}$   & $182 $ \\
$(A_2,D_{22})$  & $\frac{2}{5}$    & $22 $ &  $(A_4,D_8)$     & $\frac{16}{19}$   & $16 $  & $(A_4,D_{92})$     & $\frac{184}{187}$ & $184 $ \\
$(A_2,D_{23})$  & $\frac{23}{47}$  & $23 $ &  $(A_4,D_9)$     & $\frac{6}{7}$     & $18 $  & $(A_4,D_{93})$     & $\frac{62}{63}$   & $186 $ \\
$(A_2,D_{24})$  & $\frac{24}{49}$  & $24 $ &  $(A_4,D_{10})$  & $\frac{20}{23}$   & $20 $  & $(A_4,D_{94})$     & $\frac{188}{191}$ & $188 $ \\
$(A_2,D_{25})$  & $\frac{7}{17}$   & $25 $ &  $(A_4,D_{11})$  & $\frac{2}{5}$     & $22 $  & $(A_4,D_{95})$     & $\frac{190}{193}$ & $190 $ \\
$(A_2,D_{26})$  & $\frac{26}{53}$  & $26 $ &  $(A_4,D_{12})$  & $\frac{8}{9}$     & $24 $  & $(A_4,D_{96})$     & $\frac{12}{13}$   & $192 $ \\
$(A_2,D_{27})$  & $\frac{27}{55}$  & $27 $ &  $(A_4,D_{13})$  & $\frac{26}{29}$   & $26 $  & $(A_4,D_{97})$     & $\frac{194}{197}$ & $194 $ \\
$(A_2,D_{28})$  & $\frac{8}{19}$   & $28 $ &  $(A_4,D_{14})$  & $\frac{28}{31}$   & $28 $  & $(A_4,D_{98})$     & $\frac{196}{199}$ & $196 $ \\
$(A_2,D_{29})$  & $\frac{29}{59}$  & $29 $ &  $(A_4,D_{15})$  & $\frac{10}{11}$   & $30 $  & $(A_4,D_{99})$     & $\frac{66}{67}$   & $198 $ \\
$(A_2,D_{30})$  & $\frac{30}{61}$  & $30 $ &  $(A_4,D_{16})$  & $\frac{4}{7}$     & $32 $  & $(A_4,D_{100})$    & $\frac{200}{203}$ & $200 $ \\
$(A_2,D_{31})$  & $\frac{3}{7}$    & $31 $ &  $(A_4,D_{17})$  & $\frac{34}{37}$   & $34 $  & $(A_6,D_3)$        & $\frac{9}{11}$    & $9   $ \\
$(A_2,D_{32})$  & $\frac{32}{65}$  & $32 $ &  $(A_4,D_{18})$  & $\frac{12}{13}$   & $36 $  & $(A_6,D_4)$        & $\frac{12}{13}$   & $12  $ \\
$(A_2,D_{33})$  & $\frac{33}{67}$  & $33 $ &  $(A_4,D_{19})$  & $\frac{38}{41}$   & $38 $  & $(A_6,D_8)$        & $0$               & $24  $ \\
$(A_2,D_{34})$  & $\frac{10}{23}$  & $34 $ &  $(A_4,D_{20})$  & $\frac{40}{43}$   & $40 $  & $(A_6,D_{15})$     & $\frac{3}{5}$     & $45  $ \\
$(A_2,D_{35})$  & $\frac{35}{71}$  & $35 $ &  $(A_4,D_{21})$  & $\frac{2}{3}$     & $42 $  & $(A_6,D_{22})$     & $\frac{6}{7}$     & $66  $ \\
$(A_2,D_{36})$  & $\frac{36}{73}$  & $36 $ &  $(A_4,D_{22})$  & $\frac{44}{47}$   & $44 $  & $(A_8,D_3)$        & $\frac{12}{13}$   & $12  $ \\
$(A_2,D_{37})$  & $\frac{11}{25}$  & $37 $ &  $(A_4,D_{23})$  & $\frac{46}{49}$   & $46 $  & $(A_8,D_4)$        & $\frac{4}{5}$     & $16  $ \\
$(A_2,D_{38})$  & $\frac{38}{77}$  & $38 $ &  $(A_4,D_{24})$  & $\frac{16}{17}$   & $48 $  & $(A_8,D_{10})$     & $0$               & $40  $ \\
$(A_2,D_{39})$  & $\frac{39}{79}$  & $39 $ &  $(A_4,D_{25})$  & $\frac{50}{53}$   & $50 $  & $(A_8,D_{19})$     & $\frac{4}{5}$     & $76  $ \\
$(A_2,D_{40})$  & $\frac{4}{9}$    & $40 $ &  $(A_4,D_{26})$  & $\frac{8}{11}$    & $52 $  & $(A_{10},D_{12})$  & $0$               & $60  $ \\
$(A_2,D_{41})$  & $\frac{41}{83}$  & $41 $ &  $(A_4,D_{27})$  & $\frac{18}{19}$   & $54 $  & $(A_{12},D_{14})$  & $0$               & $84  $ \\
$(A_2,D_{42})$  & $\frac{42}{85}$  & $42 $ &  $(A_4,D_{28})$  & $\frac{56}{59}$   & $56 $  & $(A_{14},D_{16})$  & $0$               & $112 $ \\
$(A_2,D_{43})$  & $\frac{13}{29}$  & $43 $ &  $(A_4,D_{29})$  & $\frac{58}{61}$   & $58 $  & $(A_{16},D_{18})$  & $0$               & $144 $ \\
$(A_2,D_{44})$  & $\frac{44}{89}$  & $44 $ &  $(A_4,D_{30})$  & $\frac{20}{21}$   & $60 $  & $(A_{18},D_{20})$  & $0$               & $180 $ \\
$(A_2,D_{45})$  & $\frac{45}{91}$  & $45 $ &  $(A_4,D_{31})$  & $\frac{10}{13}$   & $62 $  & $(A_{20},D_{22})$  & $0$               & $220 $ \\
$(A_2,D_{46})$  & $\frac{14}{31}$  & $46 $ &  $(A_4,D_{32})$  & $\frac{64}{67}$   & $64 $  & $(A_{22},D_{24})$  & $0$               & $264 $ \\
$(A_2,D_{47})$  & $\frac{47}{95}$  & $47 $ &  $(A_4,D_{33})$  & $\frac{22}{23}$   & $66 $  & $(A_{24},D_{26})$  & $0$               & $312 $ \\
$(A_2,D_{48})$  & $\frac{48}{97}$  & $48 $ &  $(A_4,D_{34})$  & $\frac{68}{71}$   & $68 $  & $(A_{26},D_{28})$  & $0$               & $364 $ \\
$(A_2,D_{49})$  & $\frac{5}{11}$   & $49 $ &  $(A_4,D_{35})$  & $\frac{70}{73}$   & $70 $  & $(A_{28},D_{30})$  & $0$               & $420 $ \\
$(A_2,D_{50})$  & $\frac{50}{101}$ & $50 $ &  $(A_4,D_{36})$  & $\frac{4}{5}$     & $72 $  & $(A_{30},D_{32})$  & $0$               & $480 $ \\
$(A_2,D_{51})$  & $\frac{51}{103}$ & $51 $ &  $(A_4,D_{37})$  & $\frac{74}{77}$   & $74 $  & $(A_{32},D_{34})$  & $0$               & $544 $ \\
$(A_2,D_{52})$  & $\frac{16}{35}$  & $52 $ &  $(A_4,D_{38})$  & $\frac{76}{79}$   & $76 $  & $(A_{34},D_{36})$  & $0$               & $612 $ \\
$(A_2,D_{53})$  & $\frac{53}{107}$ & $53 $ &  $(A_4,D_{39})$  & $\frac{26}{27}$   & $78 $  & $(A_{36},D_{38})$  & $0$               & $684 $ \\
$(A_2,D_{54})$  & $\frac{54}{109}$ & $54 $ &  $(A_4,D_{40})$  & $\frac{80}{83}$   & $80 $  & $(A_{38},D_{40})$  & $0$               & $760 $ \\
$(A_2,D_{55})$  & $\frac{17}{37}$  & $55 $ &  $(A_4,D_{41})$  & $\frac{14}{17}$   & $82 $  & $(A_{40},D_{42})$  & $0$               & $840 $ \\
$(A_2,D_{56})$  & $\frac{56}{113}$ & $56 $ &  $(A_4,D_{42})$  & $\frac{28}{29}$   & $84 $  & $(A_{42},D_{44})$  & $0$               & $924 $ \\
$(A_2,D_{57})$  & $\frac{57}{115}$ & $57 $ &  $(A_4,D_{43})$  & $\frac{86}{89}$   & $86 $  & $(A_{44},D_{46})$  & $0$               & $1012$ \\
$(A_2,D_{58})$  & $\frac{6}{13}$   & $58 $ &  $(A_4,D_{44})$  & $\frac{88}{91}$   & $88 $  & $(A_{46},D_{48})$  & $0$               & $1104$ \\
$(A_2,D_{59})$  & $\frac{59}{119}$ & $59 $ &  $(A_4,D_{45})$  & $\frac{30}{31}$   & $90 $  & $(A_{48},D_{50})$  & $0$               & $1200$ \\
$(A_2,D_{60})$  & $\frac{60}{121}$ & $60 $ &  $(A_4,D_{46})$  & $\frac{16}{19}$   & $92 $  & $(A_{50},D_{52})$  & $0$               & $1300$ \\
$(A_2,D_{61})$  & $\frac{19}{41}$  & $61 $ &  $(A_4,D_{47})$  & $\frac{94}{97}$   & $94 $  & $(A_{52},D_{54})$  & $0$               & $1404$ \\
$(A_2,D_{62})$  & $\frac{62}{125}$ & $62 $ &  $(A_4,D_{48})$  & $\frac{32}{33}$   & $96 $  & $(A_{54},D_{56})$  & $0$               & $1512$ \\
$(A_2,D_{63})$  & $\frac{63}{127}$ & $63 $ &  $(A_4,D_{49})$  & $\frac{98}{101}$  & $98  $ & $(A_{56},D_{58})$  & $0$               & $1624$ \\
$(A_2,D_{64})$  & $\frac{20}{43}$  & $64 $ &  $(A_4,D_{50})$  & $\frac{100}{103}$ & $100 $ & $(A_{58},D_{60})$  & $0$               & $1740$ \\
$(A_2,D_{65})$  & $\frac{65}{131}$ & $65 $ &  $(A_4,D_{51})$  & $\frac{6}{7}$     & $102 $ & $(A_{60},D_{62})$  & $0$               & $1860$ \\
$(A_2,D_{66})$  & $\frac{66}{133}$ & $66 $ &  $(A_4,D_{52})$  & $\frac{104}{107}$ & $104 $ & $(A_{62},D_{64})$  & $0$               & $1984$ \\
$(A_2,D_{67})$  & $\frac{7}{15}$   & $67 $ &  $(A_4,D_{53})$  & $\frac{106}{109}$ & $106 $ & $(A_{64},D_{66})$  & $0$               & $2112$ \\
$(A_2,D_{68})$  & $\frac{68}{137}$ & $68 $ &  $(A_4,D_{54})$  & $\frac{36}{37}$   & $108 $ & $(A_{66},D_{68})$  & $0$               & $2244$ \\
$(A_2,D_{69})$  & $\frac{69}{139}$ & $69 $ &  $(A_4,D_{55})$  & $\frac{110}{113}$ & $110 $ & $(A_{68},D_{70})$  & $0$               & $2380$ \\
$(A_2,D_{70})$  & $\frac{22}{47}$  & $70 $ &  $(A_4,D_{56})$  & $\frac{20}{23}$   & $112 $ & $(A_{70},D_{72})$  & $0$               & $2520$ \\
$(A_2,D_{71})$  & $\frac{71}{143}$ & $71 $ &  $(A_4,D_{57})$  & $\frac{38}{39}$   & $114 $ & $(A_{72},D_{74})$  & $0$               & $2664$ \\
$(A_2,D_{72})$  & $\frac{72}{145}$ & $72 $ &  $(A_4,D_{58})$  & $\frac{116}{119}$ & $116 $ & $(A_{74},D_{76})$  & $0$               & $2812$ \\
$(A_2,D_{73})$  & $\frac{23}{49}$  & $73 $ &  $(A_4,D_{59})$  & $\frac{118}{121}$ & $118 $ & $(A_{76},D_{78})$  & $0$               & $2964$ \\
$(A_2,D_{74})$  & $\frac{74}{149}$ & $74 $ &  $(A_4,D_{60})$  & $\frac{40}{41}$   & $120 $ & $(A_{78},D_{80})$  & $0$               & $3120$ \\
$(A_2,D_{75})$  & $\frac{75}{151}$ & $75 $ &  $(A_4,D_{61})$  & $\frac{22}{25}$   & $122 $ & $(A_{80},D_{82})$  & $0$               & $3280$ \\
$(A_2,D_{76})$  & $\frac{8}{17}$   & $76 $ &  $(A_4,D_{62})$  & $\frac{124}{127}$ & $124 $ & $(A_{82},D_{84})$  & $0$               & $3444$ \\
$(A_2,D_{77})$  & $\frac{77}{155}$ & $77 $ &  $(A_4,D_{63})$  & $\frac{42}{43}$   & $126 $ & $(A_{84},D_{86})$  & $0$               & $3612$ \\
$(A_2,D_{78})$  & $\frac{78}{157}$ & $78 $ &  $(A_4,D_{64})$  & $\frac{128}{131}$ & $128 $ & $(A_{86},D_{88})$  & $0$               & $3784$ \\
$(A_2,D_{79})$  & $\frac{25}{53}$  & $79 $ &  $(A_4,D_{65})$  & $\frac{130}{133}$ & $130 $ & $(A_{88},D_{90})$  & $0$               & $3960$ \\
$(A_2,D_{80})$  & $\frac{80}{161}$ & $80 $ &  $(A_4,D_{66})$  & $\frac{8}{9}$     & $132 $ & $(A_{90},D_{92})$  & $0$               & $4140$ \\
$(A_2,D_{81})$  & $\frac{81}{163}$ & $81 $ &  $(A_4,D_{67})$  & $\frac{134}{137}$ & $134 $ & $(A_{92},D_{94})$  & $0$               & $4324$ \\
$(A_2,D_{82})$  & $\frac{26}{55}$  & $82 $ &  $(A_4,D_{68})$  & $\frac{136}{139}$ & $136 $ & $(A_{94},D_{96})$  & $0$               & $4512$ \\
$(A_2,D_{83})$  & $\frac{83}{167}$ & $83 $ &  $(A_4,D_{69})$  & $\frac{46}{47}$   & $138 $ & $(A_{96},D_{98})$  & $0$               & $4704$ \\
$(A_2,D_{84})$  & $\frac{84}{169}$ & $84 $ &  $(A_4,D_{70})$  & $\frac{140}{143}$ & $140 $ & $(A_{98},D_{100})$ & $0$               & $4900$ \\
$(A_2,D_{85})$  & $\frac{9}{19}$   & $85 $ &  $(A_4,D_{71})$  & $\frac{26}{29}$   & $142 $ &                    &                   &        \\
$(A_2,D_{86})$  & $\frac{86}{173}$ & $86 $ &  $(A_4,D_{72})$  & $\frac{48}{49}$   & $144 $ &                    &                   &                             
\end{longtable}
\end{center}

\begin{center}
 \renewcommand{\arraystretch}{1.25}
\begin{longtable}{c|c|c||c|c|c}
\caption{$(E_n,A_m)$ and $(E_n,D_m)$, with $n=6,7,8$ and $1\leq m\leq 100$.}\\
$(E_n,A_m)$  & $24(c-a)$ & $\text{rank}$ & $(E_n,D_m)$  & $24(c-a)$ & $\text{rank}$ \\
    \hline
\endhead
\label{tab:GGpNHC} 
$(E_6,A_1)$        & $\frac{3}{7}$     & $3   $ & $(E_6,D_3)$        & $0$               & $9   $ \\
$(E_6,A_3)$        & $0$               & $9   $ & $(E_6,D_4)$        & $0$               & $12  $ \\
$(E_6,A_4)$        & $\frac{12}{17}$   & $12  $ & $(E_8,D_3)$        & $\frac{12}{17}$   & $12  $ \\
$(E_6,A_6)$        & $\frac{18}{19}$   & $18  $ & $(E_8,D_4)$        & $0$               & $16  $ \\
$(E_6,A_7)$        & $\frac{3}{5}$     & $21  $ & $(E_8,D_6)$        & $0$               & $24  $ \\
$(E_7,A_2)$        & $\frac{5}{7}$     & $7   $ &                    &                   &        \\
$(E_7,A_4)$        & $\frac{14}{23}$   & $14  $ &                    &                   &        \\
$(E_7,A_6)$        & $\frac{21}{25}$   & $21  $ &                    &                   &        \\
$(E_7,A_8)$        & $0$               & $28  $ &                    &                   &        \\
$(E_8,A_1)$        & $\frac{1}{2}$     & $4   $ &                    &                   &        \\
$(E_8,A_2)$        & $\frac{8}{11}$    & $8   $ &                    &                   &        \\
$(E_8,A_3)$        & $\frac{12}{17}$   & $12  $ &                    &                   &        \\
$(E_8,A_5)$        & $0$               & $20  $ &                    &                   &        \\
$(E_8,A_6)$        & $\frac{24}{37}$   & $24  $ &                    &                   &        \\
$(E_8,A_9)$        & $0$               & $36  $ &                    &                   &        \\
$(E_8,A_{10})$     & $\frac{40}{41}$   & $40  $ &                    &                   &        \\
$(E_8,A_{11})$     & $\frac{4}{7}$     & $44  $ &                    &                   &        \\
$(E_8,A_{14})$     & $0$               & $56  $ &                    &                   &        \\
$(E_8,A_{19})$     & $\frac{4}{5}$     & $76  $ &                    &                   &        
\end{longtable}
\end{center}

\FloatBarrier

\bibliographystyle{ytphys}
\bibliography{ref}

\end{document}